\documentstyle[12pt,aasms4]{article}

\begin{document}

\title{Compact Nuclei in Galaxies at Moderate Redshift: I. Imaging and
Spectroscopy\footnote{Based on observations made with NASA/ESA
{\it Hubble Space Telescope}, which is operated by the Association of
Universities for Research in Astronomy, Inc., under contract with
NASA}}
\author{Vicki L. Sarajedini \footnote
{current address is UCO/Lick Observatory, University of California, Santa
Cruz, CA 95064}}
\affil{Steward Observatory, University of Arizona, Tucson, AZ 85721}
\authoremail{vicki@as.arizona.edu}
\author{Richard F. Green}
\affil{NOAO\footnote{The National Optical Astronomy Observatories
are operated by the Association of Universities for Research in
Astronomy, Inc., under Cooperative Agreement with the National
Science Foundation.}, P.O. Box 26732, Tucson, AZ 85726}
\authoremail{green@noao.edu}
\author{Richard E. Griffiths and Kavan Ratnatunga}
\affil{Carnegie Mellon University, Department of Physics,
5000 Forbes Ave., Pittsburgh, PA 15213-3890}
\authoremail{griffith@astro.phys.cmu.edu, kavan@astro.phys.cmu.edu}

\begin{abstract}

This study explores the space
density and properties of active galaxies to z$\simeq$0.8.
We have investigated the frequency and nature of unresolved
nuclei in galaxies at moderate redshift as indicators of nuclear
activity such as Active Galactic Nuclei (AGN) or starbursts.
Candidates are selected by fitting imaged galaxies
with multi-component models using maximum likelihood estimate techniques
to determine the best model fit.  We select those
galaxies requiring an unresolved, point source component in the galaxy
nucleus, in addition to a disk and/or bulge component, to adequately model
the galaxy light.

We have searched 70 WFPC2 images primarily from the Medium Deep Survey
for galaxies containing compact nuclei.
In our survey of 1033 galaxies, the fraction
containing an unresolved nuclear component $\geq$3\% of the
total galaxy light is 16$\pm$3\% corrected for incompleteness
and 9$\pm$1\% for nuclei $\geq$5\% of the galaxy light.
Spectroscopic redshifts have been obtained for 35 of our AGN/starburst
candidates and photometric redshifts are estimated to an accuracy of
$\sigma_z$$\simeq$0.1 for the remaining sample.
In this paper, the first of two for this study, we present
the selected HST imaged galaxies having unresolved nuclei and discuss the
selection procedure.  We also present the ground-based spectroscopy
for these galaxies as well as the photometric redshift predictions
for those galaxies without spectra.  

\end{abstract}


\section{Introduction}

Accurate knowledge of the luminosity function (LF) of active galactic
nuclei over a wide range of absolute magnitudes is necessary to
understand the nature and evolution of these objects.
The behavior of the LF as a function of
redshift especially at the faint end is crucial for understanding the 
manner in which AGN evolve.
There are several models for quasar evolution which can be
much better constrained with accurate knowledge of the shape
at the faint end.  Koo (1986) explains that many of the models
are difficult to discriminate when only bright quasars are included.
This is because the bright end LF shape is close to that of a 
power-law with a slope that is almost identical at all redshifts.

Additionally,
AGN are likely contributors to the X-ray background
and several studies have determined the contribution by bright
quasars to the diffuse background (e.g. Schmidt \& Green 1986).
The overall contribution of low luminosity AGNs (LLAGNs) to
the diffuse X-ray background has been a question of interest for some
time (Elvis et al. 1984; Koratkar et al. 1995).
An accurate understanding of the behavior of the faint AGN LF and its
evolution would aid in addressing this question.
Specifically, X-ray-to-optical flux ratios from
the {\it Einstein Medium Sensitivity Survey} for AGNs (Stocke et al. 1991)
allow for determination of the contribution of AGNs to the
soft X-ray background based upon their optical intensity.

The faint end of the AGN LF (M$_B$$\geq$-23) has been
determined locally using Seyfert galaxy nuclei which are considered to be the
intrinsically fainter counterparts of more distant, brighter QSOs (Cheng
et al. 1985; Huchra \& Burg 1992).
There have been many techniques employed to select Seyferts and
other galaxies with unusual activity.
Markarian et al. (1981) used the UV-excess technique to select
galaxies containing a very blue continuum.
The presence of emission lines (Salzer et al. 1989; 1995) has also been
used as an indicator of galactic activity.   
These techniques typically find that $\sim$5\% to 10\% of the 
selected galaxies host AGNs.

It is difficult, however, to obtain a Seyfert sample which is
free from biases.   
Seyferts selected by UV or X-ray excess are likely to favor type 
1 nuclei.  Selection based on IR properties is biased toward galaxies with high
star formation rates.
One of the most prevalent biases is toward
nuclei which dominate the light of the host galaxy.
Almost all Seyfert samples selected on the basis of spectroscopic
properties of the nucleus have this selection effect. 
This is true for Seyferts selected on the basis of UV excess 
or the presence of emission lines.
The Huchra \& Burg (1992) Seyfert sample from the CfA
redshift survey was obtained
through spectroscopic selection of galaxies based on the
presence of broad emission lines, indicative of Seyfert 1 activity,
or emission line flux ratios indicating Seyfert 2 or LINER activity.
The presence of such emission lines differentiates between a
thermal and non-thermal energy source in the nucleus.  Again, this selection
technique requires that the nucleus dominate the galaxy
light or have adequate spatial resolution for subtraction of the galaxy light
from the nucleus.  Maiolino \& Rieke (1995) show that the nuclear
luminosities of CfA Seyferts are closely related to the 
integrated galaxy luminosities
indicating a bias towards brighter, nucleus dominated Seyferts.
For this reason, the local AGN LF does not extend below M$_B$$\simeq$-17.5.
 
Likewise, little is currently known about how the faint end of the LF
evolves.  Even at modest redshifts, LLAGNs become virtually
impossible to observe from the ground.  In ground-based images
the unresolved nuclei cannot be distinguished from central regions of
enhanced star formation or finite central density cusps of spheroidal
components.  This problem, however, is greatly diminished by the Hubble 
Space Telescope's unique high resolution imaging capabilities.

The Medium Deep Survey (MDS) (Griffiths et al. 1994) yielded 2 to 4
parallel WFPC2 exposures per week where reliable classification was
possible for galaxies down to V$\sim$ 23.5 mag.
This survey provides an ideal sample of distant field galaxies for which
morphology and galaxy light profiles can be studied for the first time at
sub-kiloparsec resolution.  Typical galaxy redshifts are at z$\leq$0.6 (Mutz
et al. 1994).  The set of Cycle 4, 5 and 6
images consists of $\sim$150 fields with both V(F606W) and I(F814W)
exposures.

This database provides a unique opportunity to search for morphological
evidence of AGN or other compact nuclear activity such as starbursts.
The nuclear activity of Seyfert galaxies manifests itself morphologically
as an unresolved stellar-like point source in the nucleus of the galaxy.
This is due to the fact that most of the emission is originating from a small
region typically a few parsecs in diameter.  For Seyfert 2 nuclei the
emission is probably originating over the broader narrow-line region but
is still highly concentrated at the nucleus having nuclear
FWHM$\lesssim$200 pc (Nelson et al. 1996).
The physical size of an unresolved region in a WFPC2 image varies
with redshift according to Figure 1.  The unresolved region
at redshifts less than 0.8 obviously encompasses that for Seyfert-like
nuclei.  Other enhancements of this size in a galaxy profile would
include starburst regions and nuclear HII regions.  Nuclear starbursts
can have sizes of a few hundred parsecs (e.g. Weedman et al. 1981
for NGC 7714) and nuclear HII regions may be even smaller.
These stellar enhancements in a galaxy light profile would appear unresolved
over the redshift range of interest (0.2$\lesssim$z$\lesssim$0.8) unless
H$_o$ is very large.
The typical late-type spiral bulge, however, has a diameter
of $\sim$1 kpc (Boroson 1981) which is resolved in the HST images.

In this study, all galaxies
in 70 WFPC2 fields have been modeled to search for unresolved nuclei
likely to be AGN or compact regions of
star formation, i.e. starbursts.  The galaxy modeling is based on maximum
likelihood estimates used to extract quantitative morphological and
structural parameters of the faint galaxy images.  All galaxies to
I$\lesssim$21.5 in 64 MDS fields and 5 Groth survey strip fields
(Groth et al. 1998)
as well as the Hubble Deep Field to I$\lesssim$23.5 (Williams et al.
1996) have been modeled
in this way to reveal an unresolved nuclear component when present.
The selection of AGN using this technique results in a
magnitude-limited sample which is
not biased towards galaxies dominated by the nucleus.  Many other selection
techniques, such as spectroscopic selection, require the nucleus to be
the dominant galaxy component.  In this way, we probe the intrinisically
fainter population of AGN and starbursts out to intermediate redshifts
for the first look at how this population of objects evolves.

In this first of two papers, we investigate the process of selecting
galaxies containing a nuclear component.  The details
of processing HST images and issues concerning incompleteness in the sample are
also discussed.  Additionally, we discuss the ground-based spectroscopic
follow-up for the selected galaxies and present the spectra.  We also
outline a photometric redshift technique used to estimate redshifts
for selected galaxies without spectroscopic redshifts.
The analysis of the results from this study are presented in the
second paper in this series, Sarajedini et al. (1998) (hereafter referred
to as Paper II).

\section{Selection and Reduction of Survey Fields}

The Medium Deep Survey (MDS) is composed of images obtained in parallel
while HST observes a primary target with one of the
other instruments.  The fields were observed in the I (F814W) and V
(F606W) filters primarily, with some fields also being imaged with the
B (F450W) filter.  The data used for this study are comprised entirely
of post-refurbishment HST images obtained with the WFPC2 camera from
1994, January to 1996, July.  The fields were chosen to lie at a
range of high galactic latitudes away from known nearby galaxy clusters
and other bright objects.

During this timeframe, the MDS obtained images of
209 survey fields.  From these data the fields for this study were chosen
to have images in both the V and I filters so that color information
would be available.  Each field contained at least one exposure in both
V and I with total exposure times ranging from 2000 to 23100 seconds in I and
300 to 16500 seconds in V with a median exposure time of $\sim$5000 seconds.
The number of exposures per field ranges from 1 to 12 with a median number
of exposures at $\sim$3.  Table 1 lists the field name, RA and DEC,
Galactic latitude, number of exposures and total exposure time in seconds in
the V and I filters for each field selected from the MDS for this
study.  Also included in this table are the six additional non-MDS HST fields
used in this study; five fields from the Groth Survey
strip (Groth et al. 1998) and the Hubble Deep Field (Williams et al.
1996).

Many of the details of the calibration process can be found in
Ratnatunga et al. (1998)\footnote{A draft of this paper is available
at http://astro.phys.cmu.edu/mds/mle/index.html} and references therein.  
Here we will summarize
the basic information regarding calibration and reduction of the HST
images. The WFPC2 images are calibrated using the STScI static mask, super-Bias,
super-Dark and flat field calibration files.  These were created by STScI
to calibrate the Hubble Deep Field. 
Flux in fluctuating warm pixels are corrected by hot pixel tables from 
STScI for the period of observation.  The correction is made to ensure 
that the noise from any residual warm current is smaller than the read 
noise. Hot pixels which cannot be corrected to that accuracy are rejected.

A corrected version of the IRAF/STSDAS COMBINE task is used to combine the
images and generate a ``sigma" image from the statistical errors.  
See Ratnatunga et al. (1994) for a detailed discussion of the stacking 
procedure and the corrections made to the statistical error determination
in the COMBINE algorithm.
The sigma image is an estimate of the rms error of every pixel in the
calibrated stack and reflects the cosmic ray pixels rejected in the stacking
procedure, the flat field response, and any subtracted background
sky gradient or scattered Earth light.
The amount of shifting necessary between images is determined from ``jitter
files", i.e. the HST aspect solution for the WFPC2.  Exposures are stacked
with shifts corresponding to the nearest integer number of pixels.
We allow a maximum 50 pixel shift.  The orientation usually
remains constant although a maximum difference in rotation of less than
0.03 degrees is allowed, ensuring a 1-to-1 mapping of the pixels.

Cosmic rays affect about 7 pixels per second per CCD chip during the length
of the exposure time. 
When three or more exposures in the same filter are
available along the same direction with the same orientation, cosmic rays
can be effectively removed by stacking the images with a 3-sigma clip.
When only two images were available, the stacking operation will leave
the fainter cosmic rays on the output image which could be mistakenly
detected as faint objects.  Cosmic rays can be rejected
in images with only one exposure per filter when at least one exposure has been
obtained in the other filter.  The probability of a cosmic ray hitting 
the nucleus of one galaxy in the WF chip is appx 20% based on typical 
values for the exposure time and the number of galaxies per chip.
This corresponds to $\sim$4\% of the galaxies in our study.
We are careful, however,
to check that none of our detections of nuclear point sources are based
on detections in one filter when only one exposure was obtained in that
filter.

Once the images are stacked, objects are located independently
on each image using an algorithm developed for HST-WFPC data.  It is
based on finding local maxima and mapping nearby pixels, which are
significantly above the noise, to the central object.
Objects near the edge or only partially contained on the CCD chip
are still be detected.
After the initial finding algorithm has been applied, we have the 
only interactive part of the operation.  The image is examined to confirm
that it satisfies being part of the Medium Deep Survey and does not contain
parts of bright, resolved galaxy fields or globular clusters.
Any significant gradient in the sky background, caused by a nearby
bright object or scattered Earth light, is subtracted from the stacked
image and the object detection algorithm is re-executed.
Ghost images or extremities of bright stellar defraction spikes which 
have been spuriously detected as objects are deleted during this
step.  Object detections in the two filters are matched by software to create
a single catalog for each image so that corresponding
pixels in different filters are associated with the same object.

\section{Galaxy Fitting Software}

The software for modeling the galaxy light profile has been
developed by K. Ratnatunga (Ratnatunga et al. 1998).  The empirical model used
is scale free and axisymmetric with possible galaxy component profiles
consisting of an exponential disk, an r$^{1/4}$ bulge, and an exp(-r$^2$)
Gaussian profile for a point source.
Disk and bulge models have long been used to fit the broad distribution of
normal galaxies (e.g. Kormendy 1977) and have been shown to successfully
model moderately redshifted galaxies (Schade et al. 1995).  A point source
component is added to represent an unresolved nuclear AGN or starburst
region.   

For each galaxy imaged in the HST WFPC2 field, a contour is defined around
the object by selecting those pixels which are greater than
one-sigma above the estimated local sky level.  The signal-to-noise 
ratio of these pixels are then integrated to produce
a good measure of the information content of the image.  The decimal
logarithm of this integral, called SNRIL, is linearly correlated with
the galaxy magnitude as shown in Figure 3 (see Ratnatunga et al. 1998
for a detailed explanation of this parameter).  SNRIL behaves mathematically 
like a signal-to-noise ratio with exposure time.
The pixels within the one-sigma contour are
used to estimate each galaxy's center, magnitude, size, orientation,
and axis ratio using simple moments of the flux above the mean.
An elliptical annular region around the object is selected to define the mean
sky background to 0.5\% accuracy.

The maximum likelihood parameter estimation starts by using these moments
to estimate initially the model parameters.  The software creates a model
image of the object, convolves this model with the WFPC2 stellar Point
Spread Function (PSF) and compares it with the observations
within the selected region.
The likelihood function is defined as the probability for each
model pixel value with respect to the observed pixel value and its error
distribution.  The function is evaluated as the integral sum of the logarithm
of these probabilities. The likelihood function is maximized
using the modified IMSL minimization routine (Ratnatunga \& Casertano 1991).

The number and choice of parameters fitted to an image is clearly
important and defines the light profile model.  In this study,
each galaxy is fit down to a limiting
SNRIL with 3 different model choices to determine the best fit
to the data.  The first model is a 2-component disk+bulge model.
This version of the fitting software is flexibile so that a pure disk,
pure bulge, pure Gaussian point source, or bulge+disk model can be
chosen as the best fit to the data.
The parameters which can be fit in this model are: 1) sky magnitude,
2) x-position, 3) y-position, 4) total magnitude, 5) half-light radius
of the bulge+disk model
(the radius within which half the light of the unconvolved model would
be contained if it were radially symmetric),
6) disk axis ratio, 7) orientation of the galaxy, 8) bulge/(disk+bulge)
luminosity ratio, 9) bulge axis ratio and 10) bulge/disk half-light radius
ratio.

The first step utilizes a special quick mode of the minimization
routine which attempts a 10-parameter disk+bulge fit.  It does not check
for full convergence but does reach a
point in the multi-dimensional parameter space which is close enough to
the final answer to investigate the likelihood function and make some
intelligent decisions.  If the half-light radius of the object is
less than one pixel, a 5-parameter Gaussian model is tested to see if
the object is point-like.  We find that a 5-parameter model is well contrained
to a half-light radius of $\sim$0.2 pixel.  If the point source model 
does not improve the fit, a single-component disk or bulge model 
is attempted.  Significant
improvement in a model is achieved if the difference in the likelihood
function values (the likelihood ratio) is greater than 6.0.  This
likelihood ratio (LR)
is used throughout the modeling process to determine if significant
improvements are made with new model fits.  The 2-component bulge+disk
model is then checked to see if it improves upon the single model
bulge or disk fit.  The center of the galaxy model is then divided into
smaller sub-pixels to see whether a high resolution center will change 
the likelihood function significantly.  The best fit parameters for 
the chosen model are determined through an iterative process using
the model IMSL minimization routine.

The software creates a FITS data image for each object (Figure 2).  The
image is a grid with a row of 7 images for each filter.
Each row contains: 1) the observed 64 x 64 pixel region around
the imaged galaxy, 2) the selected pixels identified as part of the
galaxy, 3) the PSF convolved
maximum likelihood model image, 4) the maximum likelihood model image, 5)
the residual image (PSF convolved model subtracted from the real image), 
6) the sigma image, 7) the object mask image.

A second modeling option was then applied which
allows for a disk+point source fit.  The fitting procedure used
is the same as the 2 component disk+bulge version.
The main difference is that the second component is
a Gaussian point source instead of an r$^{1/4}$ bulge component.
The parameters for this model are: 1) sky magnitude,
2) x-position, 3) y-position, 4) total magnitude, 5) half-light radius
of the disk,
6) disk axis ratio, 7) orientation of the galaxy, 8)
point source/(disk+point source) luminosity ratio, 9) x difference in position
between the disk and point source and
10) y difference in position between the disk and point source.

The point source is allowed to be at a different origin from that of
the disk component.  This technique avoids the problem of not converging
on a real point source because it is not at the precise center of the disk.
Later in the selection process, restrictions can be placed on how far
off-center the point source nucleus can be and still be considered a
``nucleus".
The half-light radius of the Gaussian component representing the point source
is 0.02 arcsec; the radius before convolution with the point spread
function.  This is
consistent with that measured for stars in the WFPC2 fields and is
essentially a measurement of the telescope ``jitter''.  The axis
ratio for this component is 1.0.

A third modeling option allows for a 3-component
disk+bulge+point source fit to the galaxy light profile.
This model is somewhat more complicated than the previous versions
due to the third component.  The parameters fit for this model
are: 1) sky magnitude,
2) x-position, 3) y-position, 4) total magnitude, 5) half-light radius
of the disk+bulge portion of the model without regard to light from the
point source nucleus,
6) disk axis ratio, 7) orientation of the galaxy, 8)
bulge/(disk+bulge+point source) luminosity ratio,
9) bulge axis ratio,
10) bulge/(bulge+disk) half-light radius ratio,
11) point source/(disk+bulge+point source) luminosity ratio,
12) x difference in position,
13) y difference in position.
Parameters 12 and 13 reflect the fact that
the point source is allowed to have a different origin from the
remainder of the galaxy model.  Again, the half-light radius of
the Gaussian component is 0.02 arcsec and the axis
ratio is 1.0.

\section{Selection of Galaxies Containing Nuclear Point Sources}

Each galaxy is fitted with all three versions of the fitting software.
The model with the lowest likelihood function value indicates
the best fit or the closest model to the real data.
Those galaxies best fit with either the disk+point source model or
the disk+bulge+point source model are initially selected as nuclear
point source candidates.
It is important, however, to understand whether or not the model
containing a point source component is substantially better
than one without the point source component.  This is important since
a model with more parameters is more likely to have a lower likelihood
function value and be chosen as the best fit.  In the following sections
we describe various simulations and tests designed to determine the
criteria for selecting a galaxy containing a nuclear point source.

\subsection{Spurious Point Source Detections in Simulated Galaxies}

To determine the uniqueness of a fit with a galaxy model containing
a point source, simulated galaxies with no point source component
were fit in the same way as galaxies in the survey.  These simulated
galaxies have a range of magnitudes, signal-to-noise ratios, bulge-to-total
light ratios and half
light radii and were produced in both V and I images where the V magnitude
of each galaxy is one magnitude fainter than the I magnitude.
Each of the 1152 galaxies in this simulation were fit with a
disk+bulge model, disk+point source
model, and a disk+bulge+point source model.

The model with the lowest likelihood function
value is chosen as the best fit.  The simulations are used to determine
what minimum likelihood ratio must be met by the model
containing a point source over the model not containing a point source.
In 30\% of the cases, the simulated galaxies without a 
nuclear point source component, were found to have a best fit 
model which {\it did} include a point source nucleus.  
The likelihood ratio between
the non-point source component model and the point source component model
for these fits determines the range in likelihood ratios for spurious
point source detections.

The likelihood ratio (LR) values for the simulated galaxies
where the point source component model gave the best fit are found
to occur at small point source-to-total values.  This simply means
that the spurious detections tend to be faint nuclei ($\lesssim$1\% of the
total galaxy luminosity).  Applying a 1\% point source-to-total luminosity 
ratio cut-off to these data eliminates 75\% of the spurious detections.
The vast majority of the remaining spurious detections were found to
have LR values $\leq$50.  We remind the reader that the LR value is
a measure of the difference in the likelihood function values for
two different model fits.  The larger the LR, the greater the difference
in the two model fits being compared.  In this case, we are comparing
the model fit including a point source to one that does not include a
point source and the LR reflects the degree to which the model including
a point source component is a better fit.  
   
These simulations indicate that choosing a best fit galaxy model 
where the point source-to-total luminosity
ratio is greater than 1\% and the LR value is greater than 50 will avoid almost
all spurious point source detections.  Of the 1152 galaxies in this
simulation, only 4 would still be selected after applying these criteria
resulting in 0.3\% of fitted galaxies which could contain spurious
point source detections.

\subsection{Measurement of Detected Point Source Nuclei}

Not only is it important to determine how often spurious detections
will occur in our galaxy sample using simulated galaxies, we also
need to define when and how often inaccurately measured point source
nuclei are detected in the sample.  For the purposes of determining
completeness and to quantify the frequency of inaccurately measured point
source nuclei, a partially simulated galaxy dataset was produced.
The dataset consists of real galaxy images from several MDS fields where
simulated point sources of various magnitudes were added at the nucleus.
A total of 98 galaxies from 10 MDS fields were used in this Monte-Carlo
type simulation.
Each galaxy image was reproduced 10 times and had added to it point sources
of apparent magnitudes evenly distributed between I=21.5 to 26.0 and
V=22.5 to 27.0.  A total of 1960 simulated galaxy images were created in
this way (98 galaxies x 10 point source magnitudes x 2 filters (V and I)).

The galaxies with simulated point source nuclei had a variety of
bulge-to-total ratios,
apparent magnitudes, and SNRIL values (Figure 3).
The 2-component disk+bulge model, the 2-component
disk+point source model, and the 3-component
disk+bulge+point source model were applied to each of the simulated galaxies.
The best fit model was determined as that with the lowest likelihood function
value.  To be selected as a galaxy containing a nuclear point source
component, we required that the point source be at least 1\% of the
galaxy light and that the likelihood ratio for the model containing
the point source component be lower by a value of 50 than the model
without a point source component.  These criteria were applied to
avoid spurious point source detections as described above and will also
be applied to the real galaxy sample.
To provide a check on our
likelihood ratio requirement, we examined the increase in detected real
point sources in this set of simulations as compared to the increase
in spurious detections made with a lower likelihood ratio requirement.
We find that the increase in detected real point sources is negligible
($\sim$few\%) compared to the doubling of spurious detections when
the likelihood ratio is lowered from 50 to 30 and therefore maintain
the current likelihood ratio requirements for detecting nuclear point
sources.

Based on the results from this simulation, we determined
how well the input nuclear point source is detected and
how accurately its magnitude is measured.
Figure 4 shows how the measured magnitude of every detected
point source in the simulated data compares with its input
V and I magnitudes.
Although many of the points fall near the input magnitude values,
$\sim$8\% are measured brighter than the input magnitude by more
than 2 magnitudes.
Careful examination of the model fits for these cases indicate
that they are galaxies containing significant bulge contributions
($\ge$50\% of the total galaxy light).
The bulge appears to often confuse the magnitude measurement of
the point source component.
For this reason, it is necessary to set additional selection criteria
for point source detections to avoid inaccurately measuring point source
magnitudes.

Because our data contain information in both the V and I filters,
we can compare the model in the two different filters and note
any oddities.  First, many of the poorly measured point source nuclei
detected in one filter were undetected in the other filter.
This result is not unacceptable unless the measurement of the bulge is very
different in the two filters.  It is clear that the bulge is affecting
the point source measurement in the case where a large bulge contribution
is measured in one filter while a point source is detected with no bulge
component in the other filter.  We therefore try to remove from our sample
those point sources measured to be too bright due to the contamination by 
bulge light in the point source component.  We note all the cases in the
simulated data where the bulge is measured to be $\ge$25\% in one filter
and a point source with no bulge component is measured in the other filter.
In the simulated dataset, 69 point sources are detected in galaxies fulfilling
this criterion.  These point sources were neither preferentially 
bluer or redder than the underlying galaxy.
Removal of these points eliminates $\sim$80\%
of the point source nuclei in Figure 4 measured as too bright.

These simulations also show that if the galaxy contains a
large bulge component in both filters and the point source is the same
color in (V-I) as the galaxy, it is often measuring the point source magnitude
incorrectly.  We also know that many elliptical galaxies have ``cuspy"
luminosity profiles (Lauer et al. 1995) which can easily be mistaken for
unresolved nuclei in moderately redshifted galaxies.  If a small point
source is placed in these galaxies in our simulations, a bright point
source may be measured due to the inclusion of the cuspy light from the
galaxy itself.  In these cases, we would expect the color of the
point source to be the same as the galaxy, within the errors.  We
find 77 instances in this simulation where the point source color
and the total galaxy color are the same.

Finally, the matrix solution of the light profile model is examined for 
each galaxy in which a point source is detected.  Occasionally, the model
does not converge which is easily identified by rows of zero values 
in the matrix solution. A total of 56 galaxies
best fit with a point source model were removed due to the
non-convergence of the point source component in the model.

Figure 5 shows the measured magnitude vs. the input magnitude
for the remaining 597 selected point source galaxies after removing the
subsets of systematically inaccurate and non-converging point source
models described above.  These additional
selection criteria have successfully removed the majority of poorly
measured point source nuclei.  The solid line represents the one-to-one
relationship between input and output magnitudes.  The dashed vertical
errorbars are the standard deviation error in the real point source nucleus
magnitude as a function of the measured magnitude based on the spread
of the points in the figure.
We show in Figure 6 histograms of mag$_{obs}$ - mag$_{real}$ for
point sources at 23$\pm$0.25 and 24$\pm$0.25 magnitudes.  The dashed
line is the normalized Gaussian distribution for the calculated $\sigma$$_{mag}$
of each distribution.  We note that the distribution is narrower than
the normalized Gaussian distribution and therefore the adopted errors
are conservative and may overestimate the actual error in
magnitude for many of the measured point sources.
A least squares fit to these 1$\sigma$ errors as a function of observed
magnitude yields the following equation,
\begin{equation}
\sigma_{mag}=0.324-0.0685\times{(23-mag)}+0.0075\times{(23-mag)^2}
\end{equation}
which gives the error in magnitude
determination as a function of the measured magnitude.
This equation is later applied to the real point source nuclei detected
in the sample to obtain an error estimator for the true magnitude.

\section{Accuracy of the Host Galaxy Bulge Measurements}

Figure 5 shows that the point source nuclei can be accurately measured
to $\pm$0.40 magnitude for faint (24th magnitude) nuclei
and $\pm$0.26 magnitude for bright (22nd magnitude) nuclei.
In spite of fairly accurate nuclear magnitude determinations, we find
that the bulge component is often misfit in our simulations.
Figure 7a reveals how the bulge-to-bulge+disk luminosity ratio
(B/B+D) in the simulation
galaxies differs from that measured for these galaxies once a point
source nucleus has been added and detected.
It is clear from the numerous points at (B/B+D){$_{measured}$} = 0 that 
the bulge component is often not measured at all
when a point source is present.  This is most likely caused by the tendency
of the modeling software to chose a model with fewer parameters
and therefore fewer components as the best fit unless the fit is greatly
improved by additional parameters.  When a prominent point
source is present, even a large bulge component can go undetected.

Figure 7b reveals how the measured bulge-to-bulge+disk ratio (B/B+D)
varies with the point source-to-total luminosity ratio.  
For point source nuclei greater
than $\simeq$20\% of the total galaxy light, the bulge is almost always
undetected ((B/B+D){$_{measured}$} = 0) and appears to be dominated by the
point source component.  Only where the point source is below 20\%
of the galaxy light can we adequetely measure the bulge component of
the galaxy in addition to the point source component.

The results of these simulations suggest
that although the bulge is often measured incorrectly
when a point source is detected, the magnitude of the detected point
source is correct within the determined errors.   For this reason,
our main goals of determining nuclear magnitudes and colors are
unaffected by these errors in measuring the bulge contribution.
We discuss in Paper II how these results effect the determination
of the host galaxy types and colors and attempt to statistically
correct for errors in the bulge measurement in our sample.

\section{Completeness}

To determine the level of completeness in this survey, a Monte-Carlo
simulation was performed. The dataset used for this experiment is that
described above, where real galaxy images are combined
with point source nuclei of varying magnitudes.  The
fitting software is then used to determine the best fit model to the data.
After applying the criteria to ensure accurately measured, real nuclear point
source components, we determine the number of galaxies
in which the unresolved nucleus is detected.  Since each galaxy
in the simulation contains a point source for which the input apparent
magnitude is known,
the level of completeness in detecting nuclei as a function of the
point source brightness can be determined.

To eliminate spurious point source detections, each detected point source
must be at least 1\% of the total galaxy light based on modeling of
simulated galaxies described previously in section 4.1.  For this reason, 
very faint
nuclei are not detected in bright galaxies but a nucleus of the same
magnitude might be detected in a fainter galaxy if it is above the
1\% limit.  Because of the related nature of detected nuclei and the magnitude
of the host galaxy, we cannot address the issue of completeness in terms of
limiting magnitudes.  Instead, the level of completeness in detecting point
source nuclei is based on the percentage of galaxy light that the point source
comprises.

We begin by estimating completeness for those galaxies in which no bulge
component is detected.  These galaxies are adequately
modeled in our sample with a pure exponential disk.  The galaxies,
which contain simulated point source nuclei of varying point source-to-total
luminosity ratios, were fitted with the fitting software and those containing
accurately measured point source components were selected.
Figure 8a shows a histogram of the number of galaxies
in the simulation
for which there was no input bulge component as a function of the input
point source-to-total luminosity ratio (solid line).  The hatched region
represents the histogram of galaxies in which the point
source nucleus was detected and no bulge component was measured, consistent
with the input bulge parameter value.  Figure 8b is the fraction of
galaxies as a function of input point source-to-total luminosity ratio
in which the point source was accurately modeled.  The error bars
are the Poisson 1$\sigma$ error based on the number of points in each
bin.  These fractional values can be applied to those galaxies in the real
survey sample containing no measured bulge component to determine the
level of completeness in detecting point sources.

Estimating completeness for the galaxies containing a bulge component
is somewhat more complicated.  For the rest of this discussion we define
the bulge component in terms of bulge-to-bulge+disk (B/B+D) luminosity ratio
so that a galaxy with no disk has a luminosity ratio of 1.0.
Figure 7 indicates that when a point
source is detected, the
galaxy B/B+D may be quite inaccurate, especially if the point source-to-total
luminosity ratio is greater than 20\%.  For this reason, to estimate
completeness for the galaxies in our sample containing some bulge component,
it is best to separate our simulation results into bins of different
{\it input} B/B+D values as opposed to using the measured B/B+D values.
This is consistent with the null hypothesis approach we have adopted
for modeling the galaxy light profile: we assume that each galaxy does not
contain a point source and the galaxy light profile model
must be significantly improved with the point source component for the
nucleus to be detected.

Figure 9 is plotted similarly to Figure 8.
The three left panels show the histogram of galaxies in the simulation
(solid line) and those for which the point source nucleus was detected
(hatched region) for different input B/B+D
galaxy types.  The Figure 9a represents galaxies
with 0$<$(B/B+D)$\leq$0.4, 9b is for galaxies with
0.4$<$(B/B+D)$\leq$0.8 and 9c is for 0.8$<$(B/B+D)$\leq$1.0.
The adjacent panels on the right indicate the percentage
of point sources detected as a function of the point
source-to-total luminosity ratio and is an estimate of the completeness.
These percentage estimates can be applied to the real sample galaxy
B/B+D distribution to estimate completeness as a function of point source
-to-total luminosity ratio for the entire survey.  An estimate of the
total completeness in detecting point source nuclei for the survey data
is calculated in the next section based upon the outcome of this
simulation.

\section{Application to the Survey Data}

\subsection{Determining the Limiting Signal-to-Noise for Each MDS Field}

Every galaxy in each of the fields listed in Table 1 was fitted
down to an apparent magnitude limit corresponding to a SNRIL value of 2.5.
This limit is reached at an integrated galaxy magnitude of I$\simeq$22.0 
for the average MDS field but varies with exposure time as signal-to-noise.
As explained in section 3, SNRIL is the integrated signal-to-noise ratio 
of galaxy pixels above one-sigma of the sky.  The decimal logarithm of
this value is the parameter SNRIL.  The SNRIL value of 2.5 corresponds
to the signal-to-noise level required for accurate fitting of the simple
disk+bulge model (Ratnatunga et al. 1998).  For
the model requiring an additional point source component, the SNRIL
fitting limit should be higher.

To determine the SNRIL limit for the data, we again use the
simulated dataset consisting of real galaxies with added simulated
point source nuclei.  As shown in Figure 3, these galaxies
cover a range of SNRIL values.  These data can be used to determine
at what SNRIL value for the galaxy is the point source nucleus no
longer detected.  Figure 10 is the histogram of the measured SNRIL
values for the galaxies in this simulation.  To avoid some of the
incompleteness based on faint point sources as described in the preceeding
section, this histogram includes only points
measuring greater than 3\% of the total galaxy light. In Figure 10a and
10b the solid line is the total number of galaxies as a function of
SNRIL and the hatched
region is the number of galaxies where the added point source component
was detected.  The solid line in Figure 10a represents all galaxies
in the simulation
where the galaxy contained no initial bulge component and the hatched
region is the number of galaxies in this set where the 2-component
disk+point source model was the best fit.  Figure 10b represents all
of the galaxies which did contain an initial bulge component.  The
hatched region represents the galaxies in this set best fit with
a 2-component disk+point source model or a 3-component disk+bulge+point
source model.  The cross hatched region represents those galaxies for
which the 3-component model was the best fit.

It is clear that the 2-component disk+point source
model detects point sources in galaxies with lower SNRIL values
than the 3-component model.  This result
is expected since this model requires fewer fitted parameters than
the 3-component model.  The 3-component model is much more
incomplete overall than the 2-component disk+point source model.
The 3-component model fits some galaxies down to SNRIL=3.0 while the
2-component disk+point source fits some down to SNRIL=2.7.
To ensure that all galaxies in the sample can be fit with either the
2 or 3-component model, the limiting SNRIL value required for galaxies
to be included in this study is 3.0.  Setting the SNRIL value lower
would introduce a bias against detecting point source nuclei in galaxies
with significant bulge components and SNRIL$<$3.0.

The total number of galaxies in each of the 70 fields which have
SNRIL values greater than or equal to 3.0 in one or both filters
is 1033, an average of 13.5 galaxies per typical WFPC2 field (106 galaxies
lie above this limit in the HDF).  The last three columns of Table 1 show
the limiting apparent V magnitude and I magnitude corresponding
to this SNRIL value
in each field and the number of galaxies in each I band field above this
limiting value.

\subsection{Selected Galaxies Containing Nuclear Point Sources}

The output image containing the model and residual for every galaxy in the
survey was visually inspected to ensure that the model is correct
and no obvious errors have occurred in the fitting process.
The best fit model is then determined as the one with the lowest likelihood
ratio value.
The criteria for selecting a galaxy where the best model fit
contains a point source component have been discussed in detail
in the preceding sections.  Initially, the model must be the best
fit, having a likelihood ratio between the point source and non-point source
model greater than 50.  The point source component must also comprise
at least 1\% of the total galaxy light to avoid spurious point source
detections.  After applying these criteria to the data,
the galaxies in each field which are best fit with an additional point source
component are re-examined to determine if the point source location is
near the nucleus of the host galaxy.  This procedure is done to avoid
fitting bright
knots of star formation in the arms and disks of spiral galaxies or other
noisy features.  The point source is usually within $\simeq$0.2$\arcsec$
(2 pixels in the WF chip) of the host galaxy center.  Some, however, lie
further from the galaxy center if the galaxy is very asymmetric.
Occasionally, a galaxy is so irregular that it is difficult to determine
the location of the nucleus and these galaxies were never selected as having
point source nuclei because of this difficulty.  A total of 10 galaxies
in the Hubble Deep Field and 20 in the remaining 69 fields fall into this
latter catagory.

We next apply the criteria for accurately measured nuclear point
source components determined from the simulations in section 4.2.
The following is a summary of the number of galaxies which contained
a point source component in their best fit model but were rejected based
on these criteria.
Eight galaxies were removed because of non-convergence of the
point source component in the model.  Twenty-one were removed because the
point source was detected in one filter while the other filter measured
no point source (or a much smaller one) and the bulge/total luminosity ratio
was greater by 0.25 than that measured in the other filter.
Forty-nine were removed because the galaxy appeared
to be elliptical-like (bulge-to-total $\geq$0.8)
and the point source component had the
same V-I color as the galaxy, indicating that the measured point source is
likely a cuspy portion of the bulge.

After removing these galaxies from the selected sample
we have a total of 8 galaxies in the Hubble Deep Field and 93 in the
remaining 69 WFPC2 fields which are best fit with a galaxy model
containing a nuclear point source component and meet all of the
criteria described above.  This selection results in direct detection
of 7.5\% $\pm$2.7\%
of the galaxies in
the Hubble Deep Field and 10.0\% $\pm$1\% of the galaxies in the
remaining fields with nuclear point source components.
Seventy-two of the point source nuclei were detected in both
the V and I filter images, 8 were detected in the V image alone, and
21 were detected in the I image alone.  The typical reason for
non-detection in one of the filters is that the nucleus is too faint in that
filter image.  However, in six cases the point source component did not
converge properly in the other filter and the magnitude was therefore
considered unreliable.  The fraction of selected galaxies containing point 
source nuclei detected in the PC chip is appx 7% (7 galaxies) which is 
the same as that expected based on the difference in surface area between 
the PC and the 3 WF chips. 

Figure 11 is a gray-scale I band image of the selected galaxies.
The images are scaled in nuclear flux and are sorted in
decsending order of point source-to-total
luminosity ratio so that the galaxies with the brightest relative
point source nuclei are first.
The spiral structure in many of the larger galaxies is clearly visible.
Some of the nuclei are too faint to be detected by eye in this image; although,
it is useful for noting the range of sizes and galaxy types of the host 
galaxies.
Figure 12 shows the point source magnitude vs. the
integrated magnitude of the galaxy in I (a) and V (b).  The unresolved
nuclear point sources range in magnitude from 27$\lesssim$I$\lesssim$21.
The faintest galaxies (fainter than I$\simeq$21.5) are from the HDF.
The error in the point source magnitude has been adjusted to reflect
the expected error based on simulations in section 4.2.  Most of the nuclei
(87\%) comprise less than 20\% of the total galaxy light with 59\% comprising
less than 5\%.

Table 2 lists the important fitted parameters for the 101
galaxies containing nuclear point sources.  The table lists
the ID number (1), the object name (2),
the I magnitude of the galaxy model (3) and its error (4),
natural log of the I filter half-light radius in arcseconds (5)
and its error (6),
the V-I galaxy color (7) and its error (8),
the I Bulge-to-Total measurement (9),
the I Point source-to-Total measurement (10),
the V Point source-to-Total measurement (11),
and the V-I color of the point source (12) with its error (13).
The error in the point source color reflects the empirically
determined error of the point source magnitude from the simulations.
The object coordinates are given in Table 4.

The completeness in detecting point source galaxies can be determined
based on the simulations of the previous section.  Figures 8 and 9
give the completeness as a function of the point source-to-total luminosity
ratio for galaxies with various B/B+D measurements.  These completeness
estimates can be combined to determine the overall completeness of the
survey in detecting point source nuclei as a function of the point
source-to-total luminosity ratio.  To do this, we weight the completeness
estimates by the number of galaxies in the total survey with the
corresponding B/B+D measurement.  Of the 1033 galaxies in the survey,
282 have B/B+D=0, 409 have 0$<$(B/B+D)$\leq$0.4, 194 have
0.4$<$(B/B+D)$\leq$0.8, and 148 have 0.8$<$(B/B+D)$\leq$1.0.  Combining the
completeness estimates in Figures 8 and 9 weighted by the numbers
of survey galaxies in each bin yields the overall completeness estimate
for the survey illustrated in Figure 13.  For nuclei comprising
only a few percent of the galaxy light, we are $\sim$25\% complete
in detecting these over the full range of galaxy types in our sample.
For nuclei greater than 10\% of the galaxy light, we are $\gtrsim$60\%
complete.  Most of the incompleteness stems from the inability to
detect faint nuclei in galaxies with large or medium sized bulge
components.

We can correct for the apparent incompleteness by applying these
estimates to the number of point sources detected in
each point source-to-total luminosity bin.  For example, the total number of
point source nuclei in our sample comprising between 1\% and 2\% of
the total galaxy light is 19$\pm$4.4.  The apparent completeness in this
bin, according to Figure 13, is 12.1$\pm$3.6\%.  If we adjust our sample
for this level of incompleteness, the total number of nuclear point sources
in our survey could be as high as 157$\pm$59 having point source-to-total
luminosity ratios between 1\% and 2\%.  If each bin is adjusted for
incompleteness in this way, the fraction of all galaxies which contain
nuclear point sources down to 1\% of the galaxy light could be as high
as 36.4$\pm$6.7\%.  Figure 14a illustrates how the fraction of
galaxies containing unresolved nuclei varies as the limiting point
source-to-total luminosity ratio changes.  The solid line represents the
uncorrected counts and the dashed line represents the corrected
number counts according to the incompleteness estimates described above.
The high level of incompleteness at the faint end (where the point source
is 1-2\% of the total
galaxy light) causes the
adjusted fraction of galaxies to increase sharply when extending the
survey to these faint limits.
Figure 14b shows the completeness adjustment factor as a function of
point source-to-total luminosity ratio limit.  At a limit of
5\% this factor appears to level off at about 0.6.
The behavior of the completeness adjustment factor indicates that
the statistically significant point source-to-total luminosity
ratio limit appears to be from $\sim$3\% to 5\%.  Figure 14a shows that
at this limit the total fraction of galaxies containing a nuclear point
source component is $\sim$9\% to 16\% corrected for incompleteness.

\section{Spectroscopy}

Ground-based spectroscopic follow-up for the galaxies imaged with HST
is an important part of the science objectives of the Medium Deep Survey.
Spectra allow us to determine redshifts for the host galaxies for
construction of an accurate luminosity function of the unresolved nuclei.
Spectra also allow us to associate detected emission lines with
morphological properties and confirm the identification of LLAGNs or starburst
nuclei for the brightest candidates.  Additional spectra of MDS
galaxies in the field were obtained to address other scientific questions
about the nature of moderately redshifted galaxies (see Im et al. 1996;
Mutz et al. 1994).

Spectra were obtained with the Kitt Peak 4-meter telescope using the
Cryogenic Camera.  For details see "Low-to-Moderate Resolution Optical 
Spectroscopy Manual for Kitt Peak" (Massey et al. 1997).  
The detector is a dedicated
Loral (Ford) 800x1200 pixel device with relatively good cosmetics in
a fast (f/1) camera-dewar combination.  Total system throughput
(telescope + spectrograph + CCD) is typically 20\%.
This instrument allows for use of multi-slit masks so that several
targets can be exposed at the same time.
For each field, 1 to 3 masks were designed based on astrometry
taken from the WFPC2 images.  Each mask exposed between 4 and 10 galaxies
with a typical mask exposing 6 galaxy targets at once.  Using this technique,
most galaxies in each field could be observed down to a limiting magnitude
of I$\sim$21.0.

Observations of WFPC2 imaged galaxies were made during observing runs
in 1994 October, 1995 April, 1996 January, 1996 July, and 1996 September.
A total of 19 nights at the 4 meter telescope was allocated for this
project during these 5 observing runs.  Due to poor weather, 6 nights
were lost yielding a total of 13 nights.  Grism \#650 was used
during the October '94 run while Grism \#770 was used for the
subsequent runs since the longer wavelength range was desirable for
observing emission lines of higher redshift galaxies.  The wavelength range
for Grism \#650 is 4000 to 6800 $\AA$ with a resolution 
(using a 2.5$\arcsec$ slit) of 12 $\AA$.  For the more frequently used 
Grism \#770, the wavelength range is 4300 to 8500 $\AA$ with a resolution 
of 15 $\AA$ through a 2.5$\arcsec$ slit.  The gain was set to 1.5 
e$^-$/ADU and the readnoise was typically 15 electrons for all observing runs.
The wavelength range allows for coverage of the H\&K Calcium lines and
4000$\AA$ blanketing break, 3727
$\AA$ [OII], 5007/4959 $\AA$ [OIII], and
several other emission and absorption features
out to z$\simeq$0.8.  However, the actual redshift range detectable for each
object varies somewhat due to the galaxy position within the focal plane.

Typically, 2 to 3 one-hour exposures were required for
each multi-slit mask to achieve a signal-to-noise of at least 10 for
a galaxy of I$\simeq$20.0.  Calibration images consisted of quartz flats
obtained for each mask, HeNeAr lamp images obtained before and after
each mask observation, and bias frames obtained at the beginning of each
night.

The data were reduced using several IRAF routines outlined in the manual
``Multi-Slits at Kitt Peak" (DeVeny et al. 1996).  At least 20 bias frames
were averaged together for each night of observing and were subtracted from
all images taken that night using CCDPROC.  Bad CCD columns were removed
from all images by interpolation using FIXPIX.  Quartz lamp images for
each individual mask were averaged together using FLATCOMBINE.  Next,
APFLATTEN was used to flatten the flat field for each mask leaving an
image which represents the pixel-to-pixel gain variations in the CCD.
The resulting flat field no longer has spatial (slit-function) information.
This is appropriate for the quartz lamp exposures of multi-slit masks since
the short slit length ($\sim$10$\arcsec$ to 20$\arcsec$) doesn't reveal
much slow variation along the spatial direction.  Each aperture is fit with
an appropriate polynomial to remove the overall shape of the quartz lamp
flat.  Typically, this was a legendre polynomial with order = 20 to 30.
The flat field correction was then applied to each mask image (object
images as well as comparison lamp images) using CCDPROC.

Before extracting the spectra, the individual object frames for each
mask were usually combined using IMCOMBINE.  In some cases, if the
telescope had been moved slightly between exposures, each object frame
was extracted separately.  The comparison lamp frames for each mask were
also combined before extracting.  To extract the object spectra,
APALL was used to set and fit the background (sky region) of each aperture
as well as setting the object region.  The spectra were traced and
interactively fit using this task.  The resulting image contains the object
spectrum for each aperture with variance weighting, the object spectrum
without variance weighting, the sky spectrum for each aperture, and the
error spectrum for each aperture.  The HeNeAr comparison images
are then extracted using the APSUM task with the object apertures set
identically to those in the the object image extractions.  The emission
lines in each of the comparison spectra are then identified and this
dispersion solution is applied to the appropriate aperture object spectrum
in each of the images using DISPCOR.

\section{Redshift Determination}

For each spectrum with obvious emission lines and/or absorption features,
a redshift estimate was first determined by measuring the emission wavelengths
of two or more features and comparing them with the known rest wavelengths.
Cross-correlation
with a template galaxy spectrum containing similar features
was used to accurately determine the true redshift for the object.
We used FXCOR in IRAF to do the cross-correlation.  The
galaxy template spectra are from Kennicutt (1992) for various galaxy types.
At least 5 different
templates were employed depending on the galaxy type for which
the redshift was being determined.  A parabolic fit to the peak in the
cross-correlation function determined the redshift and provided a formal
error in the redshift determination based upon this fit.

In some cases only one emission or absorption line was obvious.  If the
redshift determination was based on only one line or two weak features, it is
marked as uncertain (?) in the redshift list.
Some galaxy spectra contained adequate signal but no obvious emission or
absorption features were noted at all.  The cross-correlation technique
was also employed with these spectra to determine if weaker absorption or
emission features
could be detected.  In these cases, if a redshift was determined for the
spectra, it was marked as very uncertain (??) in the final redshift list.

A total of 253 objects were observed through slitlet masks over the 5
observing runs.  Of these objects, 136 had enough signal present in the
spectra to detect emission lines or possible absorption features so that
a redshift could be determined.  This yields a success rate
of 54\% in obtaining MDS galaxy redshifts.  The majority of these galaxies
for which redshifts were determined are non-nuclear point source galaxies.

As discussed previously, a total of 101 galaxies were found
to contain nuclear point sources, including galaxies from the
Groth strip and HDF.  The spectroscopic follow-up pursued in this study
includes only MDS fields.
Other groups have obtained spectroscopic follow-up for the non-MDS
fields which we use to complement our redshift survey.

Of the 101 galaxies containing nuclear point sources, 77 are in MDS fields
with the remaining 24 observed in the Groth strip and HDF.
Of the 77 galaxies selected from the MDS fields which contain nuclear
components, 35 have been observed through slitlets at the 4 meter.
Figure 15 shows the 35 spectra where the prominent absorption and/or
emission features have been indicated.
Adequate signal along with emission or
absorption features were detected in 29 of these spectra allowing their
redshifts to be determined.  The dashed line is the arbitrarily
scaled error spectrum
indicating regions where night sky lines affected the object spectrum.  The
redshift is shown in the upper left corner of each spectrum under the
object name.  Uncertain redshifts are indicated with question marks.
We also show in the upper right corner an ID number corresponding
to the ID number in Figure 11 allowing each spectrum to be matched
to its image.  The spectra are arranged in order of decreasing
contribution of the nucleus to the galaxy light.

There are on average 1.2 galaxies containing unresolved nuclei per WFPC2
field.  For every MDS field observed at the 4 meter, slitlets were placed
on those galaxies containing nuclear components with the remainder placed
on other relatively bright (I$\lesssim$21.0) galaxies in the field.  Because
of the sparseness of the nucleated galaxies in each field, it was difficult
to obtain spectroscopic observations for the entire sample of 77 galaxies
in 13 nights of 4 meter time with the typical total exposure time per
field being 3 hours.  As mentioned above, only 35 of the compact
nuclei galaxies were observed and 29 (38\% of the 77 galaxies) yielded
good spectra for redshift determination.  We then include the compact nuclei
galaxies from the HDF (Cohen et al. 1996) and Groth strip (Koo et al. 1996)
for which redshifts are taken from the literature.  Of the 24 compact nuclei
galaxies in these 2 fields, 6 had published redshifts.  When we
include these fields we have 35 compact nuclei galaxies with
known redshifts representing 35\% of the 101 selected galaxies.

\section{Spectroscopic Identifications for Nuclear Point Source Galaxies}

The nuclei are typically too faint to contribute much light to the galaxy
spectrum.  For this reason, we do not expect to see
the spectrum of the nucleus in these observations.  The typical resolution
at the 4 meter is 1$\arcsec$ which is comparable to the galaxy size in most
cases.  This level of resolution combined with the faintness of the
nuclei makes it impossible to study the nuclear spectrum
separate from that of the host galaxy.

As mentioned previously, the spectra in Figure 15 are arranged in
order of decreasing contribution of the nucleus to the total galaxy light.
Interestingly, we find that the first two spectra displayed,
the galaxies with the brightest nuclear point source components (ua400-7
and uwy02-4), reveal broad emission lines indicative of Seyfert 1-type galaxies.
In the case of ua400-7, we see broad MgII in addition to [NeV] and [NeIII]
emission lines.  The spectrum of uwy02-4 reveals broad H$\beta$ as well as
narrow line [OII] and [OIII] emission.  In addition, usa00-35, where the
nucleus is 20\% of the galaxy V filter light, has a flux ratio of
([OIII] $\lambda$5007)/(H$\beta$ $\lambda$4861) $\simeq$ 5.
We also detect H$\alpha$ but not [SII] in the spectrum.  We determine the
([SII] $\lambda$6725)/(H$\alpha$ $\lambda$6563) flux ratio
limit to be $\simeq$0.4.  According to the line ratio diagnostics of
Veilleux \& Osterbrock (1987),
this is likely to be a Seyfert 2 nucleus.
These spectroscopic identifications
provide a check that our technique to search for unresolved nuclei
is sensitive to Seyfert-like nuclei present in the survey galaxies.

The remainder of the spectra contain
a variety of emission and absorption features.  Of the 29 spectra with
adequate signal-to-noise for redshift determination, 18 display narrow
3727 $\AA$ [OII] emission and/or 5007/4959 $\AA$ [OIII] emission.
Hydrogen emission lines of
H$\beta$ and/or H$\alpha$ are seen in 13 of the spectra.  The CaII H and
K absorption and 4000 $\AA$ blanketing break are seen in
22 of the 29 spectra.
These spectra can be used to classify the host galaxy types.
Based on spectral features, 19 have spectra consistent
with mid to late type spirals while 10 are consistent with early type
galaxies from Sa to ellipticals.
These classifications are used in Paper II in comparison with the
bulge-to-total luminosity ratio classifications for the host galaxies.

\section{Photometric Redshifts}

Our observing strategy was to place slitlets
on those galaxies containing nuclear components with the remainder placed
on other relatively bright (I$\lesssim$21.0) galaxies in the field.
As described previously, all compact nuclei galaxies
from the MDS, HDF and Groth strip for which redshifts are known
make up 35\% of the 101 compact nuclei galaxies.
For the remainder of our sample,
we can estimate redshifts based on several host galaxy parameters.

To determine redshifts ``photometrically", it is necessary to have as large
a database as possible of galaxies with measured redshifts and HST images
for high resolution photometry.  All MDS fields used in this study
are imaged in both the V and I filters.  We obtained good spectra for
136 galaxies for which redshifts could be determined.  Highly reliable
redshifts are determined for 102 of these galaxies, where the redshift is
based on at least two strong emission or absorption features.
At the time of this study, several redshifts had been published from the HDF
(Cohen et al. 1996).  Of the HDF galaxies fit with our modeling software,
47 have reliable redshifts published.
For the Groth strip, 25 modeled galaxies have reliable published redshifts
(Koo et al. 1996).
We then removed from these samples any galaxies with poorly fit parameters
such as non-convergence of the model in one filter or very irregular morphology.
Because the bulge-to-total measurement will be used in the empirical
fit, it is important to remove highly irregular galaxies for which the
bulge-to-total luminosity ratio less is meaningful.  This leaves us with
129 galaxies with excellent model fits where the bulge-to-total luminosity
ratio, magnitude and V-I color have been well determined.

Figure 16 demonstrates how the color, magnitude, and bulge-to-total
luminosity ratio vary as a function of the galaxy redshift.
If we break down our sample into
different bulge-to-total bins, we see that the relationship between
magnitude, color and redshift becomes tighter since we confine
ourselves to galaxies of a similar Hubble type.  An empirical fit to
the data in different bulge-to-total bins will allow for redshift
estimation based entirely on color, magnitude and bulge-to-total
luminosity ratio.  Initially, we also included the half-light radius
in the empirical fit to determine redshift but found that it was
not a useful parameter in constraining the redshift.

We use the fitting software GaussFit (McArthur et al. 1994) to perform
a least-squares
linear regression in 3 dimensions to determine redshift as a function
of magnitude and color according to the equation below.
\begin{equation}
z_{est}=C1\times{color}+C2\times{mag}+C3
\end{equation}
where C1 and C2 are the coefficients of color and magnitude and C3 is the
zero point of the fit.
The data were divided into various subsamples
of limited bulge-to-total ratios and the fitted z values were compared with
the input redshifts to determine the 1$\sigma$ error in the calculated
redshift.  It was found that among the galaxies having bulge-to-total ratios
less than 0.5, 2 galaxies with z$\geq$0.8 fell far
from the linear fit.  Since so few data points exist beyond z=0.8, we cannot
attempt to estimate redshifts accurately beyond this limit.
Therefore, these 2 galaxies
were removed from the dataset allowing for a much better fit to the remaining
data points.  After removing these 2 data points, we are left with 127 points
in the fit.

The GaussFit fitting procedure was applied to various subsamples of the
total galaxy dataset of 127 points.  In Figures 17 through 20 we show the
fitting results for subsamples of the data separated into bulge-to-total
luminosity ratio bins.  Large bulge-to-total bins were chosen so that
uncertainties in the determination of the bulge component in galaxies where a
nuclear point source is also detected will not significantly affect
the redshift estimation.
Figure 17 contains galaxies with  Bulge/Total$\leq$0.2,
those with Bulge/Total$\leq$0.5 are in Figure 18, those with
Bulge/Total$>$0.5 are in Figure 19, and Figure 20 contains those with
Bulge/Total$\geq$0.8.  In each figure, a) is the
I magnitude for the galaxies vs. their measured redshifts, b)
is the V-I galaxy color vs. the redshift, c)
is the estimated redshift based on the fit vs. the measured redshift,
and d) is the residual (z$_{measured}$ - z$_{est}$)
vs. the measured redshift.  The 1$\sigma$ error in the fitted redshift
value is given in the upper left corner of the residual plot in d).
Figure 21 shows the estimated redshifts for all of the galaxies
used in this calculation vs. their spectroscopically measured redshifts.
This figure is the combination of panel c) from each of the preceding
figures.  The coefficients for the fit in each subsample in addition to the
standard deviation between the fit and the data are listed in Table 3.
These results were then applied to the galaxies hosting compact nuclei to 
estimate redshifts for those without spectroscopic information.

This technique relies on the same principles as the photometric
redshift techniques using multicolor information.  Connolly et al. (1995)
found that a galaxy's position in multicolor space is a function of 
redshift, luminosity and spectral type.  With accurate photometry
in U,B,R and I, they have been able to predict redshifts with 
$\sigma$$_z$$\sim$0.05.  Because of our lack of color information,
the typical error in determining redshifts using this technique is
$\sigma$$_z$$\sim$0.1.  For the purposes of this study,
errors of this size are acceptable in determining the space density
of compact nuclei.  The redshift errors can be incorporated later in the
determination of luminosity functions for these objects (see Paper II).

Table 4 is the resulting redshift list for the 101 compact nuclei
galaxies in our sample.  The 29 galaxies with spectroscopic redshifts
obtained at the 4 meter are indicated with a star (*).  The
error in the spectroscopic redshift is
estimated from the formal error of the cross-correlation template fit
and the error in the fit to identified lines in the comparison spectrum
which was used to determine the dispersion solution.
Redshifts from the HDF and Groth strip
are reported from the literature and are indicated in the table.
For the 66 photometrically estimated galaxies, the error in
the reported redshift is based upon the standard deviation of the points
in the fit and the error in determining the galaxy magnitude and color.
The RA and DEC in J2000 are also given in this table for each object.

Four of the estimated redshifts (6\%) yielded negative values or values
less than z=0.1.  Spectroscopic measurements indicate that objects with 
redshifts less than z=0.1 are intrinsically rare in this survey.
Therefore, we assume our estimated redshifts to be in error.
Each of the four galaxies have very blue colors
(V-I$\lesssim0.2$) which caused the redshift to be estimated too low.
To achieve a more realistic estimate of the redshift
for these galaxies, we determine the average redshift of galaxies within
$\pm$0.5 magnitude of the galaxy in question.  In this way, we use only
magnitude information and no color information in the determination.  The
standard deviation of the average provides the new error estimate for
the redshift which is $\sigma$$_z$$\simeq$0.2.  This error is rather large
but applies to only a small portion of the sample galaxies.

Figure 22 is a histogram of the redshifts for the compact nuclei
galaxies.  The solid line represents all 101 galaxies with spectroscopic or
photometric redshifts.  The hatched region is the histogram of only
the 35 galaxies with spectroscopic redshifts.  There is clearly some
peakiness which appears in the spectroscopic as well
as the total redshift distribution.  The distribution peaks
near z$\sim$0.4 and extends to z=1.0.  However, only those redshifts
estimated at z$\leq$0.8 are considered reliable since very few
points beyond z=0.8 were used in the empirical
fit for the redshift estimation technique.

\section{Summary}

The purpose of this study is to understand the space
density and properties of active galaxies to z$\simeq$0.8 through
morphological identification of nuclear activity.
In this first paper of two, we describe the procedure for
selecting Hubble Space Telescope fields and modeling
the light profiles of galaxies found in these images.  We select
galaxies from this survey which require an additional nuclear
point source component to adequately model the galaxy light profile.  
Monte Carlo simulations using real and simulated data allow us to
determine the criterion for a unique galaxy model.
A total of 1033 galaxies from 70 WFPC2 fields have been modeled in
this study where 16$\pm$3\% of the galaxies contain an unresolved 
nuclear component $\geq$3\% of the total galaxy light.  
We find 9$\pm$1\% of galaxies contain nuclei $\geq$5\% of the galaxy light.
These percentages have been corrected for incompleteness effects in
our survey. 

To determine the space density of these galaxies, it is necessary to
obtain redshift information for the selected galaxies in our study.
Spectroscopic redshifts have been obtained for 35 of our AGN/starburst
candidates and photometric redshifts are estimated to an accuracy of
$\sigma_z$$\simeq$0.1 for the remaining sample. We have presented the
ground-based spectroscopy
for these galaxies as well as the photometric redshift predictions
for those galaxies without spectra.  

In Paper II, we focus on the analysis of the data presented here.
We investigate the properties of the host galaxies and nuclei themselves such
as colors, magnitudes, sizes, and Hubble types.  We compare these properties
with those of local Seyferts and starburst galaxies.  The colors
of the nuclei are used to differentiate between Seyfert-like nuclei and young
starburst nuclei based on colors from representative spectra of these
objects.  Based on this color selection, we present the upper limit 
luminosity function
for Low-Luminosity AGN (LLAGN) in the magnitude range 
-20$\leq$M$_B$$\leq$-14 in two redshift
bins out to z=0.8 and compare it with the LFs of local Seyferts and
moderate redshift QSOs.  We also comment on the likely contribution of
these nuclei to the soft X-ray background.

%Acknowledgements

\acknowledgements

We would like to thank the referee and editor for comments and suggestions
which have improved the quality of this paper.
The authors would like to thank John Huchra and
Rogier Windhorst for helpful discussions and guidance on this project.
We would also like to thank Eric Ostrander who provided us with 
processed HST images through the MDS pipeline and aided in selection
of MDS fields for this project.
In addition, we thank Judy Cohen for providing the
redshift list from her study of the Hubble Deep Field.
This work is based
on observations taken with the NASA/ESA {\it Hubble Space Telescope},
obtained at the Space Telescope Science Institute, operated by the
Association of Universities for Research in Astronomy, Inc.  This work
was supported in part by STScI grant GO-02684.06-87A to R. F. G. for
the Medium Deep Survey project.

%References

%Begin Figure Captions
\newpage

\centerline{FIGURE CAPTIONS}

\figcaption{The diameter of an unresolved region in a WFPC2 image as a function
of the object's redshift.  The lines represent values of H$_o$ = 50,
75, and 100 km/s/Mpc.}

\figcaption{Output from the 2-dimensional modeling software.  From left
to right the boxes are: 1) the observed 64 x 64 pixel area around the galaxy,
2) the selected region for analysis, 3) the PSF convolved
maximum likelihood model image, 4) the maximum likelihood model image, 5)
the residual image (model subtracted from real image), 6) the sigma
image, 7) the object mask image.}

\figcaption{Parameters of galaxies used in the Monte-Carlo simulation.
The galaxies cover a range of Bulge-to-Total luminosity ratios,
magnitudes and SNRIL values.  SNRIL is the integrated signal-to-noise
measurement described in the text.}

\figcaption{The input vs. the measured magnitude for the point source
nuclei detected in the Monte-Carlo simulation.}

\figcaption{The input vs. the measured magnitude for the point source
nuclei detected in the Monte-Carlo simulation after applying the selection
criteria outlined in the text.}

\figcaption{The histogram of mag$_{obs}$ - mag$_{real}$ for
point sources at 23$\pm$0.25.  The dotted line represents the gaussian
curve with $\sigma$=0.26.  b) Same histogram for point sources
at 24$\pm$0.25 magnitudes.  The gaussian represents $\sigma$=0.56.}

\figcaption{a) The Bulge-to-Bulge+Disk measured in galaxies within
which point source nuclei were detected in the simulations vs. the
input Bulge-to-Bulge+Disk for the host galaxy.  b) The measured
Bulge-to-Bulge+Disk as a function of the point source-to-total
luminosity ratio for galaxies in the simulations in which point
source nuclei were detected. Note that 
nuclear point source detections greater than 
20\% of the total galaxy light are associated with
galaxies having no detected bulge ((B/B+D) = 0).}

\figcaption{a) The histogram of galaxies in the simulation containing no
bulge component as a function
of the point source-to-total luminosity ratio.  The solid line represents
all galaxies in the simulation with no bulge component and the hatched
region represents those galaxies in which the nuclear point source was
detected. b) The fractional completeness or success rate in detecting
point source nuclei in galaxies containing no bulge component.  The
errorbars are the Poisson statistics based on the number of
objects in each bin.}

\figcaption{a) The histogram of galaxies in the simulation containing
bulge components with 0$<$(B/B+D)$\leq$0.4 as a function
of the point source-to-total luminosity ratio.  The hatched region
represents those in which the point source nucleus was detected.
The panel to the right is the fractional completeness or success rate
in detecting point source nuclei for these galaxies.
b) Same as above except for galaxies with 0.4$<$(B/B+D)$\leq$0.8.
c) Same as above except for galaxies with 0.8$<$(B/B+D)$\leq$1.0.}

\figcaption{a) The histogram of all galaxies in the simulation containing no
bulge component as a function of the SNRIL value.  The hatched region
represents those galaxies where a 2-component point source+disk model
was the best fit thereby detecting the point source nucleus.
b) The histogram of all galaxies in the simulation containing some
bulge component as a function of the SNRIL value.  The hatched region
represents those galaxies where a 2-component point source+disk model
or a 3-component point source+disk+bulge model was the best fit.
The cross-hatched region represents those galaxies where the
3-component model was the best fit.}

\figcaption{I band image of galaxies containing unresolved nuclear
point sources down to 1\% of the total galaxies light.  The
galaxies are arranged in descending order from the greatest
point source-to-total luminosity ratio from left to right, top to bottom.
The ID number is used to indentify each object throughout the paper.}

\figcaption{The point source nucleus magnitude vs. the integrated galaxy
magnitude in a) I and b) V.  The dashed line represents the locus
where the point source is 1\% of the total galaxy light.}

\figcaption{The overall completeness estimate for detecting nuclei
in the survey galaxies as a function of the point source-to-total
luminosity ratio.}

\figcaption{a) The fraction of total survey galaxies containing unresolved
nuclei as a function of the limiting point source-to-total luminosity
ratio.  The dashed line represents this fraction after correcting for
completeness according to Figure 2.15.  b) The completeness adjustment
factor as a function of the limiting point source-to-total luminosity
ratio.  This is the amount by which the fraction must be multiplied to
correct for incompleteness.}

\figcaption{Spectra of galaxies containing compact nuclei obtained at
the Kitt Peak 4-meter telescope.  The spectra are not flux calibrated
and are in units of flux (electrons) vs. wavelength (angstroms).
Prominent emission and absorption features are indicated.
The object name and redshift is in the upper left corner and
an ID number corresponding to the ID number for the galaxy image
in Figure 2.13 is in the upper right corner.  The dashed line is the
arbitrarily scaled error spectrum.}

\figcaption{a) The I magnitude for galaxies in our survey vs. the measured
redshift.  b) The V-I color vs. the redshift.  c) The Bulge/Total luminosity
ratio vs. the redshift.}

\figcaption{a) The I magnitude for galaxies in our survey with
Bulge/Total$\leq$0.2 vs. the measured
redshift.  b) The V-I color vs. the redshift.  c) The estimated redshift
based on fits to the color and magnitude vs. the true redshift.  d)
The residual of the fit (z$_{measured}$ - z$_{est}$) vs. the measured
redshift.  The standard deviation of the fit is indicated.}

\figcaption{a) The I magnitude for galaxies in our survey with
Bulge/Total$\leq$0.5 vs. the measured
redshift.  b) The V-I color vs. the redshift.  c) The estimated redshift
based on fits to the color and magnitude vs. the true redshift.  d)
The residual of the fit (z$_{measured}$ - z$_{est}$) vs. the measured
redshift.  The standard deviation of the fit is indicated.}

\figcaption{a) The I magnitude for galaxies in our survey with
Bulge/Total$>$0.5 vs. the measured
redshift.  b) The V-I color vs. the redshift.  c) The estimated redshift
based on fits to the color and magnitude vs. the true redshift.  d)
The residual of the fit (z$_{measured}$ - z$_{est}$) vs. the measured
redshift.  The standard deviation of the fit is indicated.}

\figcaption{a) The I magnitude for galaxies in our survey with
Bulge/Total$>$0.8 vs. the measured
redshift.  b) The V-I color vs. the redshift.  c) The estimated redshift
based on fits to the color and magnitude vs. the true redshift.  d)
The residual of the fit (z$_{measured}$ - z$_{est}$) vs. the measured
redshift.  The standard deviation of the fit is indicated.}

\figcaption{The photometric redshift vs.
the measured spectroscopic redshift for those galaxies
used in the empirical fit.}

\figcaption{The histogram of redshifts for the compact nuclei galaxies.
The hatched region represents those galaxies where the redshift was
determined spectroscopically.}

%Begin Tables

\begin{deluxetable}{crrrccccccc}
%\small
\footnotesize
%\scriptsize
\tablewidth{0pt}
\tablecaption{MDS Fields}
\tablecomments{Column 1 - MDS Field name, Columns 2 \& 3 - J2000 coordinates,
Column 4 - galactic latitude, Column 5 - \# of V exposures, Column 6 - total
V exposure time, Column 7 - \# of I exposures, Column 8 - total I exposure
time, Columns 9 \& 10 - the limiting V and I integrated galaxy magnitude for
fitting galaxies with a 3-component model, Column 11 - the number of
galaxies in the I image above this limiting magnitude.
Additional information on MDS fields can be obtained
at http://archive.stsci.edu/mds/mds.cgi.}
\tablehead{
\colhead{Field Name} &
\colhead{RA(J2000)} & \colhead{DEC(J2000)}  &
\colhead{b}         & \colhead{V$_{\#}$} &
\colhead{V$_{exp}$ (s)}      & \colhead{I$_{\#}$} &
\colhead{I$_{exp}$ (s)} &
\colhead{V$_{lim}$} &
\colhead{I$_{lim}$} &
\colhead{\#$_{gal}$}}

\startdata
uad01 &  0 15 47.8 & -16 19 04.9 & -76.40 &   2 &   1200 &   2 &   2000 	 &      20.1 &      20.7 &  4 \nl
uad00 &  0 15 55.4 & -16 18 06.2 & -76.40 &   2 &   1200 &   2 &   2000 	 &      21.1 &      20.1 &  3 \nl
ua400 &  0 24 53.6 & -27 16 23.4 & -84.10 &   4 &   8000 &   4 &   8000 	 &      22.6 &      21.9 & 24 \nl
ubz01 &  0 50 32.5 & -52 07 25.3 & -65.00 &   2 &   1200 &   2 &   2000 	 &      21.2 &      20.8 &  8 \nl
ueh00 &  0 53 23.2 &  12 33 57.7 & -50.30 &   3 &   5400 &   3 &   6300 	 &      21.6 &      21.0 & 15 \nl
ueh02 &  0 53 36.6 &  12 49 49.4 & -50.04 &   2 &   3300 &   2 &   4200 	 &      22.2 &      21.3 & 23 \nl
ua-30 &  0 58 06.8 & -28 11 40.7 & -88.18 &   3 &   3900 &   5 &   4900 	 &      22.5 &      22.0 & 19 \nl
ua-00 &  1 02 26.5 & -27 11 52.8 & -87.55 &   2 &   2100 &   2 &   4200 	 &      21.7 &      21.5 & 27 \nl
ua-01 &  1 04 36.0 & -27  5 17.1 & -87.07 &   3 &   8700 &   4 &  10700 	 &      22.8 &      22.0 & 22 \nl
ujh01 &  1 09 03.5 &  35 36 25.2 & -27.13 &   1 &   1200 &   2 &   4200 	 &      21.3 &      21.2 & 10 \nl
ubi01 &  1 09 59.8 &  -2 27 23.8 & -64.93 &   2 &   3300 &   3 &   6300 	 &      21.5 &      21.0 & 20 \nl
ubi00 &  1 10 03.0 &  -2 26 22.8 & -64.91 &   1 &   1200 &   2 &   4200 	 &      20.8 &      20.8 &  6 \nl
uci10 &  1 24 42.4 &   3 51 27.6 & -57.99 &   3 &   4800 &   4 &  10800 	 &      22.2 &      21.9 & 18 \nl
ufj00 &  2 07 05.8 &  15 25 18.3 & -43.66 &   1 &   1200 &   2 &   4200 	 &      21.4 &      21.4 & 23 \nl
ugk00 &  2 38 51.6 &  16 44 38.2 & -38.97 &   1 &   2700 &   2 &   5400 	 &      21.4 &      21.0 &  8 \nl
udm00 &  2 42 39.5 &   0 48 49.3 & -51.32 &   1 &   1200 &   2 &   3000 	 &      22.1 &      21.1 &  3 \nl
udm10 &  2 42 51.7 &   0 04 25.0 & -51.95 &   4 &   4000 &   3 &   5400 	 &      21.8 &      21.6 & 14a \nl
ucs01 &  2 56 22.0 & -33 22 25.4 & -62.42 &   2 &   1500 &   2 &   4200 	 &      21.5 &      21.2 & 13 \nl
uem00 &  3 05 03.2 &   0 11 13.1 & -48.11 &   2 &   2400 &   5 &   6600 	 &      21.8 &      21.4 & 14 \nl
uim01 &  3 55 31.4 &   9 43 32.0 & -32.15 &   6 &   3600 &  10 &   6600 	 &      21.5 &      21.0 & 14 \nl
uko01 &  4 56 45.5 &   3 52 40.8 & -23.28 &   1 &   1200 &   2 &   4200 	 &      21.4 &      21.3 & 12 \nl
uqk11 &  7 24 46.6 &  60 31 02.3 &  27.41 &   1 &   1000 &   5 &   3100 	 &      21.3 &      21.1 & 16 \nl
uop00 &  7 50 47.1 &  14 40 44.2 &  19.63 &   5 &   7200 &   2 &   4200 	 &      22.0 &      21.3 &  9 \nl
urp03 &  8 47 21.5 &  17 57 29.8 &  33.42 &   1 &   1200 &   2 &   3000 	 &      21.4 &      20.8 & 12 \nl
urp01 &  8 47 24.1 &  17 56 22.9 &  33.42 &   1 &    600 &   1 &   2400 	 &      20.4 &      21.1 &  8 \nl
usp00 &  8 54 16.1 &  20 03 41.9 &  35.68 &   2 &   3300 &   2 &   4200 	 &      21.8 &      21.2 &  8 \nl
ust01 & 10 05 16.9 &  -7 47 35.1 &  36.74 &   1 &   1200 &   3 &   2380 	 &      21.6 &      20.4 &  8 \nl
ust00 & 10 05 46.3 &  -7 41 30.3 &  36.90 &  10 &  16500 &  11 &  23100 	 &      23.1 &      22.2 & 17 \nl
uui00 & 11 42 04.7 &  71 37 43.9 &  44.47 &   3 &   5400 &   3 &   6300 	 &      22.5 &      21.6 & 32 \nl
uzp01 & 11 50 29.8 &  28 48 29.6 &  76.45 &   2 &   3300 &   3 &   6300 	 &      22.0 &      21.7 & 22 \nl
uyj00 & 11 53 25.2 &  49 31 12.9 &  64.98 &   1 &    300 &   3 &   2700 	 &      20.2 &      21.0 & 15 \nl
uzk02 & 12 11 13.1 &  39 26 56.0 &  75.11 &   1 &    600 &   3 &   7200 	 &      21.2 &      21.8 & 17 \nl
uzx00 & 12 30 16.5 &  12 21 47.3 &  74.42 &   3 &   4500 &   1 &   2100 	 &      22.2 &      20.9 & 22 \nl
uzx07 & 12 30 51.0 &  12 19 03.0 &  74.42 &   4 &   5200 &   3 &   2700 	 &      22.1 &      20.5 &  8 \nl
uzx01 & 12 30 54.2 &  12 19 05.6 &  74.43 &   5 &   3480 &   4 &   6200 	 &      22.0 &      21.4 &  8 \nl
uxy00 & 12 32 31.6 &  -2 21 48.6 &  60.16 &   1 &   1200 &   2 &   4200 	 &      21.7 &      21.4 & 11 \nl
uxy10 & 12 36 38.9 &   0 41 54.9 &  61.95 &   3 &    980 &   5 &   3780 	 &      21.0 &      21.1 &  7 \nl
HDF   & 12 36 49.4 &  62 12 58.0 &  54.83 & 103 &   1051 &  58 &   2137 	 &      24.2 &      23.5 &106 \nl
uzy00 & 12 38 14.1 &  11 52 30.6 &  74.44 &   1 &   1200 &   2 &   2700 	 &      21.2 &      20.7 &  4 \nl
uzy01 & 12 38 15.8 &  11 51 18.2 &  74.42 &   2 &   3000 &   2 &   2700 	 &      21.9 &      20.7 &  4 \nl
uwy02 & 12 40 22.8 & -11 31 29.7 &  51.25 &   6 &  11700 &   5 &   9600 	 &      23.0 &      22.1 & 21 \nl
urz00 & 12 53 01.9 & -29 14 21.4 &  33.63 &   3 &   5400 &   4 &   8400 	 &      22.3 &      22.2 & 24 \nl
uz-00 & 13 00 23.6 &  28 20 13.2 &  87.68 &   2 &   1200 &   2 &   2000 	 &      21.3 &      20.8 &  4 \nl
uzd10 & 13 55 18.3 &  40 20 30.6 &  71.33 &   1 &   3500 &   2 &   6100 	 &      22.1 &      21.5 & 21 \nl
uy000 & 14 16 18.1 &  11 32 22.7 &  64.70 &   4 &   6000 &   4 &   6900 	 &      22.3 &      21.6 & 16 \nl
u26x9 & 14 17 23.7 &  52 25 13.0 &  59.77 &   4 &    700 &   4 &   1000 	 &      22.0 &      21.6 & 21 \nl
u26x8 & 14 17 30.2 &  52 26 22.8 &  59.79 &   4 &    700 &   4 &   1000 	 &      21.9 &      21.4 & 22 \nl
u26x7 & 14 17 36.8 &  52 27 32.7 &  59.82 &   4 &    700 &   4 &   1000 	 &      21.8 &      21.3 & 16 \nl
u26x6 & 14 17 49.9 &  52 29 52.3 &  59.87 &   4 &    700 &   4 &   1000 	 &      22.1 &      21.4 & 12 \nl
u26x5 & 14 17 56.4 &  52 31 02.1 &  59.89 &   4 &    700 &   4 &   1000 	 &      22.0 &      21.2 & 13 \nl
uy400 & 14 34 57.8 &  25 11 45.4 &  66.73 &   6 &   5200 &   6 &   6000 	 &      22.1 &      21.5 &  8 \nl
uy402 & 14 35 16.9 &  24 59 04.0 &  66.62 &   2 &   1400 &   3 &   5400 	 &      21.5 &      21.5 &  2 \nl
uy401 & 14 35 33.1 &  25 18 15.9 &  66.62 &   4 &   2400 &   8 &   8000 	 &      21.9 &      21.9 & 15 \nl
ux400 & 15 19 41.2 &  23 52 05.5 &  56.51 &   2 &   3300 &   4 &   7500 	 &      21.9 &      21.6 & 24 \nl
ux401 & 15 19 55.0 &  23 44 46.0 &  56.43 &   2 &   3300 &   3 &   6000 	 &      21.9 &      21.5 & 11 \nl
uvd01 & 15 43 23.7 &  53 52 46.4 &  48.77 &   3 &   9000 &   2 &   6000 	 &      23.0 &      21.8 & 18 \nl
ut201 & 16 01 12.3 &   5 36 02.8 &  40.06 &   6 &   3600 &  12 &  12000 	 &      22.2 &      21.8 & 21 \nl
ut200 & 16 01 27.4 &   5 23 55.6 &  39.90 &   6 &   5200 &   6 &   6000 	 &      22.3 &      21.7 & 15 \nl
usa02 & 17 12 23.2 &  33 35 49.3 &  34.25 &   3 &   5400 &   3 &   6300 	 &      22.6 &      21.7 &  9 \nl
usa01 & 17 12 23.9 &  33 36 03.9 &  34.25 &   3 &   5400 &   3 &   6300 	 &      22.4 &      21.8 &  7 \nl
usa00 & 17 12 24.6 &  33 36 15.8 &  34.25 &   3 &   5400 &   3 &   6300 	 &      22.7 &      21.9 & 11 \nl
uqa02 & 17 36 22.5 &  28 00 58.7 &  27.84 &   2 &   1200 &   2 &   2000 	 &      21.3 &      21.2 &  3 \nl
uqa01 & 17 36 38.6 &  28 04 08.8 &  27.80 &   2 &   1200 &   4 &   2280 	 &      21.3 &      20.7 &  4 \nl
uj000 & 19 39 22.9 & -46 13 46.2 & -27.20 &   1 &   1200 &   2 &   4500 	 &      21.2 &      21.4 & 11 \nl
uj700 & 19 40 40.2 & -69 16 01.8 & -29.58 &   3 &   5400 &   3 &   6300 	 &      22.1 &      21.6 &  9 \nl
umd08 & 21 50 34.9 &  28 49 41.6 & -19.23 &   4 &   1200 &   1 &   2400 	 &      21.4 &      21.0 & 12 \nl
umd09 & 21 50 38.5 &  28 55 56.5 & -19.16 &   2 &   2400 &   2 &   4200 	 &      22.0 &      21.2 & 12 \nl
umd05 & 21 51 07.2 &  29 00 00.5 & -19.18 &   1 &   1200 &   2 &   3900 	 &      21.5 &      21.5 &  6 \nl
umd0a & 21 51 13.1 &  29 00 04.6 & -19.20 &   2 &   3300 &   3 &   8700 	 &      22.1 &      22.0 & 22 \nl
uec00 & 23 04 28.6 &   3 04 38.2 & -50.28 &   1 &   1200 &   2 &   3000 	 &      21.4 &      21.1 &  9 \nl
\enddata
\end{deluxetable}

\begin{deluxetable}{clccccccccccc}
\scriptsize
\tablewidth{0pt}
\tablecaption{Model Parameters for Galaxies Containing Unresolved Nuclei}
\tablehead{
\colhead{ID\#} &
\colhead{Name} &
\colhead{I$_{mag}$} &
\colhead{$\sigma$$_I$} &
\colhead{hlr} &
\colhead{$\sigma$$_{hlr}$} &
\colhead{V-I} &
\colhead{$\sigma$$_{V-I}$} &
\colhead{B/T$_I$} &
\colhead{P/T$_I$} &
\colhead{P/T$_V$} &
\colhead{V-I$_{P}$} &
\colhead{$\sigma$$_{V-I}$}}

\startdata
  1& ua400-7 &   20.530 &   0.008 &  -0.767 &   0.020 &   0.602 &   0.011 &   0.000 &   0.485 &   0.687 &   0.224 &   0.333 \nl
  2& uwy02-4 &   19.041 &   0.003 &  -0.367 &   0.006 &   1.071 &   0.004 &   0.000 &   0.414 &   0.489 &   0.890 &   0.285 \nl
  3& ua-30-13 &   21.249 &   0.012 &  -1.016 &   0.024 &   0.810 &   0.016 &   0.000 &   0.333 &   0.410 &   0.584 &   0.436 \nl
  4& ust01-11 &   19.760 &   0.017 &  -0.937 &   0.039 &   0.867 &   0.021 &   1.000 &   0.000 &   0.293 & \nodata  & \nodata  \nl
  5& u26x9-20 &   21.231 &   0.011 &  -1.049 &   0.019 &   1.995 &   0.035 &   0.000 &   0.285 &   0.262 &   2.086 &   0.555 \nl
  6& uy401-12 &   20.390 &   0.007 &  -0.605 &   0.013 &   0.980 &   0.017 &   0.000 &   0.276 &   0.214 &   1.256 &   0.413 \nl
  7& ut200-28 &   20.897 &   0.017 &  -0.687 &   0.026 &   1.416 &   0.039 &   0.000 &   0.264 &   0.256 &   1.449 &   0.478 \nl
  8& ufj00-13 &   19.841 &   0.006 &  -0.857 &   0.012 &   1.047 &   0.013 &   0.000 &   0.264 &   0.230 &   1.197 &   0.371 \nl
  9& udm10-24 &   21.646 &   0.033 &  -1.254 &   0.043 &   1.275 &   0.059 &   0.000 &   0.250 &   0.366 &   0.861 &   0.528 \nl
 10& umd0a-47 &   20.912 &   0.008 &  -1.013 &   0.014 &   1.709 &   0.040 &   0.000 &   0.190 &   0.165 &   1.862 &   0.549 \nl
 11& ua400-26 &   21.448 &   0.013 &  -0.534 &   0.020 &   1.754 &   0.028 &   0.000 &   0.184 &   0.000 & \nodata  & \nodata  \nl
 12& uvd01-26 &   21.157 &   0.012 &  -0.522 &   0.020 &   1.368 &   0.018 &   0.000 &   0.183 &   0.162 &   1.500 &   0.550 \nl
 13& ua-01-10 &   21.153 &   0.011 &  -0.731 &   0.014 &   1.502 &   0.024 &   0.000 &   0.175 &   0.187 &   1.430 &   0.550 \nl
 14& uzx01-37 &   21.190 &   0.021 &  -0.890 &   0.032 &   1.209 &   0.051 &   0.000 &   0.173 &   0.279 &   0.690 &   0.511 \nl
 15& ut201-18 &   20.827 &   0.009 &  -0.763 &   0.014 &   1.351 &   0.026 &   0.000 &   0.169 &   0.241 &   0.966 &   0.489 \nl
 16& uvd01-14 &   20.221 &   0.009 &  -0.505 &   0.012 &   1.169 &   0.013 &   0.000 &   0.167 &   0.130 &   1.441 &   0.458 \nl
 17& uzx00-3 &   18.912 &   0.083 &  -0.498 &   0.084 &   1.257 &   0.084 &   0.291 &   0.160 &   0.089 &   1.894 &   0.402 \nl
 18& usa00-35 &   22.319 &   0.020 &  -1.084 &   0.030 &   0.416 &   0.024 &   0.000 &   0.154 &   0.202 &   0.121 &   0.623 \nl
 19& ueh02-14 &   20.715 &   0.034 &  -0.244 &   0.037 &  -0.246 &   0.358 &   0.000 &   0.144 &   0.000 & \nodata  & \nodata  \nl
 20& umd0a-25 &   19.717 &   0.012 &  -0.593 &   0.023 &   1.286 &   0.014 &   0.335 &   0.141 &   0.178 &   1.033 &   0.410 \nl
 21& uui00-31 &   21.469 &   0.028 &  -0.789 &   0.036 &   0.844 &   0.030 &   0.317 &   0.133 &   0.055 &   1.803 &   0.669 \nl
 22& u26x6-11 &   21.311 &   0.015 &  -0.978 &   0.025 &   2.144 &   0.056 &   0.000 &   0.124 &   0.139 &   2.020 &   0.676 \nl
 23& usa00-9 &   19.364 &   0.004 &  -0.502 &   0.000 &   1.007 &   0.006 &   0.150 &   0.118 &   0.117 &   1.016 &   0.392 \nl
 24& uzp01-19 &   20.859 &   0.016 &  -0.305 &   0.017 &   1.169 &   0.036 &   0.000 &   0.117 &   0.112 &   1.216 &   0.554 \nl
 25& umd05-46 &   20.913 &   0.028 &  -0.283 &   0.029 &   0.684 &   0.189 &   0.000 &   0.116 &   0.035 &   1.985 &   0.674 \nl
 26& ueh02-4 &   19.720 &   0.014 &  -0.087 &   0.015 &   0.336 &   0.095 &   0.000 &   0.115 &   0.000 & \nodata  & \nodata  \nl
 27& uy401-4 &   18.498 &   0.008 &   0.512 &   0.011 &   1.194 &   0.021 &   0.463 &   0.094 &   0.077 &   1.411 &   0.366 \nl
 28& uj700-29 &   20.014 &   0.036 &   0.219 &   0.030 &   1.240 &   0.059 &   0.152 &   0.000 &   0.087 & \nodata  & \nodata  \nl
 29& uvd01-12 &   20.014 &   0.007 &  -0.414 &   0.009 &   1.119 &   0.009 &   0.026 &   0.086 &   0.088 &   1.094 &   0.488 \nl
 30& uzp01-24 &   21.422 &   0.017 &  -0.850 &   0.021 &   0.949 &   0.032 &   0.000 &   0.086 &   0.084 &   0.975 &   0.652 \nl
 31& umd05-37 &   21.016 &   0.015 &  -0.793 &   0.020 &   0.742 &   0.029 &   0.000 &   0.082 &   0.094 &   0.594 &   0.579 \nl
 32& u26x5-6 &   20.819 &   0.014 &  -0.822 &   0.018 &   0.941 &   0.020 &   0.000 &   0.080 &   0.065 &   1.166 &   0.596 \nl
 33& uy400-16 &   21.172 &   0.031 &   0.021 &   0.028 &   1.455 &   0.062 &   0.000 &   0.078 &   0.056 &   1.815 &   0.703 \nl
 34& ust00-8 &   19.541 &   0.014 &  -0.679 &   0.018 &   1.173 &   0.021 &   0.627 &   0.074 &   0.070 &   1.233 &   0.467 \nl
 35& uzx00-4 &   18.827 &   0.026 &   0.580 &   0.021 &   1.177 &   0.028 &   0.079 &   0.074 &   0.046 &   1.693 &   0.432 \nl
 36& uzx00-20 &   20.454 &   0.016 &  -0.665 &   0.018 &   0.931 &   0.019 &   0.000 &   0.073 &   0.057 &   1.200 &   0.564 \nl
 37& u26x7-18 &   21.325 &   0.017 &  -0.904 &   0.022 &   1.256 &   0.034 &   0.035 &   0.069 &   0.093 &   0.932 &   0.666 \nl
 38& u26x8-23 &   21.384 &   0.015 &  -0.914 &   0.020 &   1.187 &   0.023 &   0.000 &   0.067 &   0.097 &   0.785 &   0.669 \nl
 39& uzd10-6 &   19.236 &   0.011 &   0.362 &   0.011 &   1.004 &   0.017 &   0.360 &   0.067 &   0.038 &   1.620 &   0.471 \nl
 40& u26x9-25 &   21.598 &   0.028 &  -0.877 &   0.032 &   1.353 &   0.055 &   0.000 &   0.059 &   0.000 & \nodata  & \nodata  \nl
 41& u26x6-9 &   21.118 &   0.018 &  -1.052 &   0.022 &   0.951 &   0.026 &   0.123 &   0.059 &   0.053 &   1.067 &   0.674 \nl
 42& ust00-27 &   21.157 &   0.008 &  -0.584 &   0.009 &   1.430 &   0.019 &   0.129 &   0.058 &   0.064 &   1.323 &   0.701 \nl
 43& uhdfk-106 &   22.976 &   0.009 &  -1.739 &   0.015 &   1.954 &   0.023 &   0.091 &   0.056 &   0.055 &   1.974 &   1.113 \nl
 44& umd09-4 &   17.900 &   0.006 &   0.591 &   0.005 &   0.963 &   0.012 &   0.000 &   0.056 &   0.044 &   1.225 &   0.354 \nl
 45& urp03-14 &   20.770 &   0.024 &  -0.153 &   0.025 &   0.608 &   0.038 &   0.000 &   0.056 &   0.055 &   0.628 &   0.614 \nl
 46& uxy10-6 &   20.276 &   0.010 &  -1.083 &   0.010 &   0.753 &   0.019 &   0.173 &   0.000 &   0.052 & \nodata  & \nodata  \nl
 47& uui00-17 &   20.335 &   0.014 &   0.145 &   0.017 &   1.270 &   0.024 &   0.167 &   0.050 &   0.043 &   1.434 &   0.623 \nl
 48& uci10-11 &   20.551 &   0.011 &   0.108 &   0.010 &   0.751 &   0.018 &   0.000 &   0.049 &   0.046 &   0.820 &   0.601 \nl
 49& uhdfk-36 &   21.484 &   0.004 &  -0.846 &   0.005 &   0.950 &   0.006 &   0.000 &   0.049 &   0.000 & \nodata  & \nodata  \nl
 50& u26x8-31 &   21.355 &   0.034 &  -0.461 &   0.031 &   1.219 &   0.069 &   0.000 &   0.049 &   0.067 &   0.879 &   0.721 \nl
 51& ufj00-17 &   20.477 &   0.011 &  -0.539 &   0.013 &   0.607 &   0.019 &   0.000 &   0.048 &   0.038 &   0.861 &   0.604 \nl
 52& u26x7-10 &   20.519 &   0.016 &  -0.496 &   0.016 &   1.224 &   0.048 &   0.000 &   0.045 &   0.035 &   1.497 &   0.665 \nl
 53& uzk02-5 &   20.630 &   0.008 &  -0.742 &   0.012 &   0.435 &   0.022 &   0.000 &   0.043 &   0.000 & \nodata  & \nodata  \nl
 54& usa02-34 &   21.318 &   0.021 &  -0.751 &   0.023 &   0.776 &   0.026 &   0.173 &   0.042 &   0.067 &   0.269 &   0.689 \nl
 55& uui00-11 &   20.176 &   0.008 &  -0.617 &   0.009 &   0.588 &   0.009 &   0.000 &   0.041 &   0.044 &   0.511 &   0.558 \nl
 56& uim01-9 &   19.678 &   0.024 &   0.213 &   0.017 &   0.637 &   0.059 &   0.000 &   0.039 &   0.020 &   1.362 &   0.568 \nl
 57& uhdfk-68 &   22.465 &   0.010 &  -0.936 &   0.013 &   1.418 &   0.021 &   0.000 &   0.039 &   0.043 &   1.312 &   0.978 \nl
 58& ua-01-9 &   20.852 &   0.016 &  -0.860 &   0.021 &   1.242 &   0.086 &   0.400 &   0.000 &   0.037 & \nodata  & \nodata  \nl
 59& usp00-10 &   21.198 &   0.016 &  -0.266 &   0.000 &   0.637 &   0.035 &   0.123 &   0.000 &   0.036 & \nodata  & \nodata  \nl
 60& uhdfk-32 &   21.315 &   0.004 &  -0.922 &   0.006 &   0.457 &   0.014 &   0.247 &   0.034 &   0.000 & \nodata  & \nodata  \nl
 61& u26x7-7 &   20.685 &   0.017 &  -0.685 &   0.020 &   0.653 &   0.021 &   0.279 &   0.033 &   0.038 &   0.500 &   0.657 \nl
 62& umd0a-63 &   21.437 &   0.028 &  -0.711 &   0.033 &   0.398 &   0.056 &   0.158 &   0.032 &   0.017 &   1.085 &   0.829 \nl
 63& uhdfk-17 &   20.409 &   0.003 &  -0.363 &   0.003 &   1.181 &   0.006 &   0.000 &   0.031 &   0.000 & \nodata  & \nodata  \nl
 64& usp00-3 &   19.001 &   0.008 &   0.344 &   0.005 &   1.022 &   0.017 &   0.027 &   0.031 &   0.024 &   1.300 &   0.513 \nl
 65& uqk11-6 &   19.203 &   0.052 &   0.386 &   0.102 &   1.628 &   0.120 &   0.933 &   0.031 &   0.045 &   1.223 &   0.557 \nl
 66& ut201-37 &   21.348 &   0.014 &  -0.739 &   0.016 &   0.903 &   0.030 &   0.000 &   0.030 &   0.019 &   1.399 &   0.875 \nl
 67& uhdfk-63 &   22.603 &   0.007 &  -1.628 &   0.010 &   0.611 &   0.009 &   0.000 &   0.028 &   0.035 &   0.369 &   0.974 \nl
 68& ua-30-20 &   21.381 &   0.027 &  -0.759 &   0.028 &   0.909 &   0.045 &   0.000 &   0.027 &   0.032 &   0.725 &   0.814 \nl
 69& ugk00-1 &   18.206 &   0.008 &   0.572 &   0.006 &   0.762 &   0.017 &   0.000 &   0.026 &   0.017 &   1.223 &   0.444 \nl
 70& u26x9-8 &   20.217 &   0.023 &   0.043 &   0.019 &   0.751 &   0.061 &   0.000 &   0.025 &   0.000 & \nodata  & \nodata  \nl
 71& u26x8-7 &   18.880 &   0.011 &   0.279 &   0.008 &   0.820 &   0.017 &   0.000 &   0.025 &   0.000 & \nodata  & \nodata  \nl
 72& ust01-7 &   19.280 &   0.017 &  -0.068 &   0.016 &   0.906 &   0.030 &   0.000 &   0.024 &   0.000 & \nodata  & \nodata  \nl
 73& uj000-28 &   20.111 &   0.019 &  -0.109 &   0.018 &   0.303 &   0.049 &   0.000 &   0.023 &   0.000 & \nodata  & \nodata  \nl
 74& u26x7-14 &   20.417 &   0.042 &   0.500 &   0.027 &   0.652 &   0.166 &   0.000 &   0.022 &   0.000 & \nodata  & \nodata  \nl
 75& uhdfk-46 &   22.199 &   0.007 &  -1.351 &   0.007 &   0.650 &   0.009 &   0.000 &   0.021 &   0.024 &   0.505 &   0.969 \nl
 76& usa01-34 &   21.149 &   0.008 &  -1.360 &   0.013 &   1.033 &   0.012 &   0.000 &   0.020 &   0.057 &  -0.104 &   0.799 \nl
 77& ut201-33 &   21.118 &   0.033 &  -0.216 &   0.029 &   0.829 &   0.093 &   0.000 &   0.020 &   0.016 &   1.071 &   0.875 \nl
 78& uy400-15 &   21.469 &   0.017 &  -1.178 &   0.018 &   1.372 &   0.038 &   0.148 &   0.019 &   0.054 &   0.238 &   0.884 \nl
 79& uy000-14 &   19.829 &   0.035 &  -0.174 &   0.042 &   1.415 &   0.040 &   0.597 &   0.019 &   0.020 &   1.359 &   0.699 \nl
 80& ua400-8 &   19.913 &   0.008 &  -0.054 &   0.008 &   0.833 &   0.012 &   0.000 &   0.018 &   0.011 &   1.368 &   0.711 \nl
 81& urp03-8 &   19.488 &   0.006 &   0.293 &   0.007 &   0.766 &   0.031 &   0.191 &   0.000 &   0.018 & \nodata  & \nodata  \nl
 82& uqa01-21 &   19.183 &   0.008 &  -0.818 &   0.008 &   0.846 &   0.014 &   0.000 &   0.018 &   0.012 &   1.286 &   0.608 \nl
 83& uj000-20 &   20.034 &   0.190 &   0.184 &   0.097 &   0.048 &   0.190 &   0.000 &   0.017 &   0.000 & \nodata  & \nodata  \nl
 84& usa00-5 &   18.945 &   0.007 &   0.259 &   0.006 &   0.569 &   0.009 &   0.000 &   0.016 &   0.014 &   0.714 &   0.544 \nl
 85& ust00-23 &   20.667 &   0.030 &   0.136 &   0.020 &   1.755 &   0.042 &   0.000 &   0.016 &   0.000 & \nodata  & \nodata  \nl
 86& ujh01-2 &   18.329 &   0.005 &  -0.088 &   0.006 &   0.790 &   0.010 &   0.002 &   0.016 &   0.011 &   1.197 &   0.556 \nl
 87& u26x6-6 &   21.076 &   0.041 &  -0.690 &   0.033 &  -0.026 &   0.042 &   0.000 &   0.016 &   0.000 & \nodata  & \nodata  \nl
 88& ux400-7 &   18.932 &   0.003 &  -0.203 &   0.004 &   0.763 &   0.006 &   0.000 &   0.016 &   0.000 & \nodata  & \nodata  \nl
 89& uko01-25 &   20.378 &   0.036 &   0.040 &   0.028 &   0.617 &   0.155 &   0.000 &   0.016 &   0.039 &  -0.350 &   0.695 \nl
 90& ueh00-2 &   18.065 &   0.016 &   0.565 &   0.010 &   0.719 &   0.023 &   0.000 &   0.016 &   0.011 &   1.126 &   0.476 \nl
 91& u26x8-5 &   19.000 &   0.010 &   0.1209 &   0.008 &   0.716 &   0.016 &   0.000 &   0.015 &   0.000 & \nodata  & \nodata  \nl
 92& urz00-8 &   19.401 &   0.005 &  -0.087 &   0.005 &   0.853 &   0.007 &   0.030 &   0.015 &   0.016 &   0.783 &   0.620 \nl
 93& uem00-4 &   18.695 &   0.011 &   0.490 &   0.008 &   1.177 &   0.022 &   0.000 &   0.015 &   0.012 &   1.419 &   0.580 \nl
 94& uwy02-5 &   19.135 &   0.010 &  -0.050 &   0.011 &   0.686 &   0.010 &   0.199 &   0.015 &   0.000 & \nodata  & \nodata  \nl
 95& uzx07-4 &   18.955 &   0.025 &   0.488 &   0.022 &   0.703 &   0.028 &   0.047 &   0.000 &   0.013 & \nodata  & \nodata  \nl
 96& uhdfk-27 &   20.972 &   0.004 &  -0.049 &   0.005 &   1.015 &   0.004 &   0.000 &   0.013 &   0.017 &   0.724 &   0.872 \nl
 97& u26x8-12 &   20.219 &   0.031 &   0.155 &   0.020 &   1.223 &   0.078 &   0.000 &   0.013 &   0.000 & \nodata  & \nodata  \nl
 98& uui00-3 &   18.417 &   0.005 &   0.076 &   0.005 &   0.812 &   0.006 &   0.012 &   0.012 &   0.027 &  -0.068 &   0.477 \nl
 99& umd08-13 &   18.085 &   0.004 &   0.222 &   0.001 &   0.620 &   0.018 &   0.059 &   0.000 &   0.012 & \nodata  & \nodata  \nl
100& usa02-6 &   18.658 &   0.016 &   0.403 &   0.015 &   0.745 &   0.027 &   0.204 &   0.012 &   0.000 & \nodata  & \nodata  \nl
101& uim01-4 &   18.603 &   0.019 &   0.578 &   0.020 &   1.155 &   0.041 &   0.530 &   0.012 &   0.027 &   0.275 &   0.524 \nl
\enddata
\end{deluxetable}

\begin{deluxetable}{lccccc}
\tablewidth{0pt}
\tablecaption{Redshift Estimation Coefficients}
\tablehead{
\colhead{Bulge/Total} & \colhead{$\sigma$$_z$} 
& \colhead{\# of gals} &  \colhead{C1 (color)} 
& \colhead{C2 (mag)} & \colhead{C3 (zero pt.)}}

\startdata
$\leq$0.20  & 0.09657  &  66 & 0.5154 &  0.0681 & -1.4084 \nl
$\leq$0.50  & 0.10043  &  88 & 0.4013 &  0.0761 & -1.4901 \nl
$>$0.50 & 0.11124 &   39 & 0.2768 & 0.1115 & -2.0903 \nl
$>$0.80 & 0.10246 &   22 & 0.0868 & 0.1560 & -2.6792 \nl
\enddata
\end{deluxetable}

\begin{deluxetable}{rlrrllc}
\tablewidth{0pt}
\tablecaption{Redshift List}
\tablecomments{Comments are as follows: (*)-spectroscopically 
determined at the KPNO 4-meter;
(1)-from Cohen et al. (1996); (2)-from Koo et al. (1996).
Additional information on individual objects can be obtained
at the MDS website at http://archive.stsci.edu/mds/mds.cgi.}
\tablehead{
\colhead{ID\#} &
\colhead{Name} & 
\colhead{RA} &
\colhead{DEC} &
\colhead{Redshift}  &
\colhead{$\sigma$$_z$}         & \colhead{Comment}}

\startdata
  1& ua400-7   &  0 24 53.4 & -27 16 43.4 &   0.9947 &  0.0003 &* \nl
  2& uwy02-4   & 12 40 20.6 & -11 32 13.5 &   0.4556 &  0.0004 &* \nl
  3& ua-30-13  &  0 58 02.7 & -28 12 43.2 &   0.5547 &  0.0975 &  \nl
  4& ust01-11  & 10 05 19.2 &  -7 46 24.8 &   0.5113 &  0.1025 &  \nl
  5& u26x9-20  & 14 17 28.3 &  52 25 56.7 &   1.0727 &  0.0995 &  \nl
  6& uy401-12  & 14 35 32.1 &  25 17 38.0 &   0.4632 &  0.0970 &  \nl
  7& ut200-28  & 16 01 08.6 &   5 12 47.7 &   0.7611 &  0.0989 &  \nl
  8& ufj00-13  &  2 07 06.5 &  15 25 00.7 &   0.4798 &  0.0969 &  \nl
  9& udm10-24  &  2 42 54.2 &  -0 04 30.9 &   0.8381 &  0.1025 &  \nl
 10& umd0a-47  & 21 51 14.3 &  29 01 07.3 &   0.8951 &  0.0990 &  \nl
 11& ua400-26  &  0 24 52.9 & -27 16 09.3 &   0.8575 &  0.0977 &  \nl
 12& uvd01-26  & 15 43 22.6 &  53 52 11.2 &   0.7382 &  0.0971 &  \nl
 13& ua-01-10  &  1 04 40.5 & -27 05 43.7 &   0.8287 &  0.0974 &  \nl
 14& uzx01-37  & 12 30 55.3 &  12 18 34.3 &   0.7486 &  0.1006 &  \nl
 15& ut201-18  & 16 01 08.4 &   5 35 25.2 &   0.7706 &  0.0977 &  \nl
 16& uvd01-14  & 15 43 20.9 &  53 51 52.1 &   0.5603 &  0.0968 &  \nl
 17& uzx00-3   & 12 30 14.0 &  12 22 29.4 &   0.2474 &  0.0003 &* \nl
 18& usa00-35  & 17 12 25.9 &  33 36 36.4 &   0.2549 &  0.0002 &* \nl
 19& ueh02-14  &  0 53 38.0 &  12 50 20.8 &   0.5021 &  0.2084 &  \nl
 20& umd0a-25  & 21 51 14.6 &  28 59 55.7 &   0.3951 &  0.0003 &* \nl
 21& uui00-31  & 11 42 14.9 &  71 38 16.9 &   0.4566 &  0.1013 &  \nl
 22& u26x6-11  & 14 17 46.0 &  52 30 32.6 &   1.1674 &  0.1012 &  \nl
 23& usa00-9   & 17 12 29.4 &  33 36 36.0 &   0.3235 &  0.0003 &* \nl
 24& uzp01-19  & 11 50 27.0 &  28 47 27.4 &   0.6206 &  0.0984 &  \nl
 25& umd05-46  & 21 51 03.1 &  29 00 16.6 &   0.3284 &  0.1373 &  \nl
 26& ueh02-4   &  0 53 37.9 &  12 49 48.8 &   0.3853 &  0.1575 &  \nl
 27& uy401-4   & 14 35 29.3 &  25 19 23.4 &   0.3938 &  0.0003 &* \nl
 28& uj700-29  & 19 40 33.8 & -69 17 13.7 &   0.6446 &  0.1013 &  \nl
 29& uvd01-12  & 15 43 16.3 &  53 52 35.8 &   0.4164? &  0.0003 &* \nl
 30& uzp01-24  & 11 50 29.6 &  28 47 45.1 &   0.5450 &  0.0981 &  \nl
 31& umd05-37  & 21 51 02.8 &  28 59 43.8 &   0.1918 &  0.0002 &* \nl
 32& u26x5-6   & 14 17 47.9 &  52 30 47.0 &   0.4915 &  0.0972 &  \nl
 33& uy400-16  & 14 35 18.6 &  24 59 05.9 &   0.7761 &  0.1018 &  \nl
 34& ust00-8   & 10 05 44.5 &  -7 41 07.2 &   0.5360 &  0.0003 &* \nl
 35& uzx00-4   & 12 30 20.9 &  12 22 47.7 &   0.2604 &  0.0002 &* \nl
 36& uzx00-20  & 12 30 20.4 &  12 21 50.5 &   0.4604 &  0.0971 &  \nl
 37& u26x7-18  & 14 17 42.2 &  52 26 45.2 &   0.7111 &  0.0982 &  \nl
 38& u26x8-23  & 14 17 35.8 &  52 25 33.1 &   0.6830 &  0.0975 &  \nl
 39& uzd10-6   & 13 55 23.0 &  40 20 49.7 &   0.3691 &  0.1007 &  \nl
 40& u26x9-25  & 14 17 23.1 &  52 25 28.9 &   0.7302 &  0.1007 &  \nl
 41& u26x6-9   & 14 17 46.4 &  52 30 43.8 &   0.5208 &  0.0976 &  \nl
 42& ust00-27  & 10 05 43.8 &  -7 42 01.5 &   0.7774 &  0.0971 &  \nl
 43& uhdfk-106 & 12 36 41.4 &  62 12 16.3 &   1.1670 &  0.0978 &  \nl
 44& umd09-4   & 21 50 40.3 &  28 56 29.1 &   0.2488 &  0.0002 &* \nl
 45& urp03-14  &  8 47 23.3 &  17 57 51.2 &   0.3231 &  0.0987 &  \nl
 46& uxy10-6   & 12 36 37.5 &  -0 42 58.8 &   0.3904 &  0.0971 &  \nl
 47& uui00-17  & 11 41 57.1 &  71 37 30.1 &   0.5513?? &  0.0003 &* \nl
 48& uci10-11  &  1 24 41.2 &   3 51 56.6 &   0.3801 &  0.0970 &  \nl
 49& uhdfk-36  & 12 36 49.6 &  62 12 58.6 &   0.4750 &  \nodata &1 \nl
 50& u26x8-31  & 14 17 35.1 &  52 25 41.5 &   0.6886 &  0.1030 &  \nl
 51& ufj00-17  &  2 07 10.4 &  15 25 37.4 &   0.2967 &  0.0971 &  \nl
 52& u26x7-10  & 14 17 40.6 &  52 27 13.1 &   0.6174 &  0.0997 &  \nl
 53& uzk02-5   & 12 11 19.0 &  39 26 27.0 &   0.1994 &  0.0973 &  \nl
 54& usa02-34  & 17 12 28.0 &  33 35 29.3 &   0.4613 &  0.0975 &  \nl
 55& uui00-11  & 11 42 02.8 &  71 37 18.5 &   0.2735 &  0.0967 &  \nl
 56& uim01-9   &  3 55 30.1 &   9 43 44.7 &   0.2520 &  0.1013 &  \nl
 57& uhdfk-68  & 12 36 49.1 &  62 11 49.6 &   0.8576 &  0.0972 &  \nl
 58& ua-01-9   &  1 04 35.0 & -27 04 54.2 &   0.6116 &  0.1063 &  \nl
 59& usp00-10  &  8 54 18.7 &  20 03 51.4 &   0.3020? &  0.0003 &* \nl
 60& uhdfk-32  & 12 36 49.5 &  62 14 07.7 &   0.7520 &  \nodata &1 \nl
 61& u26x7-7   & 14 17 37.1 &  52 27 39.1 &   0.3511 &  0.1008 &  \nl
 62& umd0a-63  & 21 51 17.7 &  29 00 54.4 &   0.2504 &  0.1008 &  \nl
 63& uhdfk-17  & 12 36 53.8 &  62 12 54.9 &   0.6420 &  \nodata &1 \nl
 64& usp00-3   &  8 54 12.4 &  20 04 15.4 &   0.4573? &  0.0003 &* \nl
 65& uqk11-6   &  7 24 48.9 &  60 31 26.3 &   0.4964 &  0.0003 &* \nl
 66& ut201-37  & 16 01 14.6 &   5 34 35.0 &   0.5067 &  0.0978 &  \nl
 67& uhdfk-63  & 12 36 43.3 &  62 11 52.8 &   0.4519 &  0.0967 &  \nl
 68& ua-30-20  &  0 58 02.9 & -28 12 29.0 &   0.5211 &  0.0994 &  \nl
 69& ugk00-1   &  2 38 52.5 &  16 43 29.3 &   0.2752 &  0.0003 &* \nl
 70& u26x9-8   & 14 17 24.6 &  52 23 57.3 &   0.3431 &  0.1015 &  \nl
 71& u26x8-7   & 14 17 31.1 &  52 25 24.0 &   0.2870 &  \nodata &2 \nl
 72& ust01-7   & 10 05 12.2 &  -7 47 15.8 &   0.3597 &  0.0978 &  \nl
 73& uj000-28  & 19 39 18.3 & -46 14 46.8 &   0.4663 &  0.1935 &  \nl
 74& u26x7-14  & 14 17 39.9 &  52 28 20.8 &   0.3072 &  0.1292 &  \nl
 75& uhdfk-46  & 12 36 45.3 &  62 13 27.0 &   0.4416 &  0.0967 &  \nl
 76& usa01-34  & 17 12 27.1 &  33 35 57.6 &   0.5873 &  0.0969 &  \nl
 77& ut201-33  & 16 01 11.3 &   5 35 32.9 &   0.4562 &  0.1079 &  \nl
 78& uy400-15  & 14 35 16.9 &  24 59 11.4 &   0.7825 &  0.0986 &  \nl
 79& uy000-14  & 14 16 15.4 &  11 32 04.6 &   0.5149 &  0.1119 &  \nl
 80& ua400-8   &  0 24 53.9 & -27 15 57.5 &   0.4310 &  0.0003 &* \nl
 81& urp03-8   &  8 47 21.3 &  17 56 52.7 &   0.3280 &  0.0003 &* \nl
 82& uqa01-21  & 17 36 41.5 &  28 04 31.2 &   0.2528? &  0.0003 &* \nl
 83& uj000-20  & 19 39 19.9 & -46 13 02.8 &   0.4797 &  0.1973 &  \nl
 84& usa00-5   & 17 12 21.3 &  33 35 56.7 &   0.2555 &  0.0002 &* \nl
 85& ust00-23  & 10 05 49.2 &  -7 41 38.5 &   0.8957 &  0.0990 &  \nl
 86& ujh01-2   &  1 09 00.4 &  35 35 38.9 &   0.2480 &  0.0002 &* \nl
 87& u26x6-6   & 14 17 47.6 &  52 29 04.1 &   0.8098? &  \nodata &2 \nl
 88& ux400-7   & 15 19 38.9 &  23 53 02.0 &   0.2663 &  0.0966 &  \nl
 89& uko01-25  &  4 56 49.5 &   3 52 06.7 &   0.5929? &  0.0003 &* \nl
 90& ueh00-2   &  0 53 22.4 &  12 32 56.0 &   0.1908 &  0.0973 &  \nl
 91& u26x8-5   & 14 17 26.5 &  52 27 05.7 &   0.2472 &  0.0969 &  \nl
 92& urz00-8   & 12 53 00.6 & -29 15 04.3 &   0.3541 &  0.0966 &  \nl
 93& uem00-4   &  3 04 59.2 &  -0 11 46.5 &   0.4775 &  0.0004 &* \nl
 94& uwy02-5   & 12 40 22.6 & -11 31 12.5 &   0.3060 &  0.0002 &* \nl
 95& uzx07-4   & 12 30 47.7 &  12 19 43.4 &   0.2681 &  0.0003 &* \nl
 96& uhdfk-27  & 12 36 49.7 &  62 13 14.0 &   0.4750 &  \nodata &1 \nl
 97& u26x8-12  & 14 17 34.4 &  52 25 25.8 &   0.5925 &  0.1047 &  \nl
 98& uui00-3   & 11 42 01.7 &  71 37 33.1 &   0.2221 &  0.0002 &* \nl
 99& umd08-13  & 21 50 32.2 &  28 50 29.4 &   0.1220 &  0.0002 &* \nl
100& usa02-6   & 17 12 20.5 &  33 34 41.4 &   0.2600 &  0.0003 &* \nl
101& uim01-4   &  3 55 33.2 &   9 44 45.6 &   0.3097 &  0.1118 &  \nl
\enddata
\end{deluxetable}

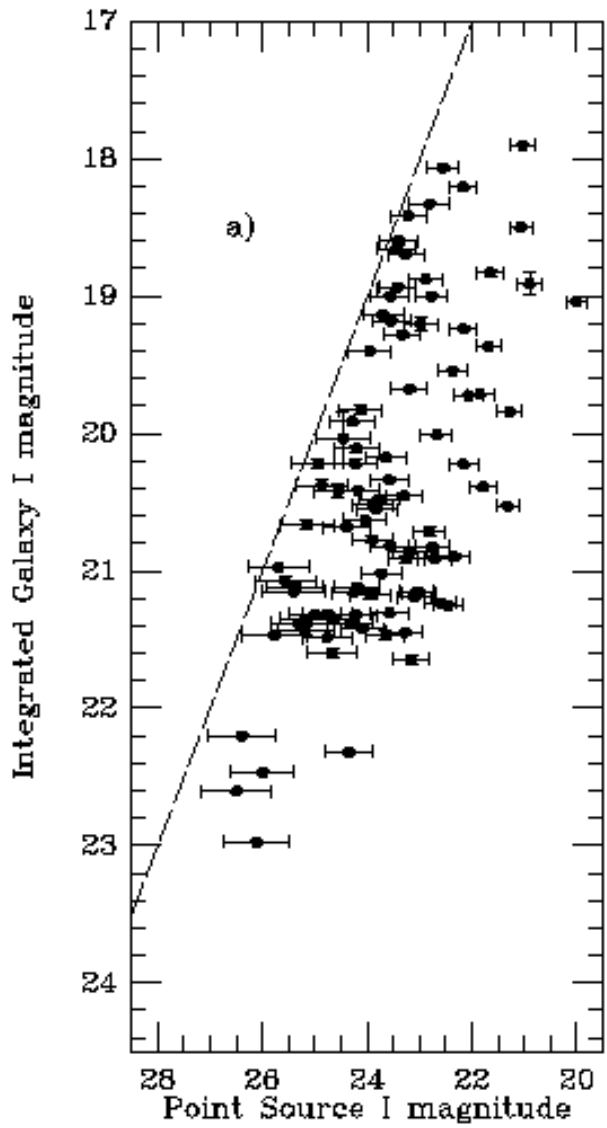
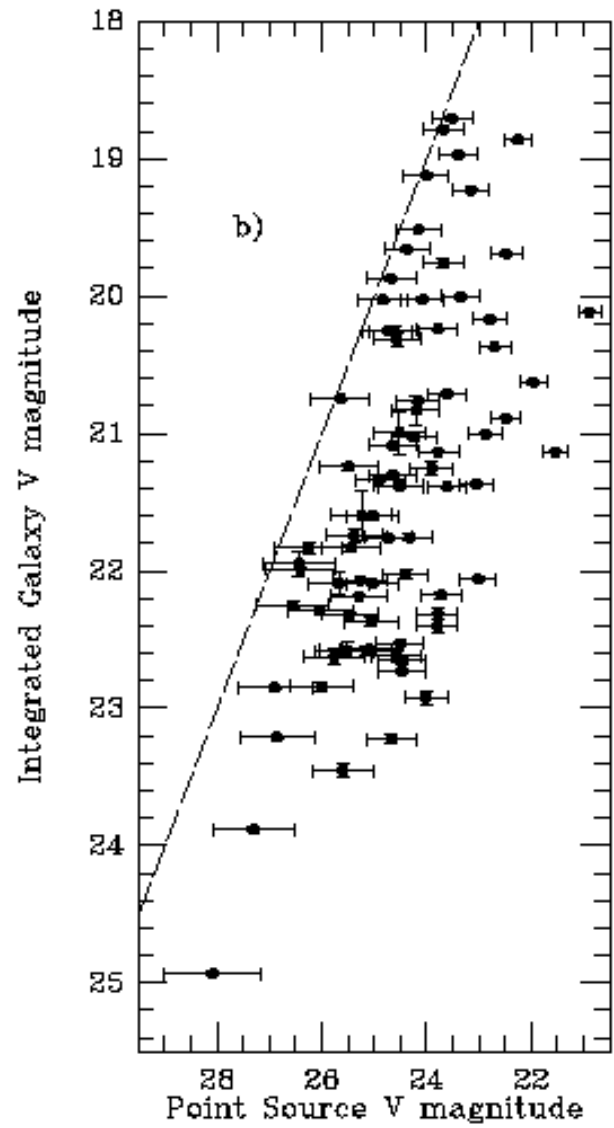
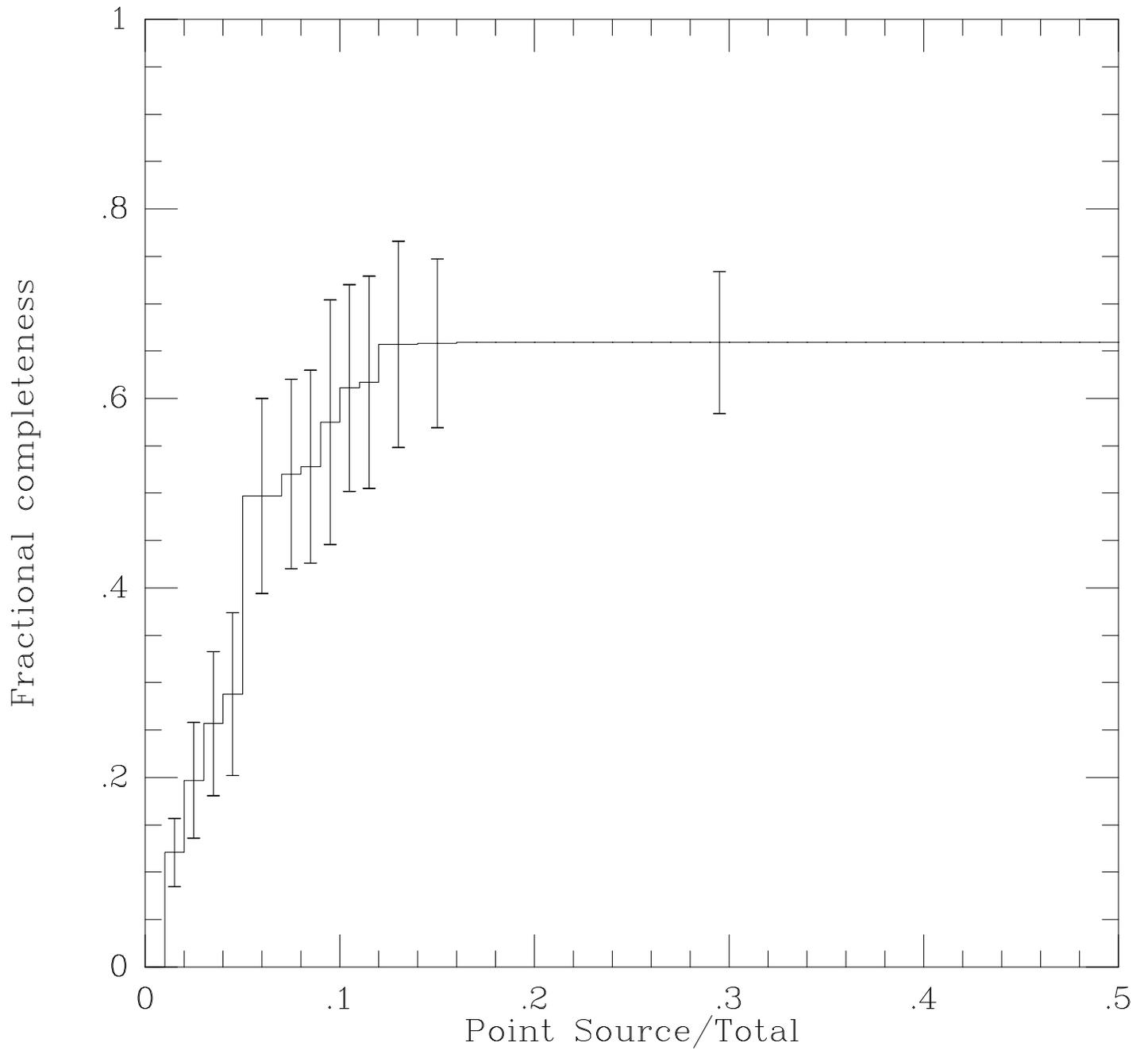
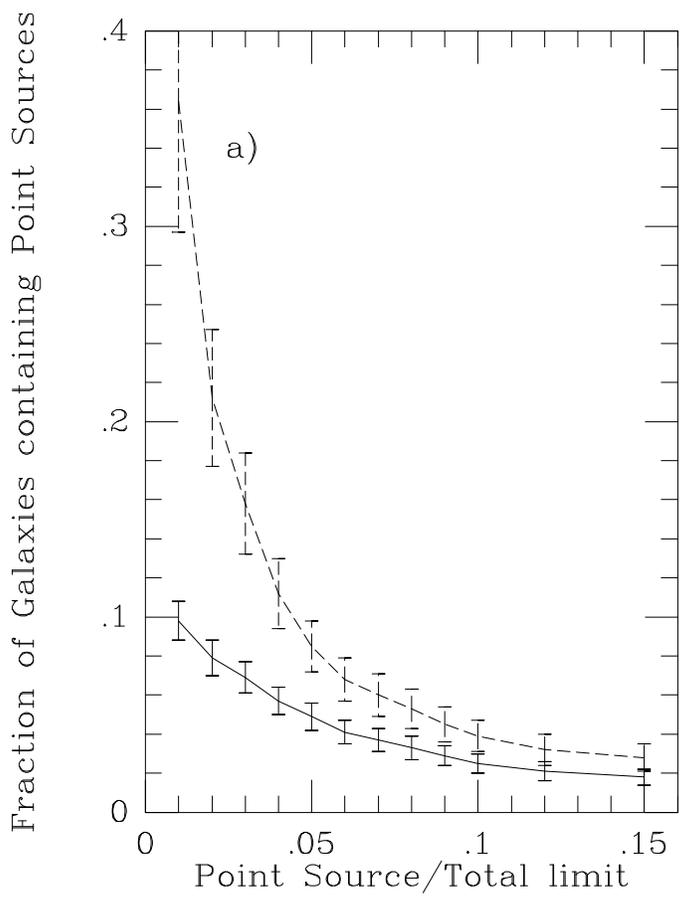
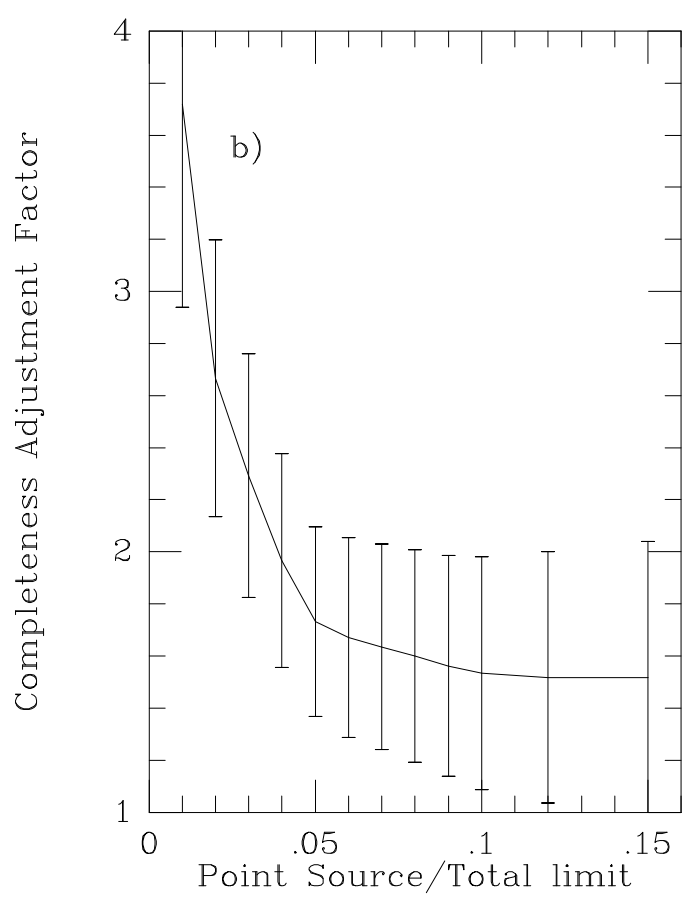
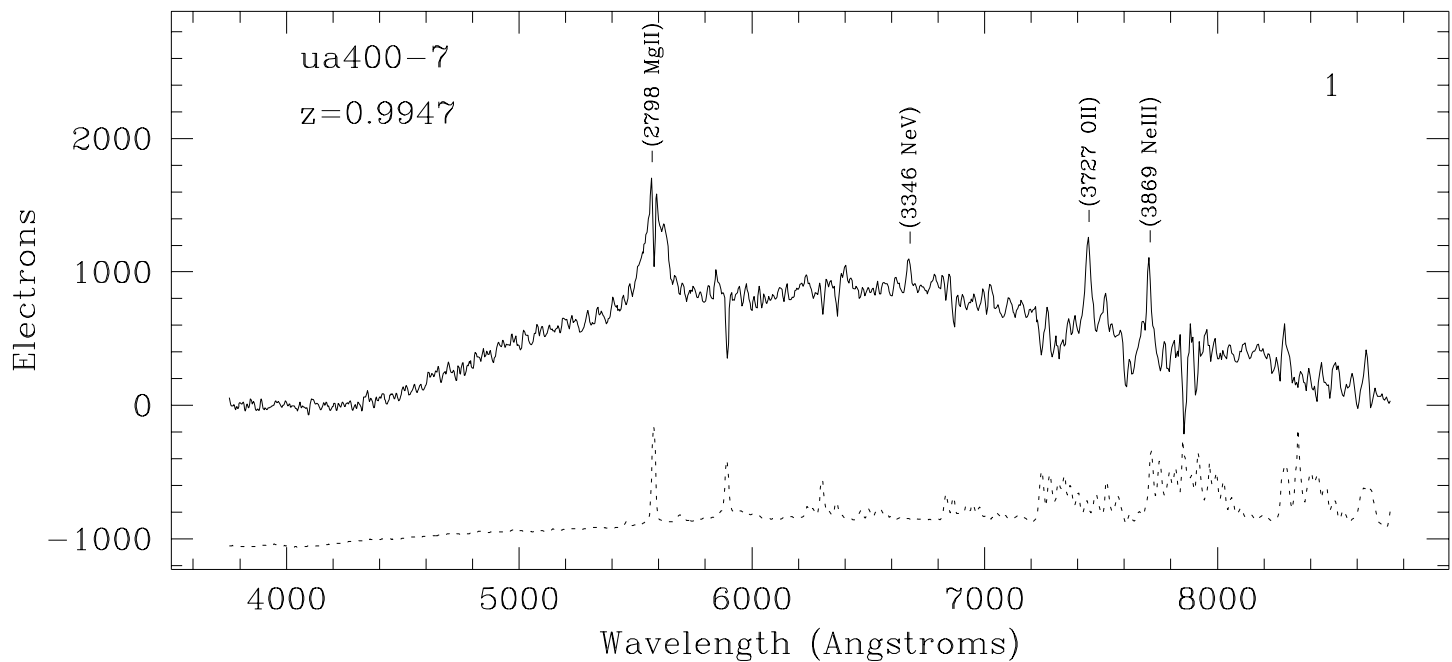
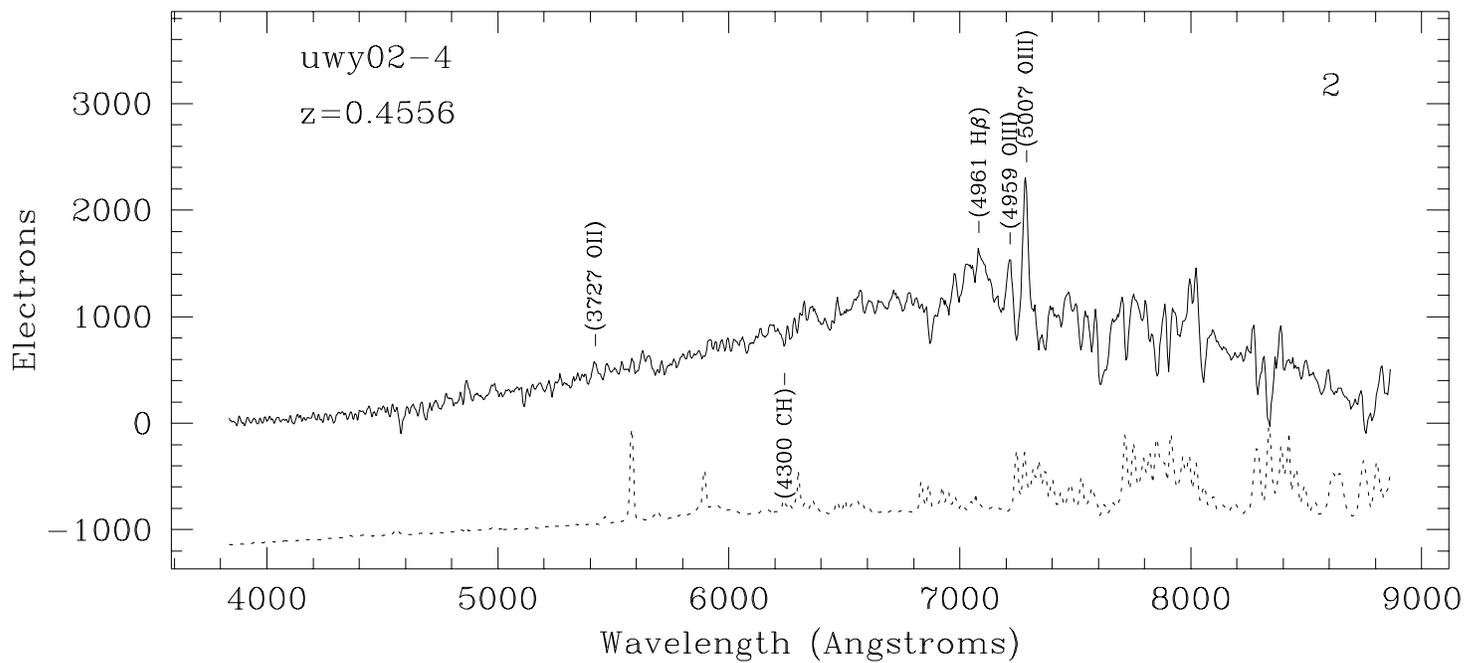
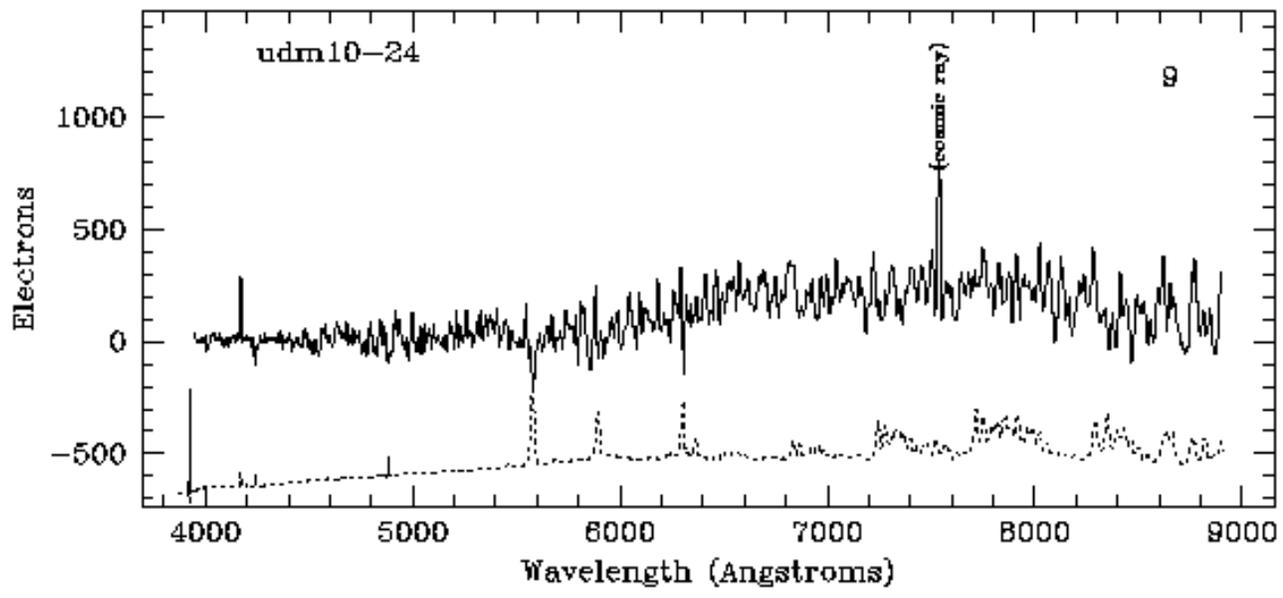
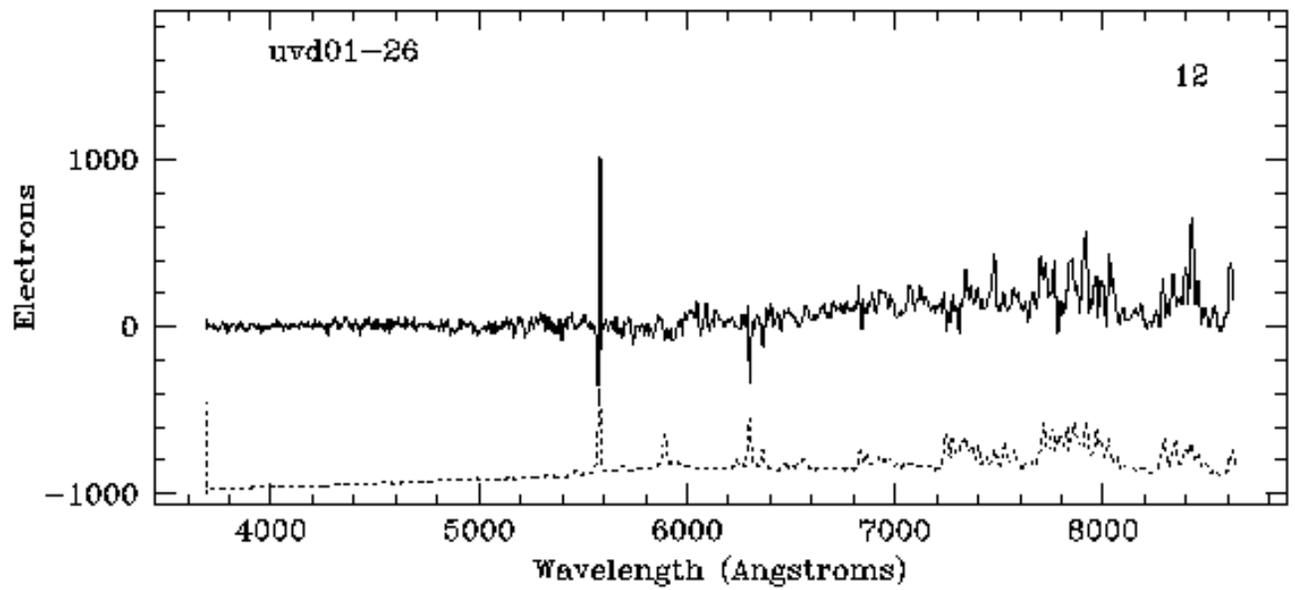
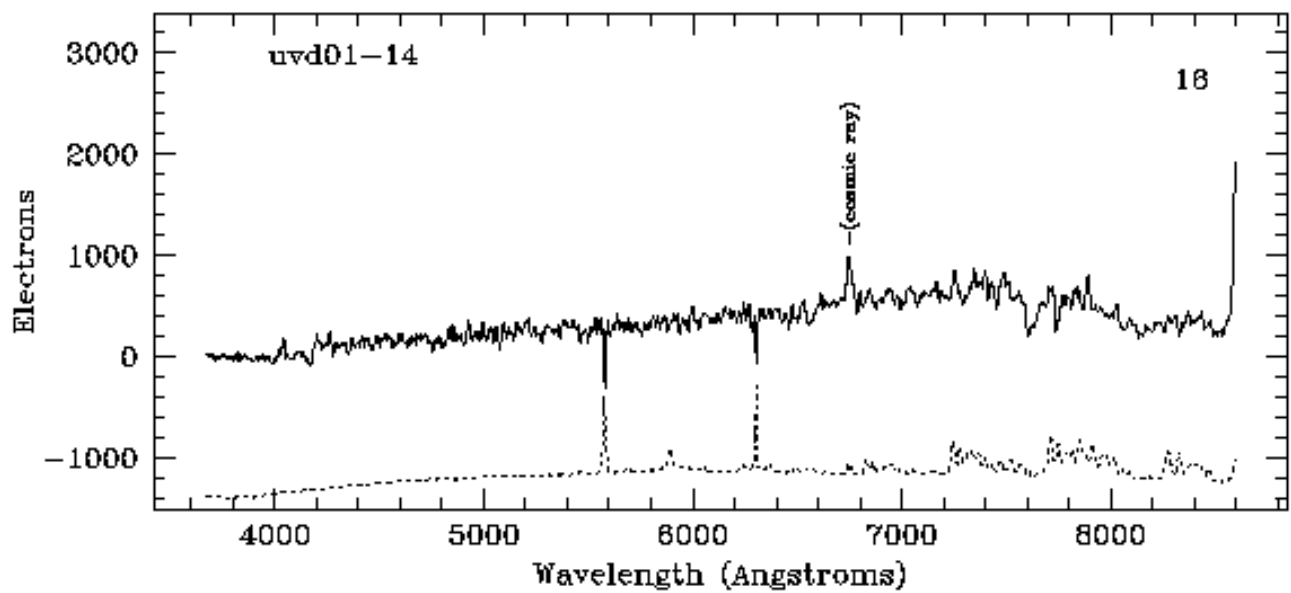
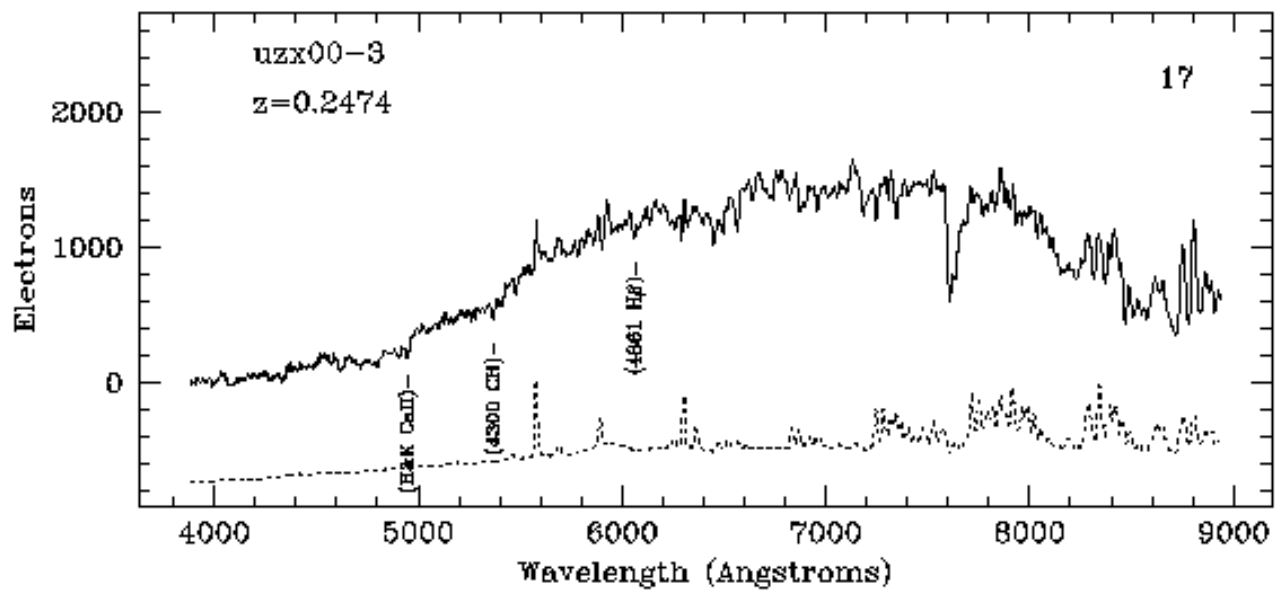
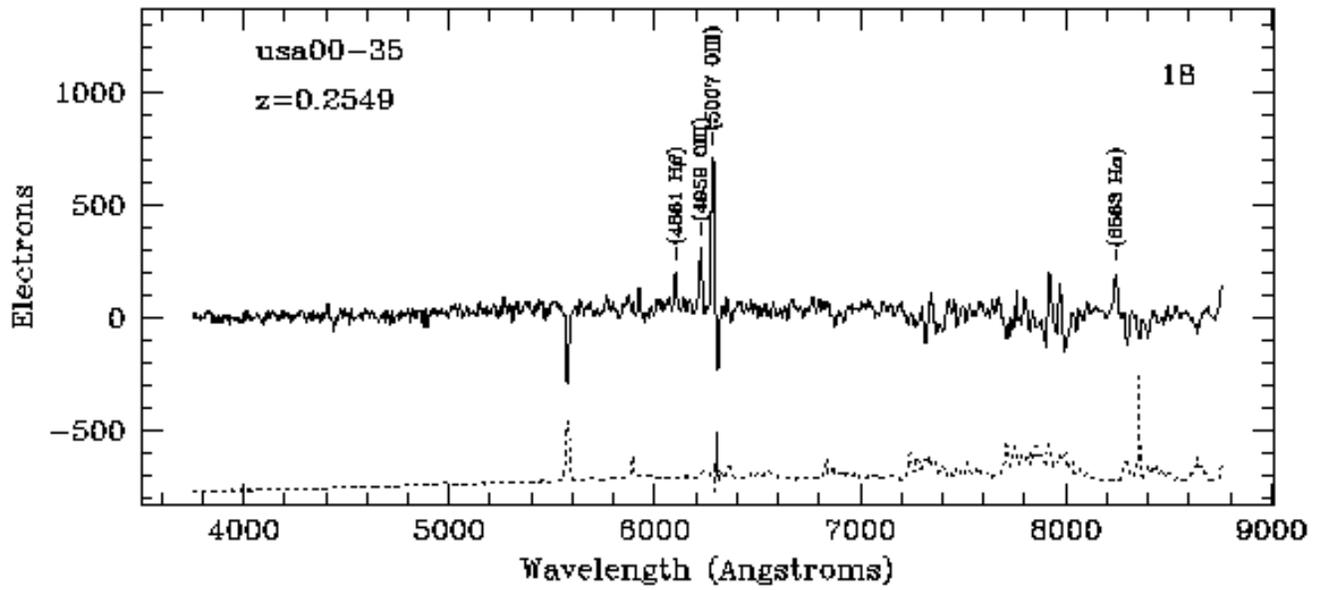
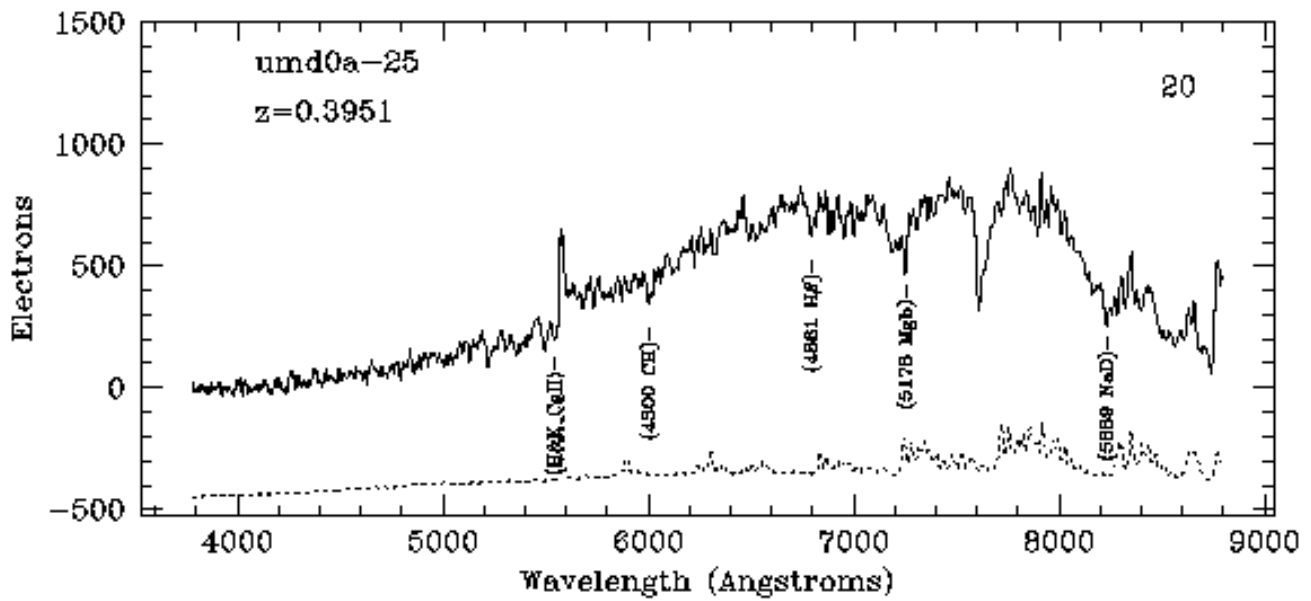
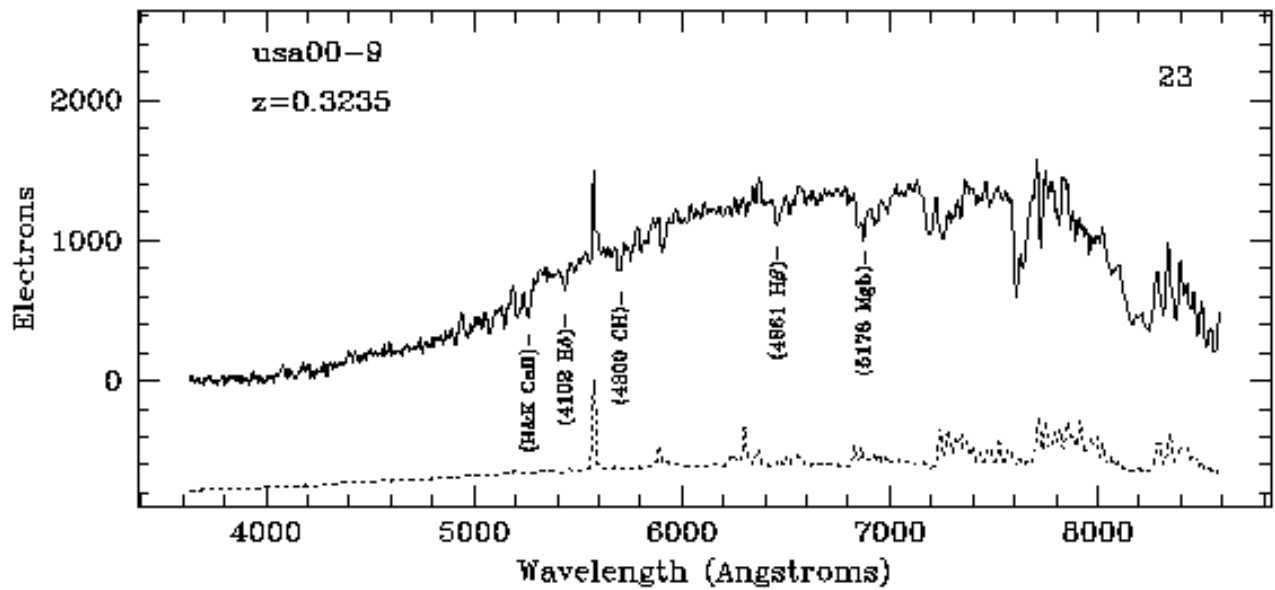
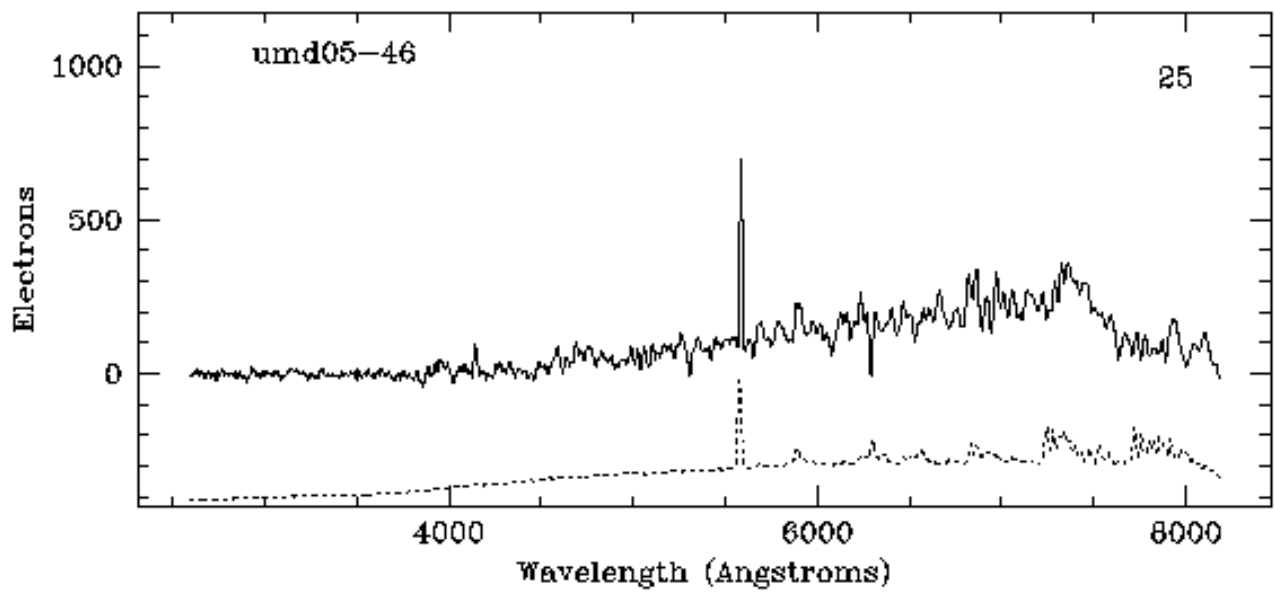
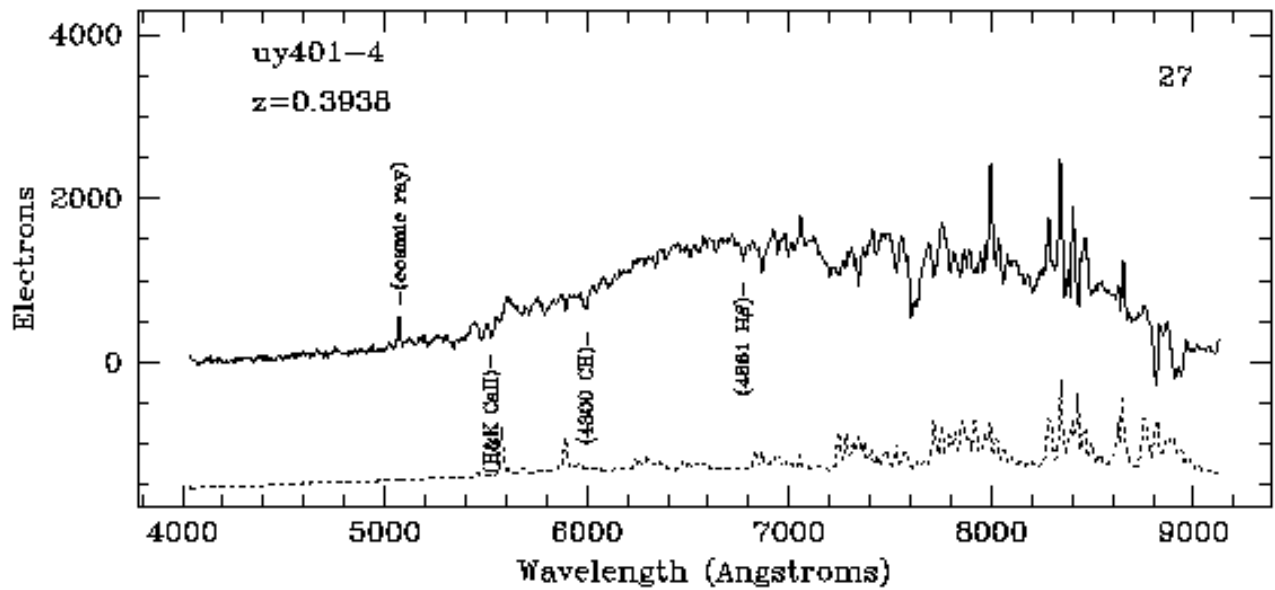
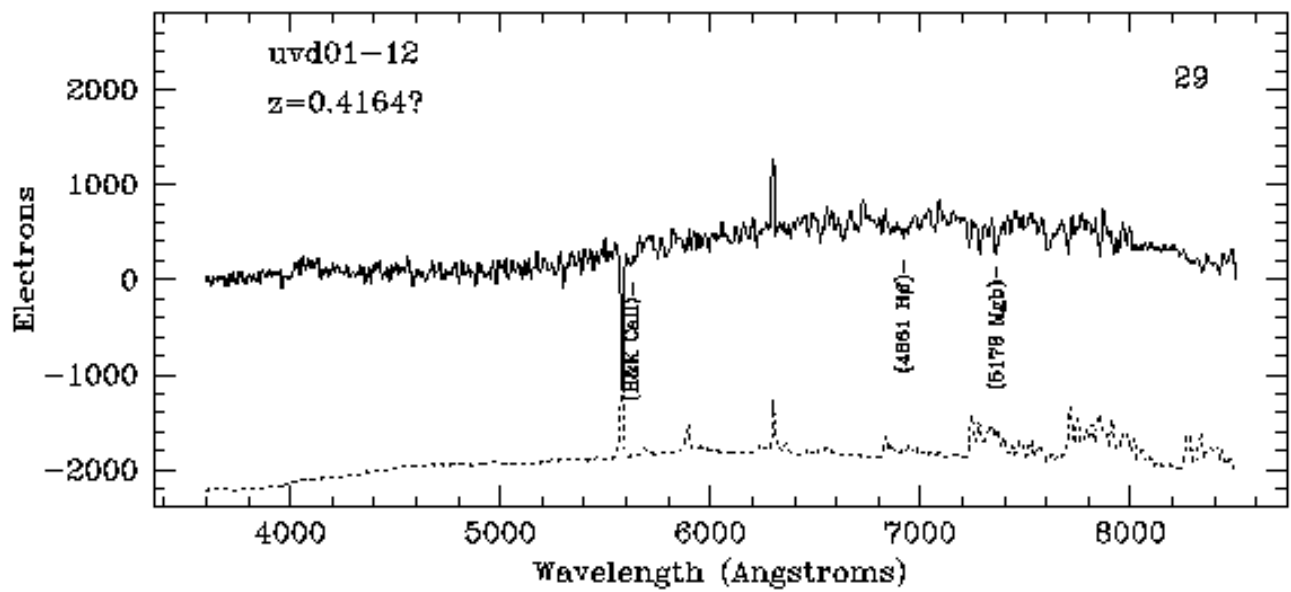
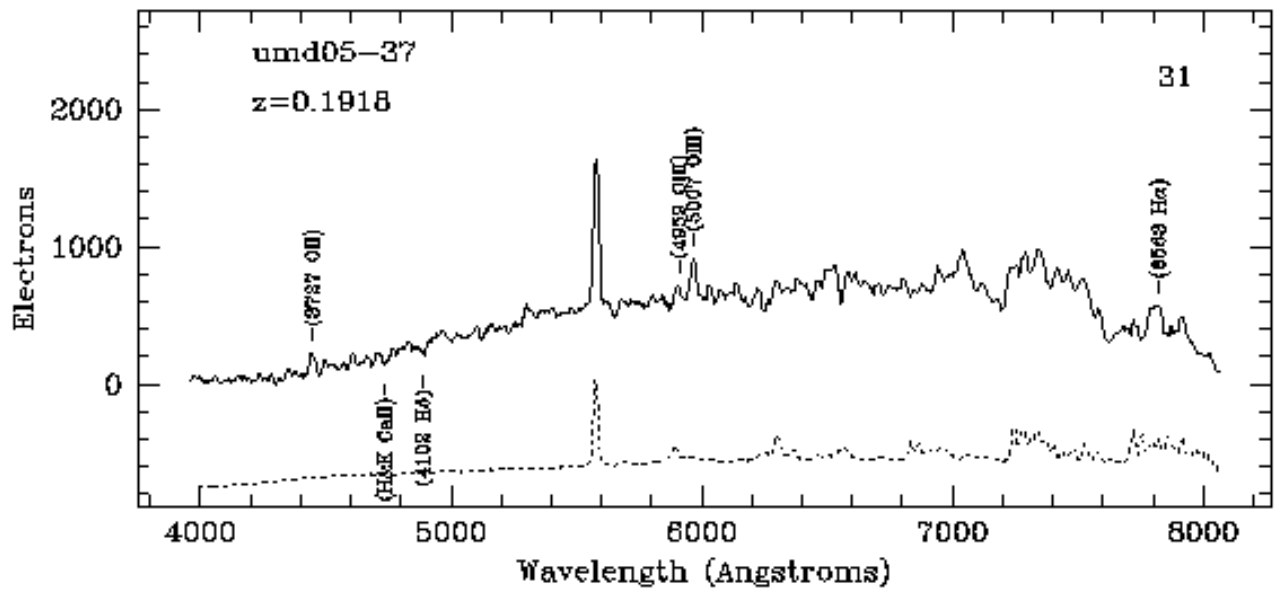
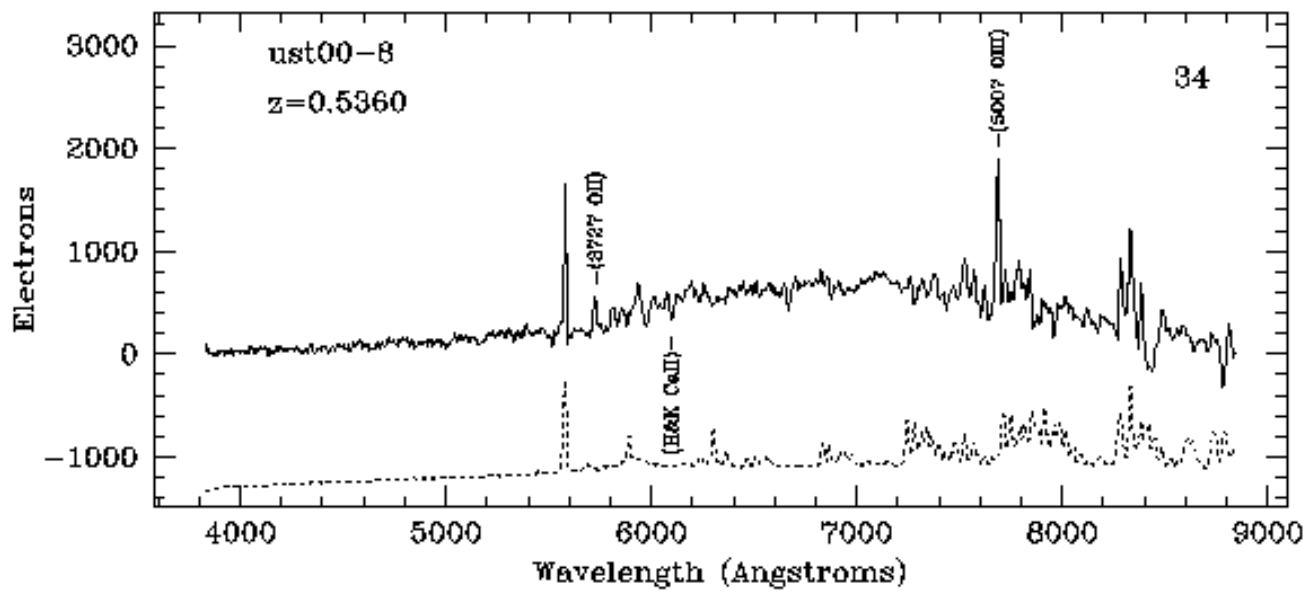
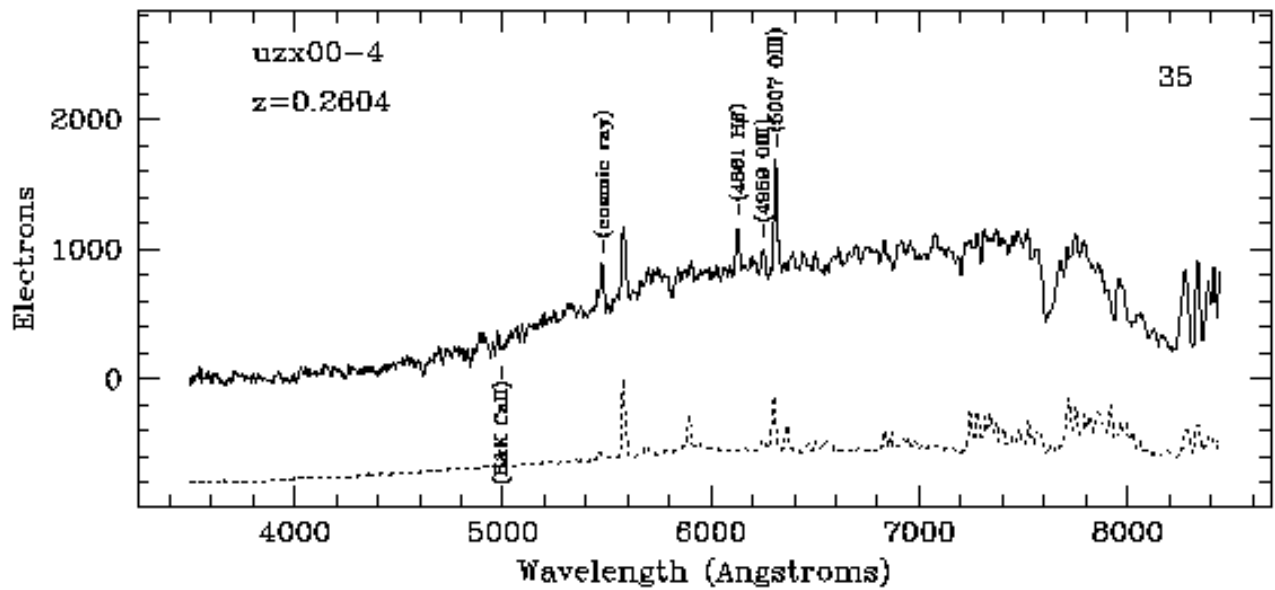
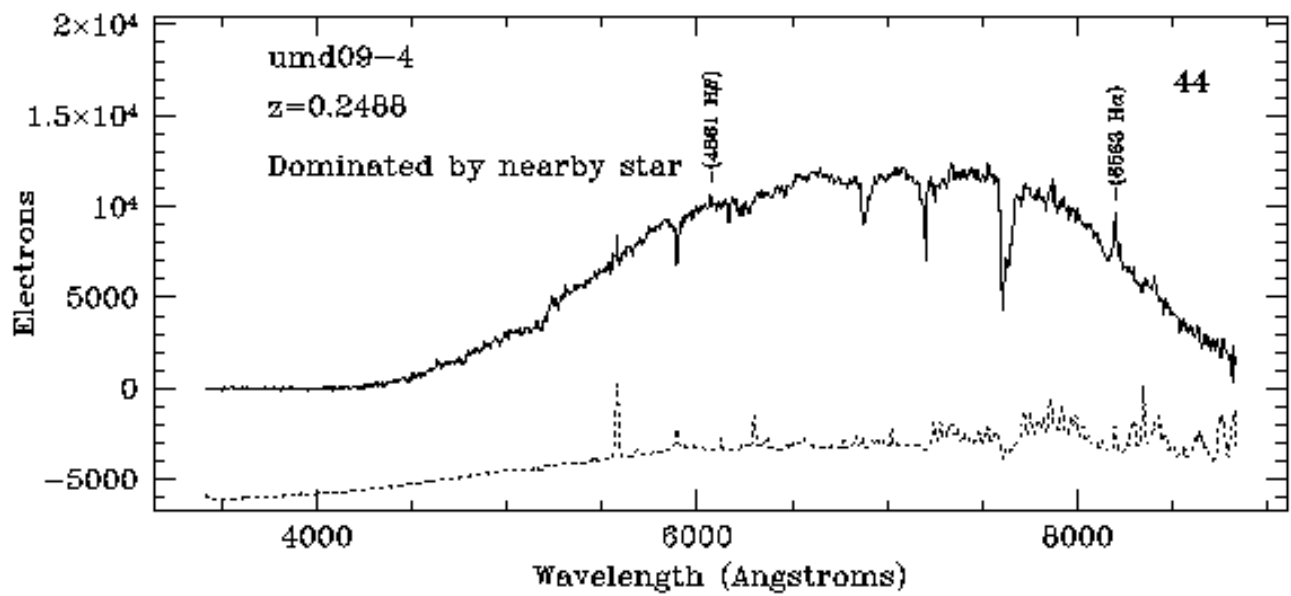
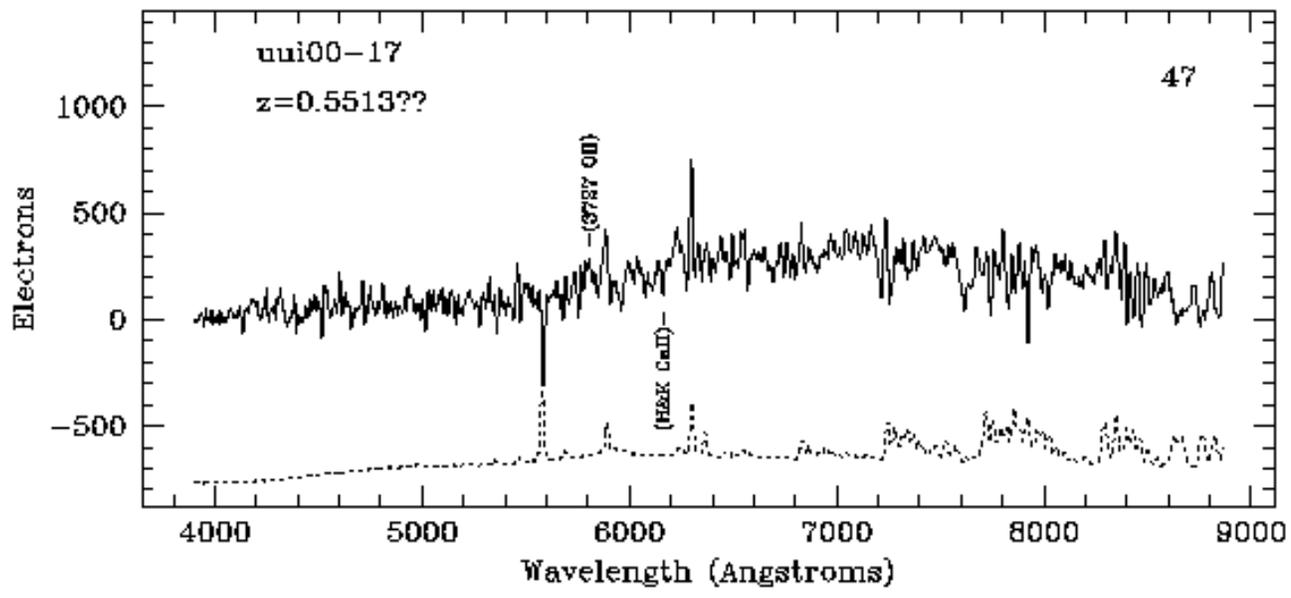
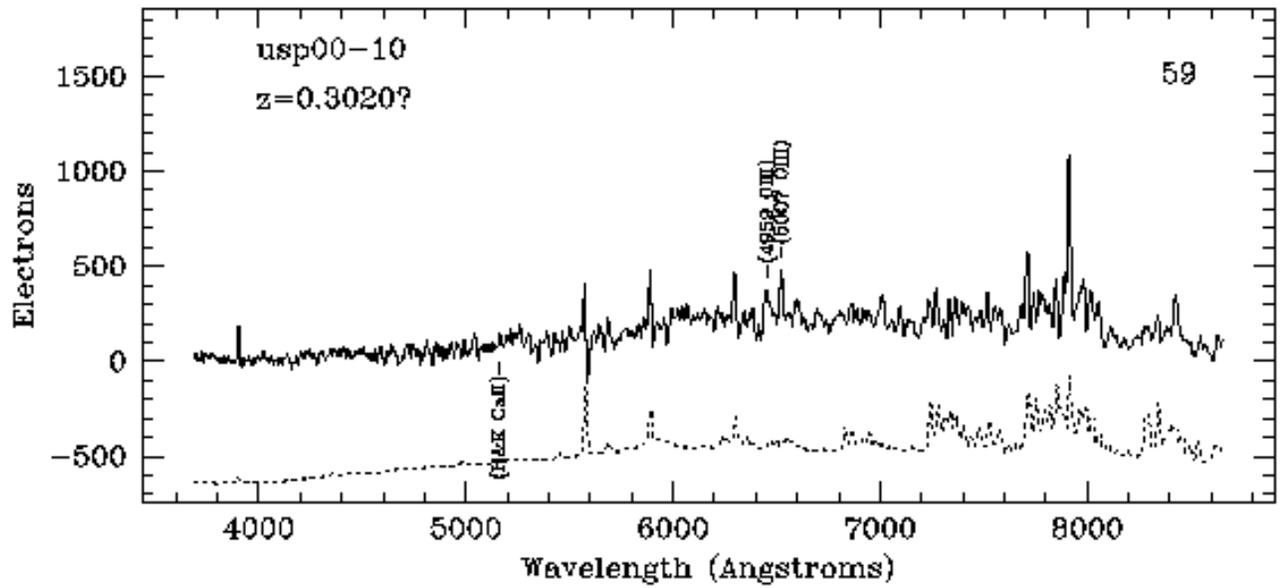
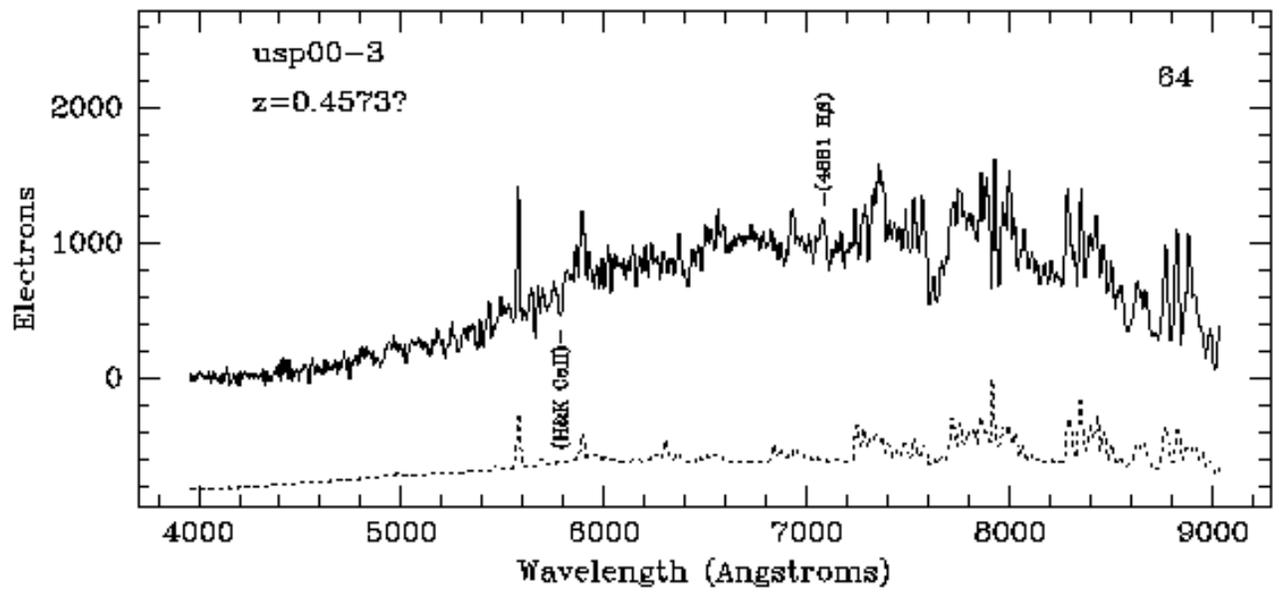
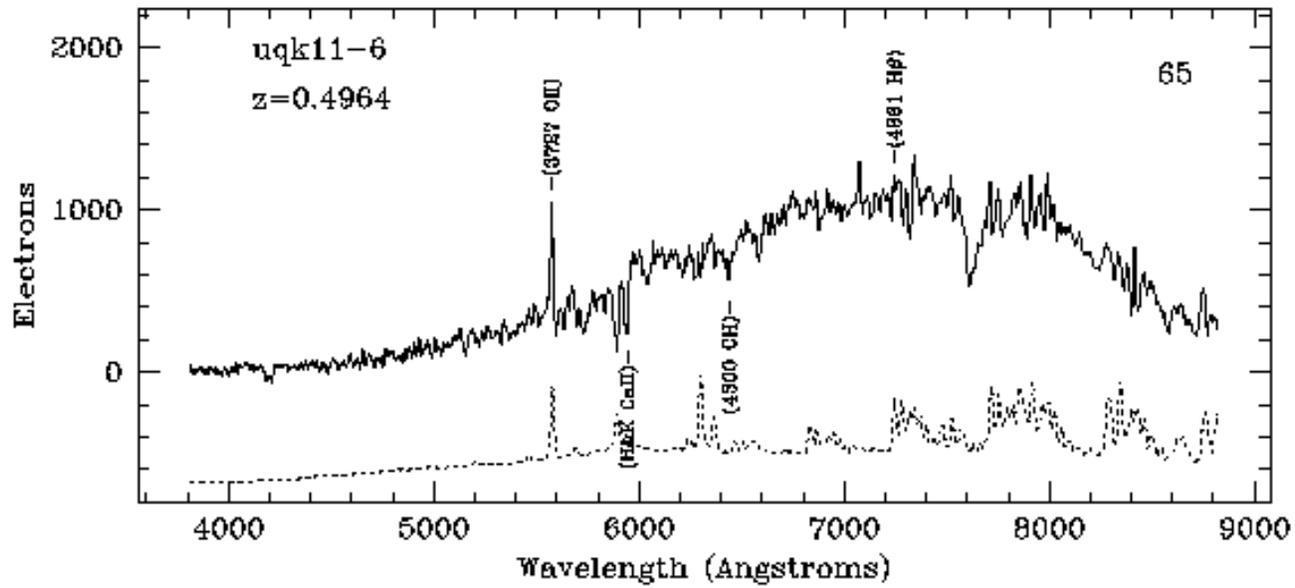
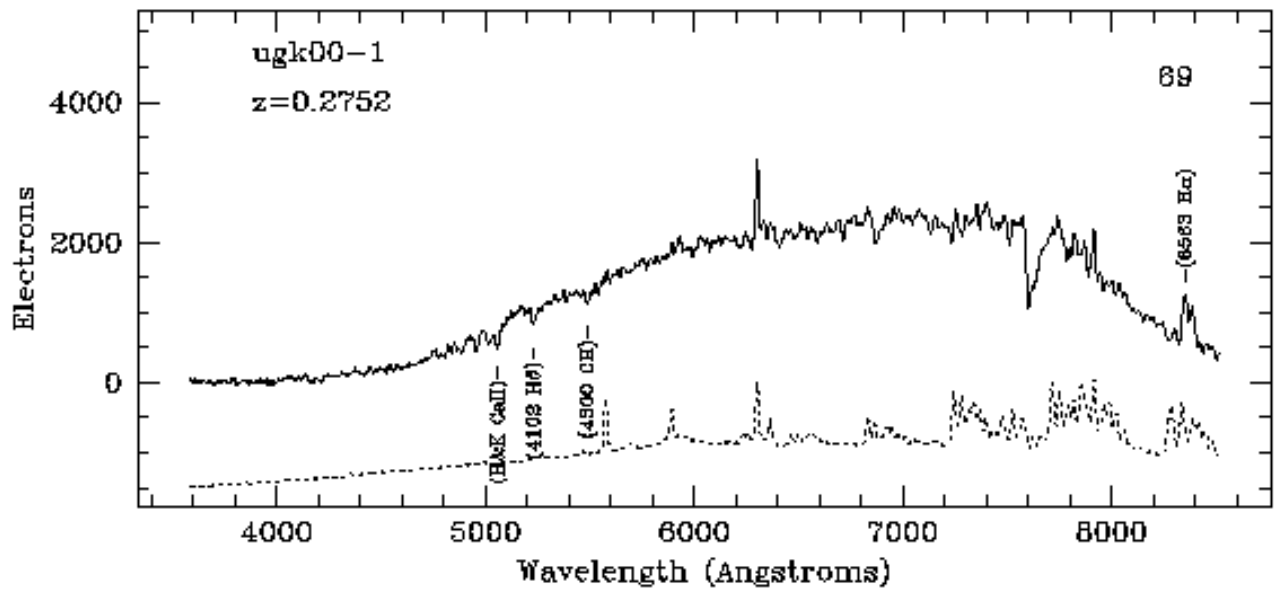
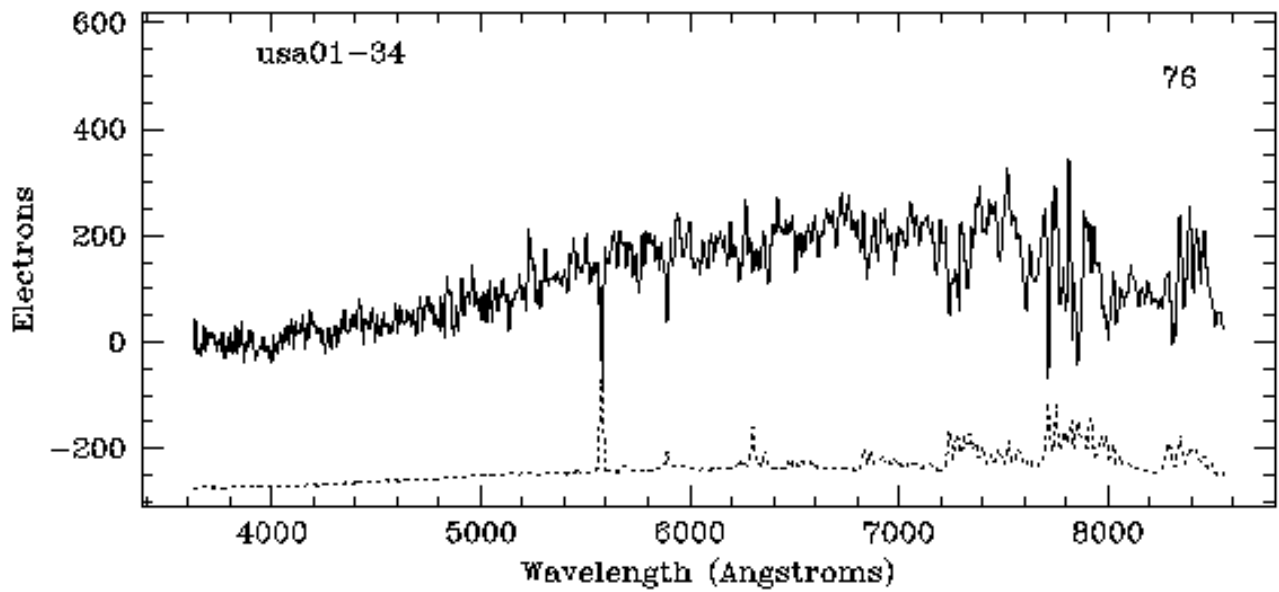
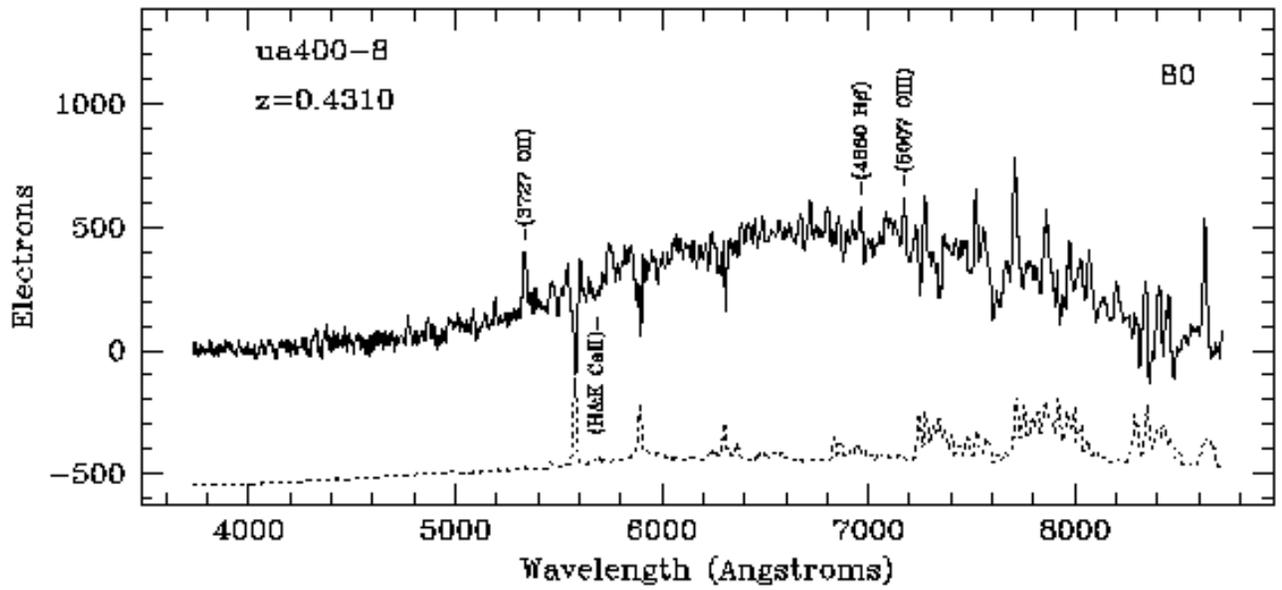
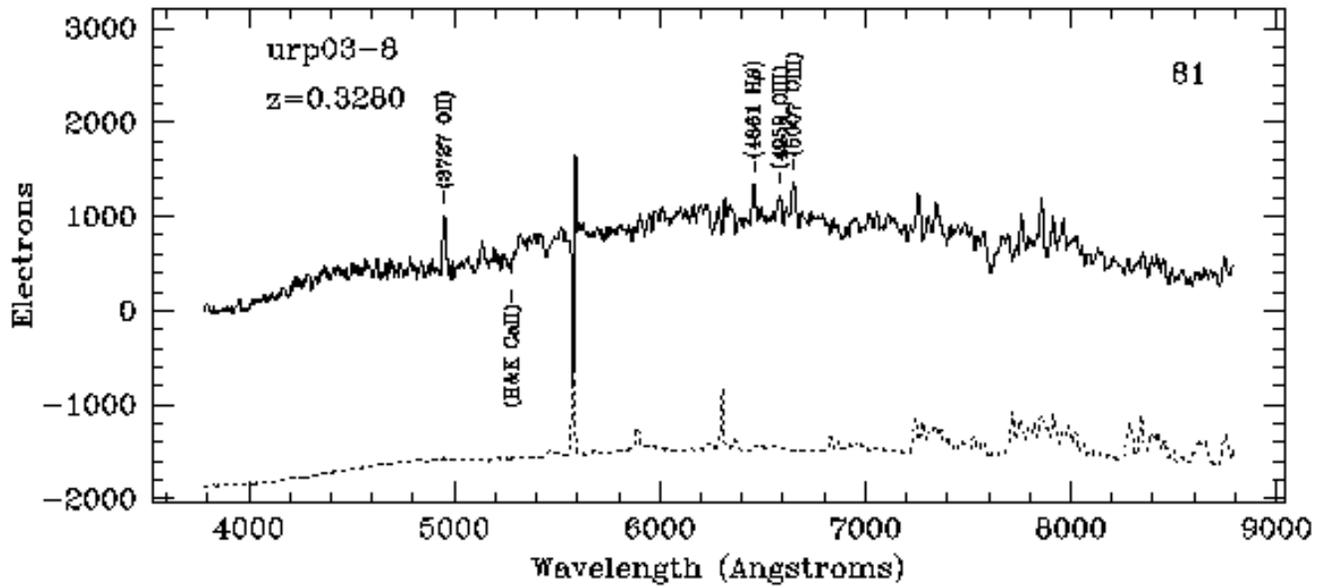
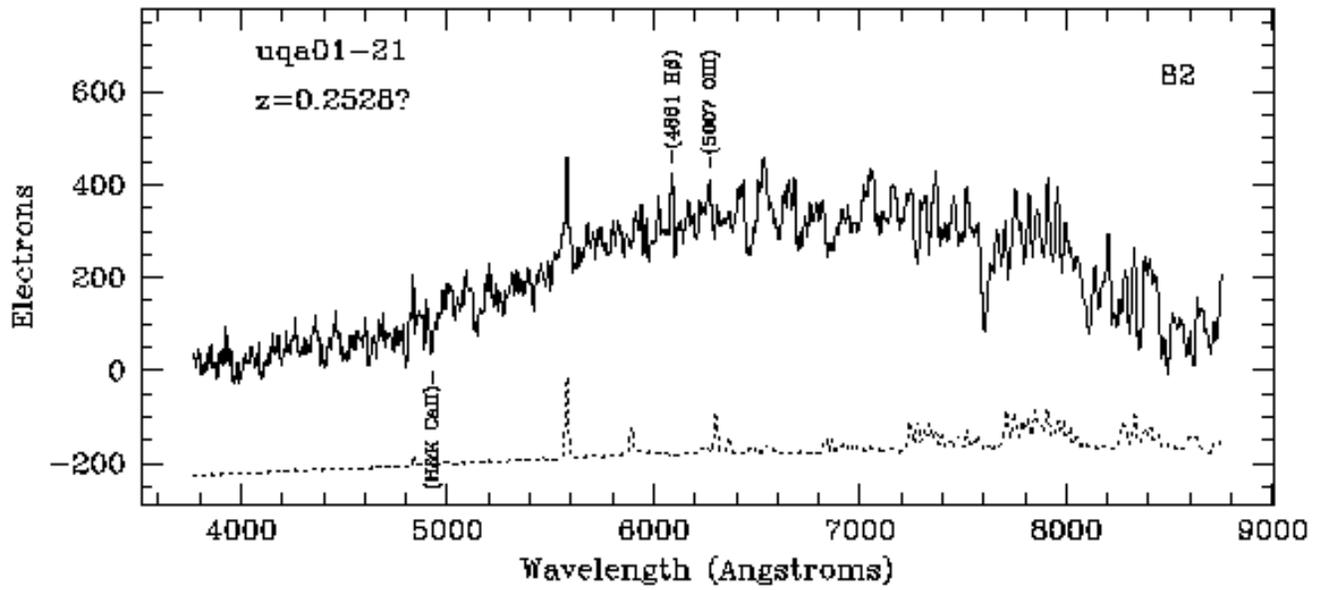
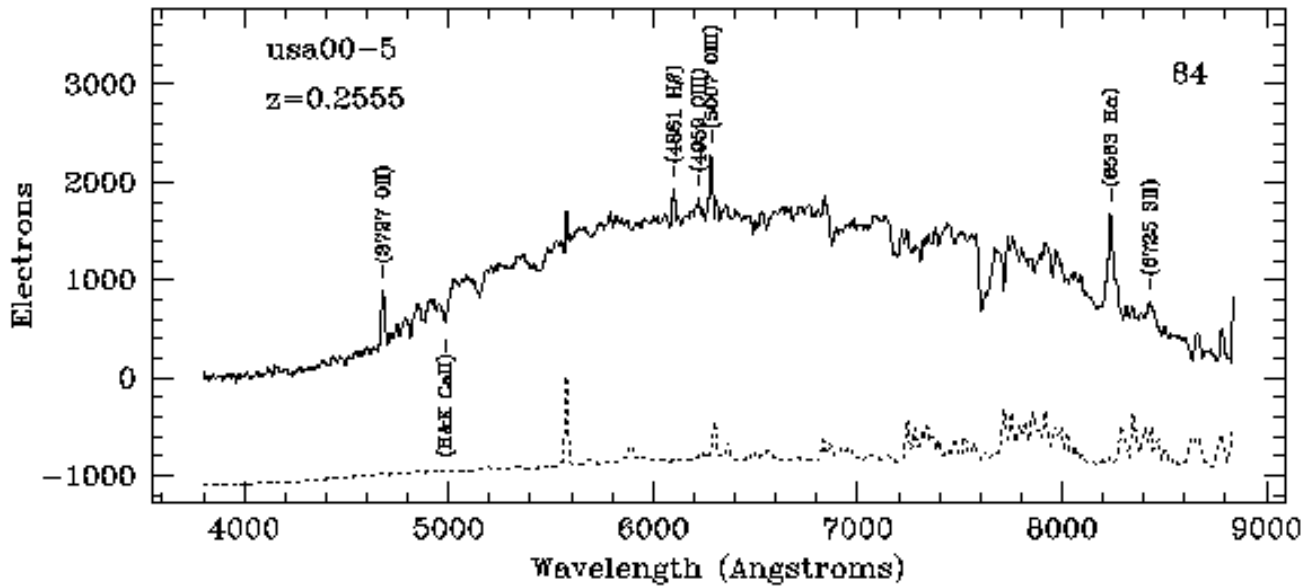
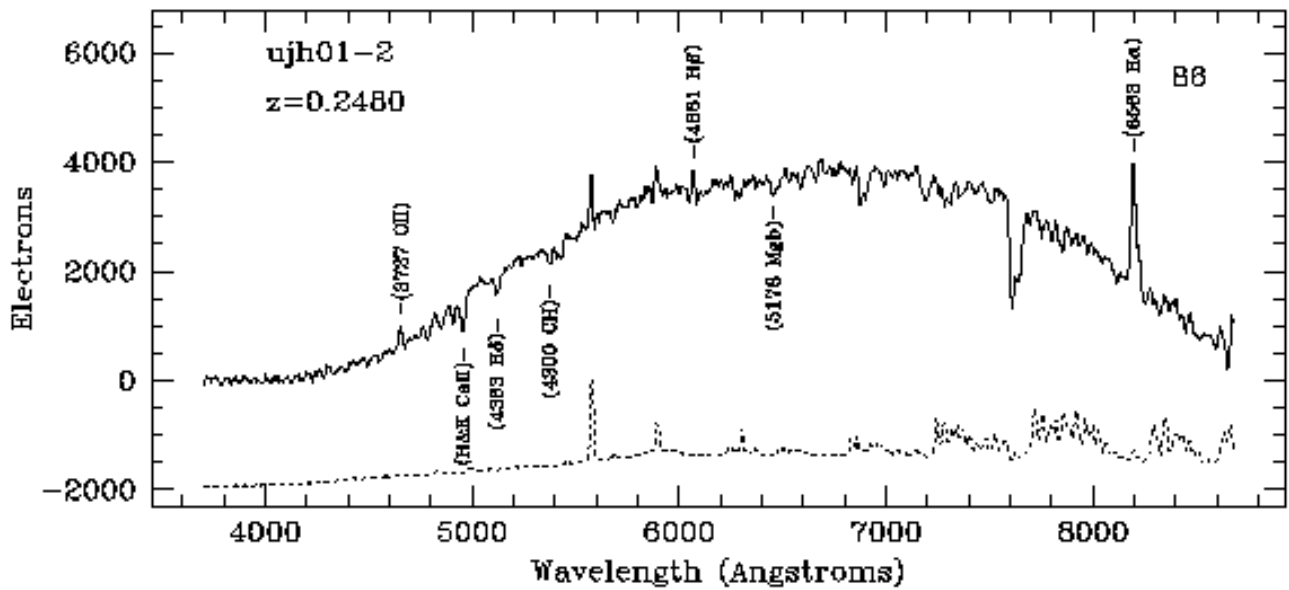
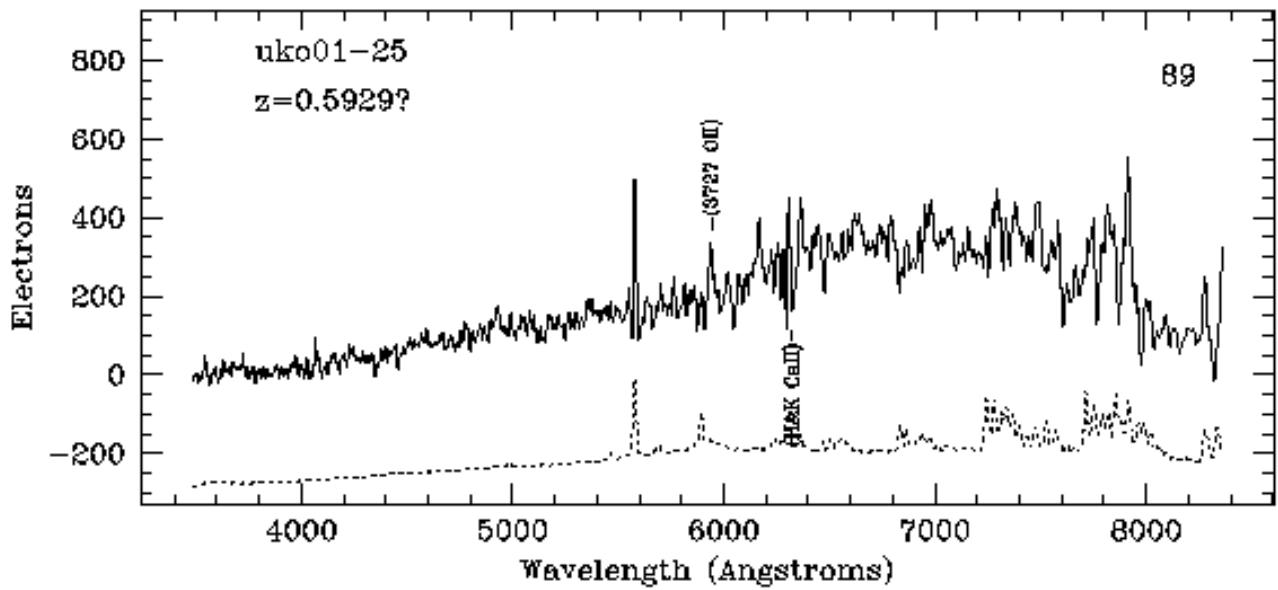
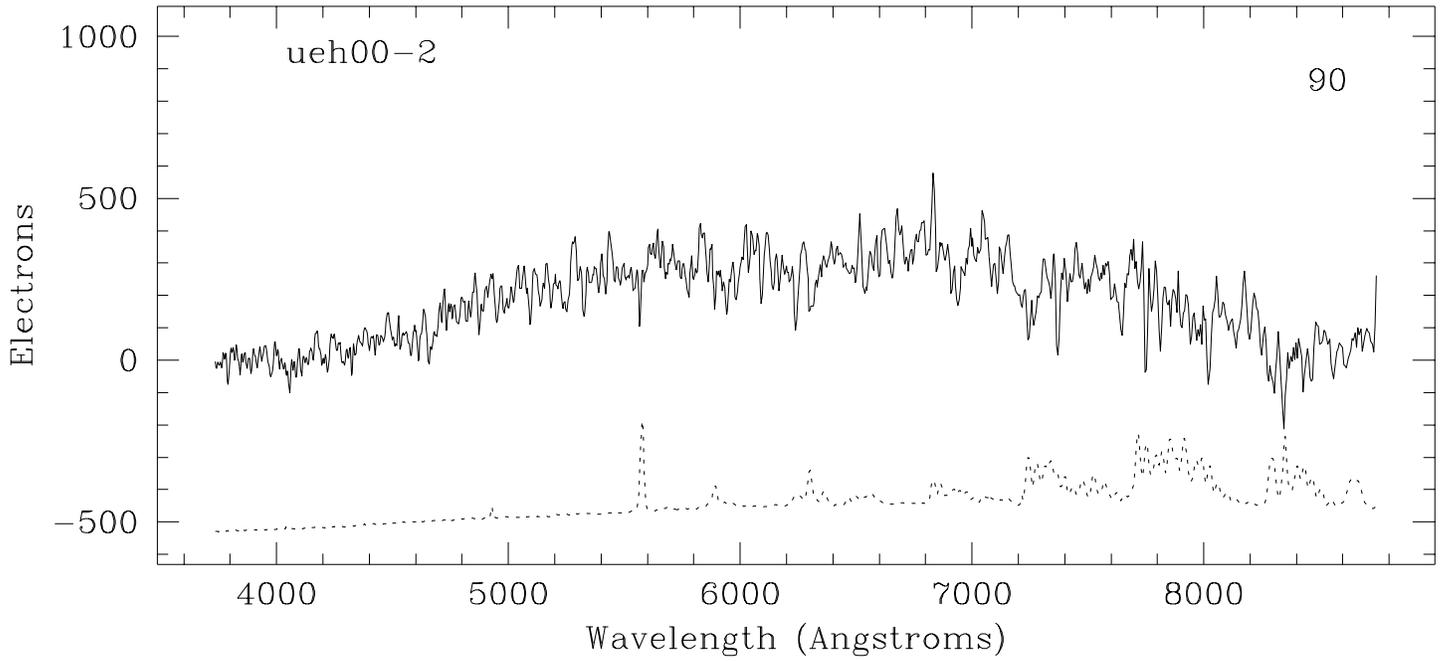
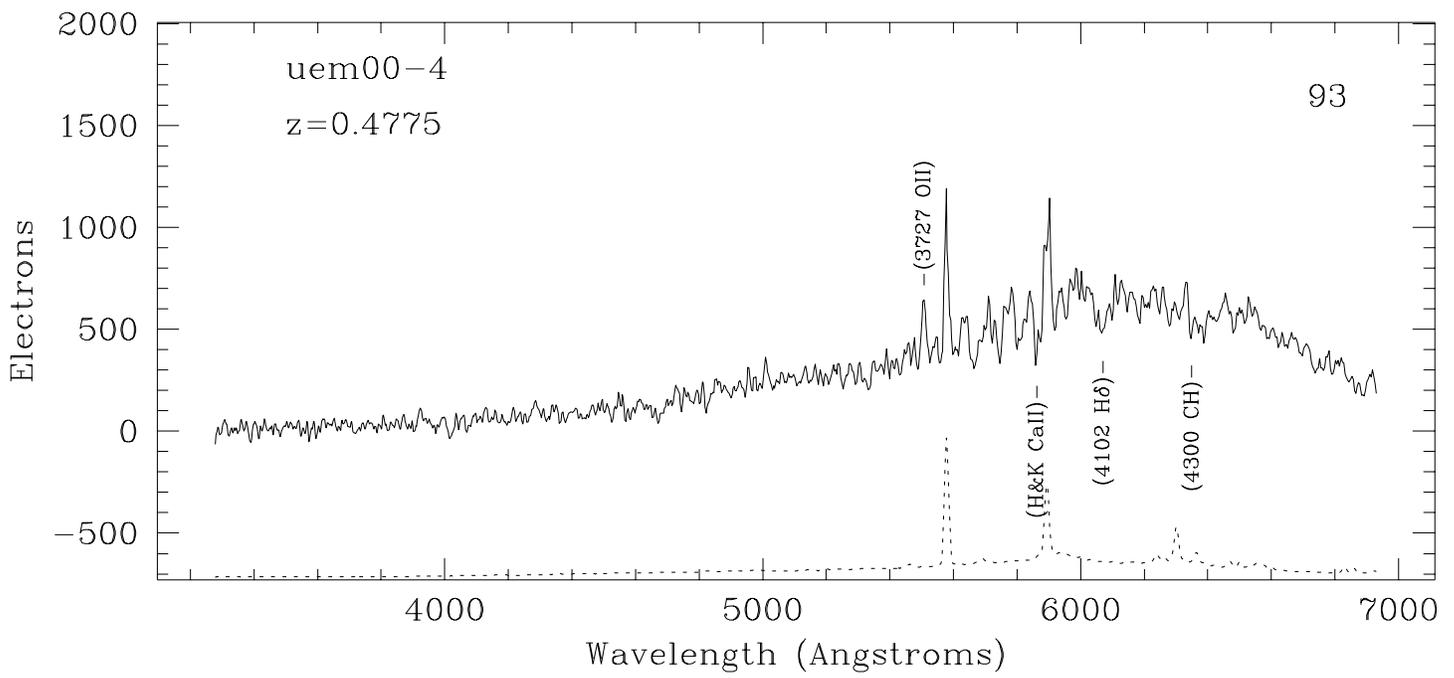
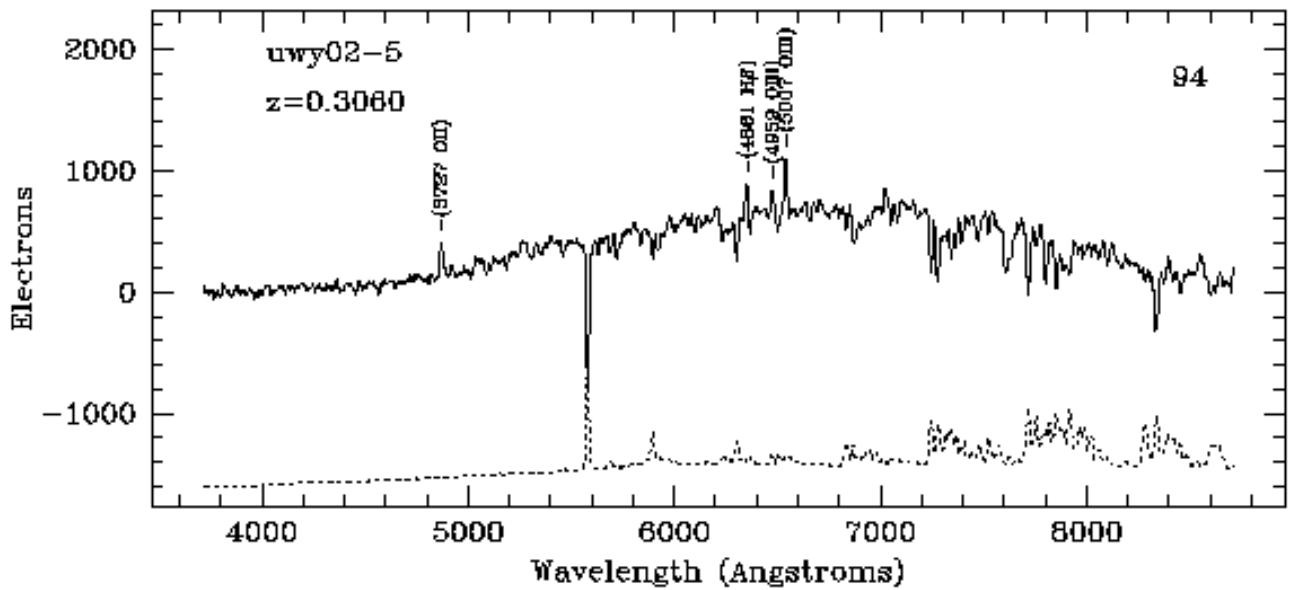
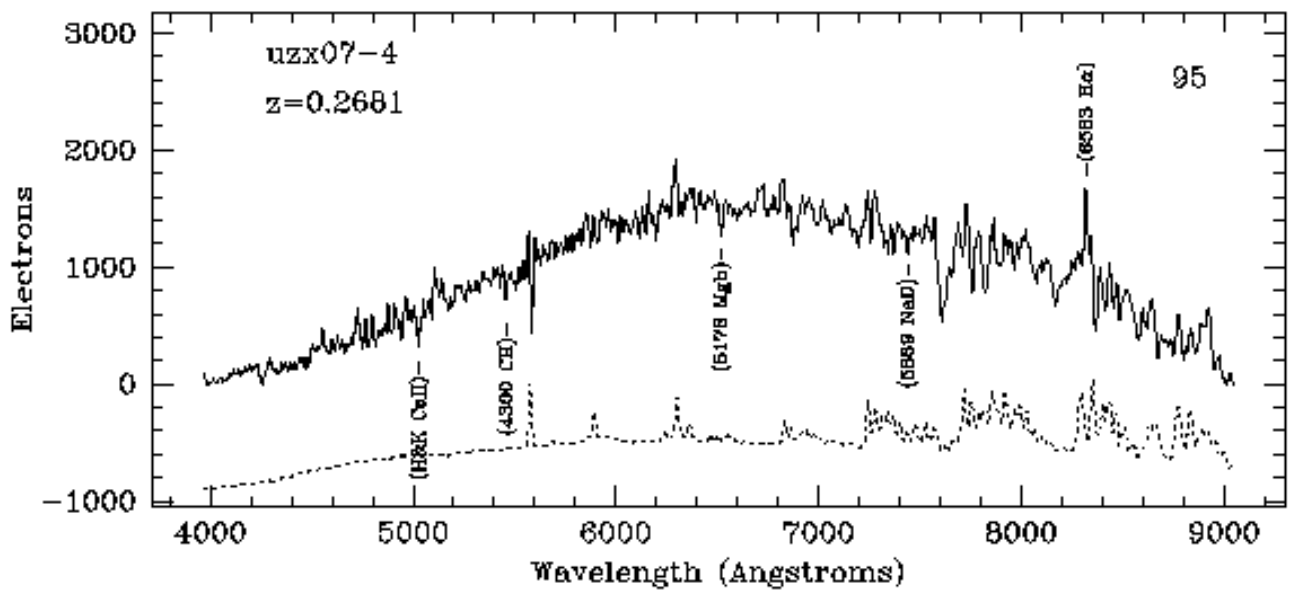
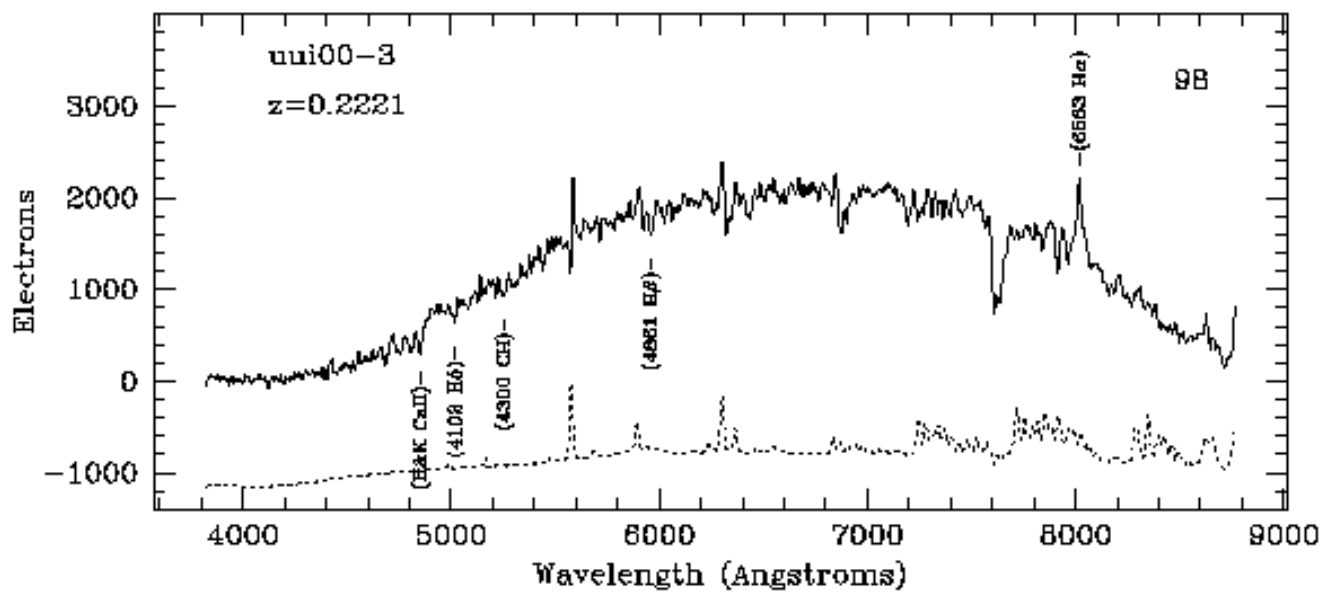
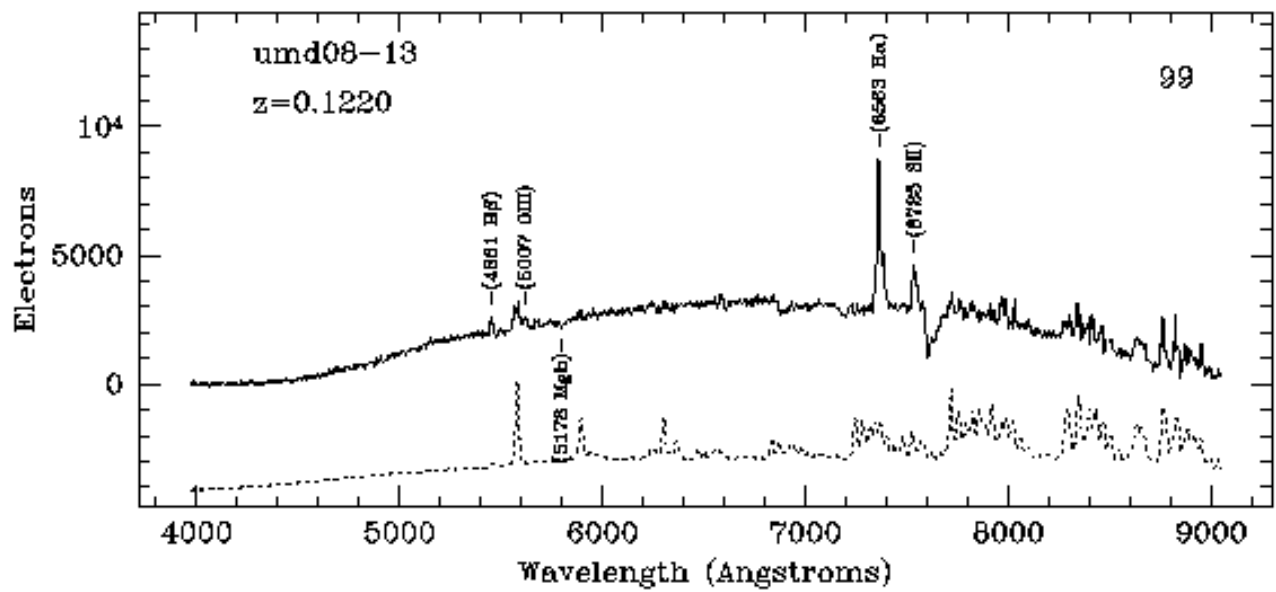
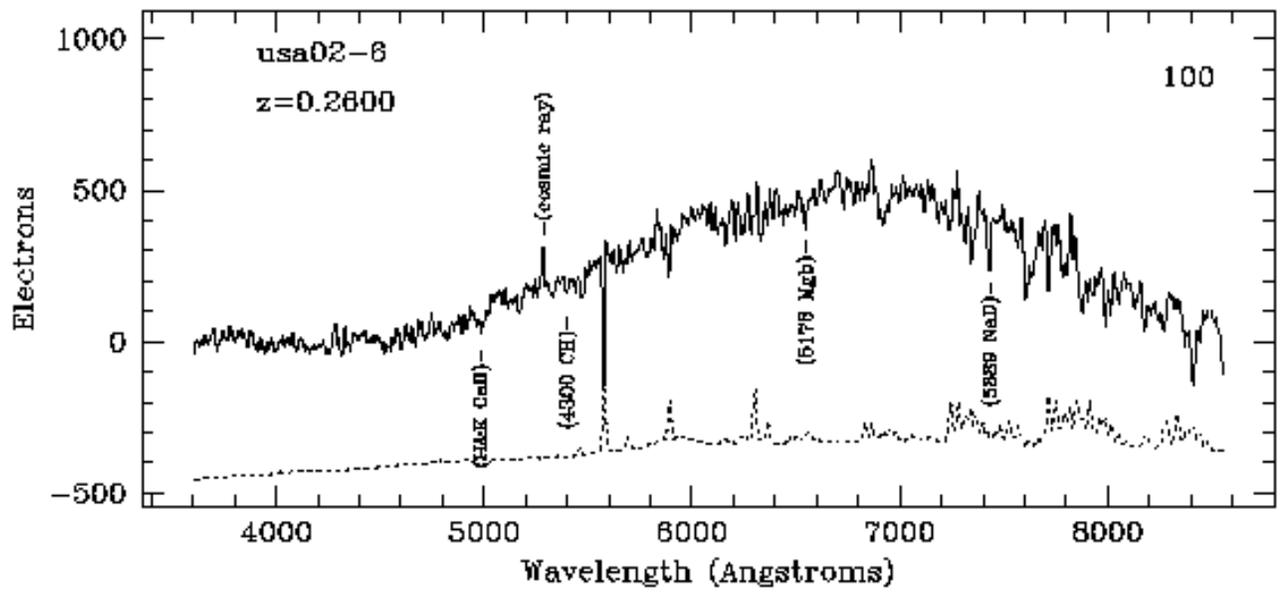
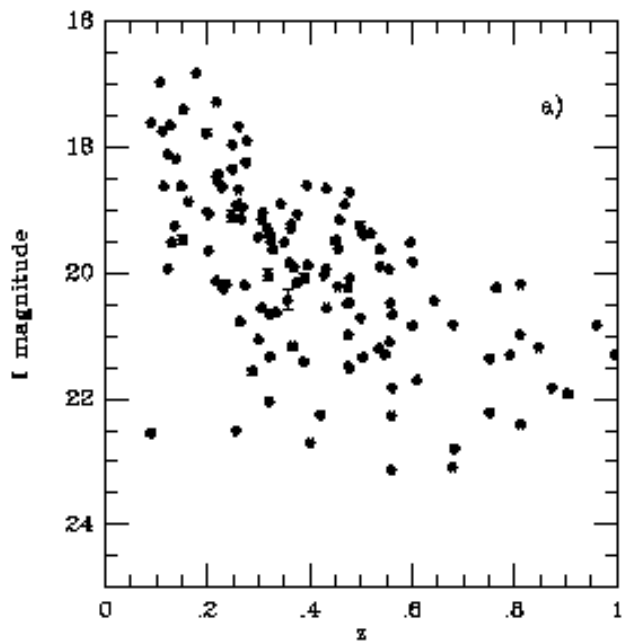
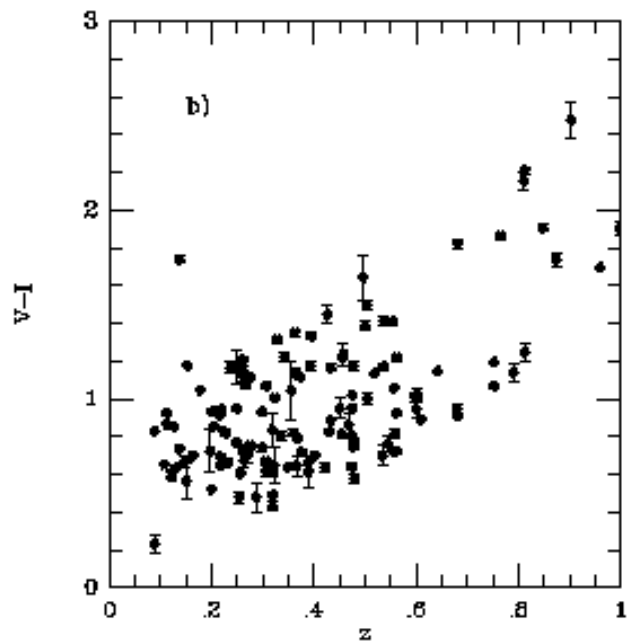
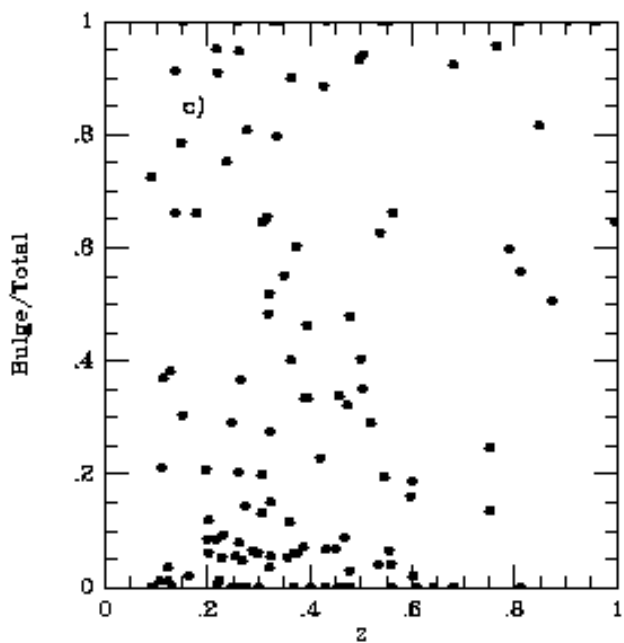
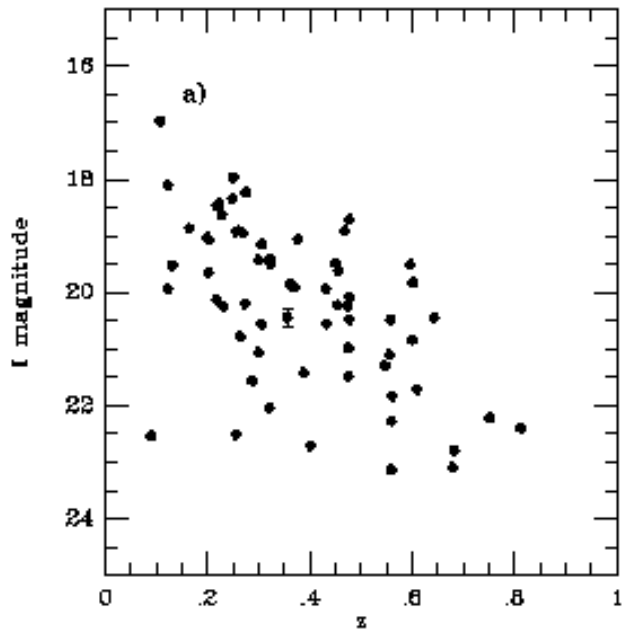
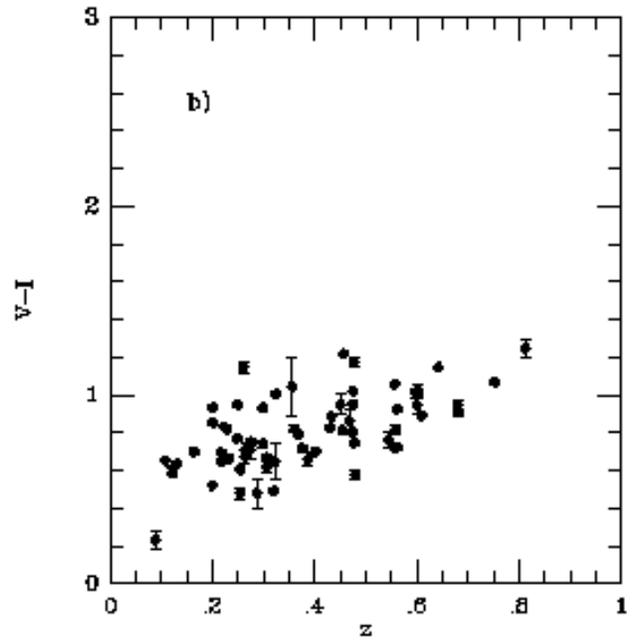
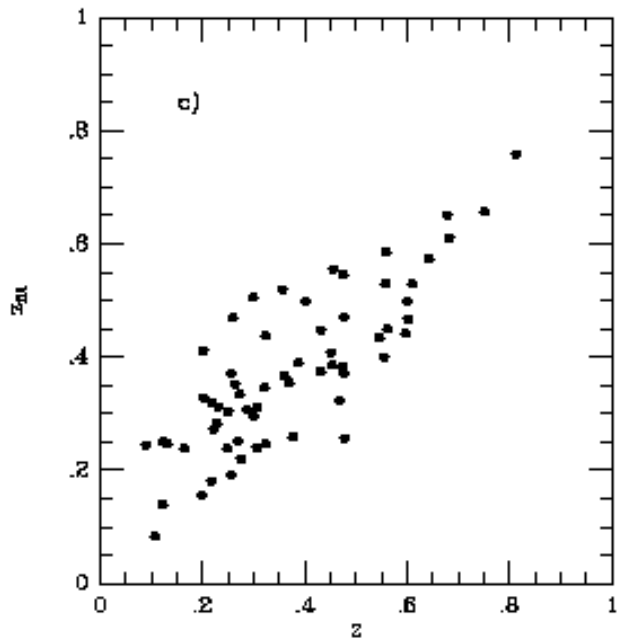
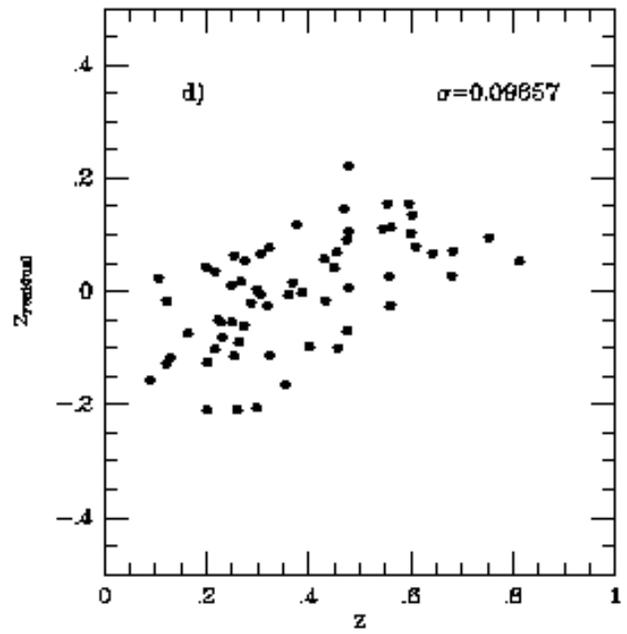
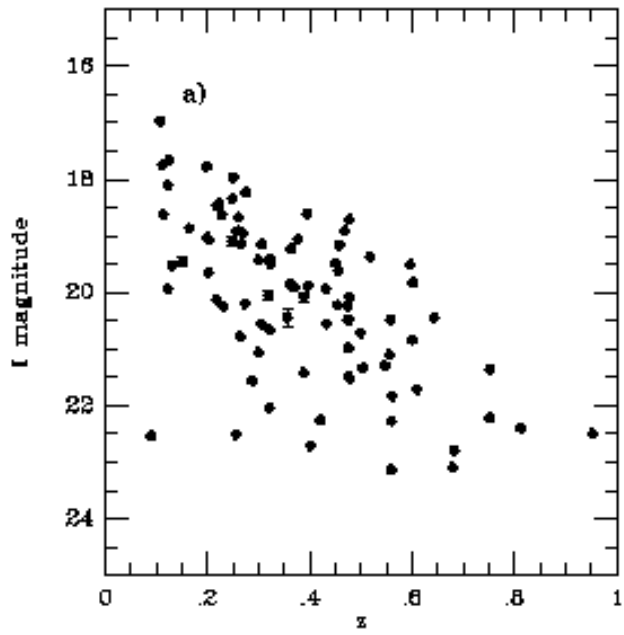
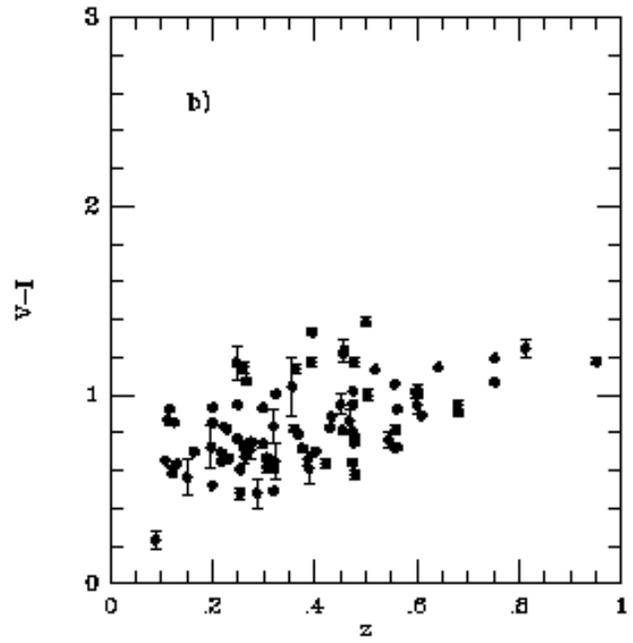
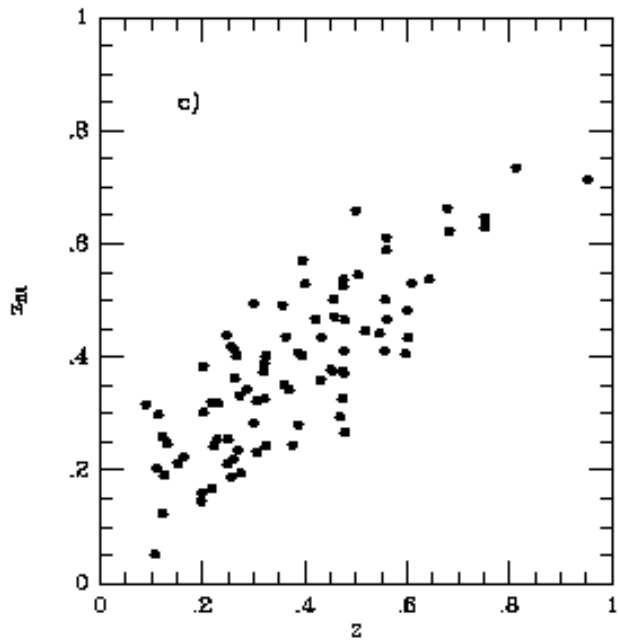
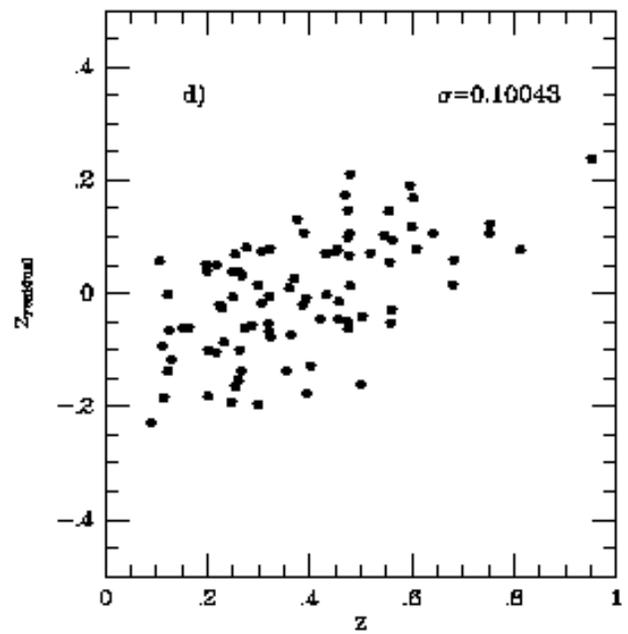
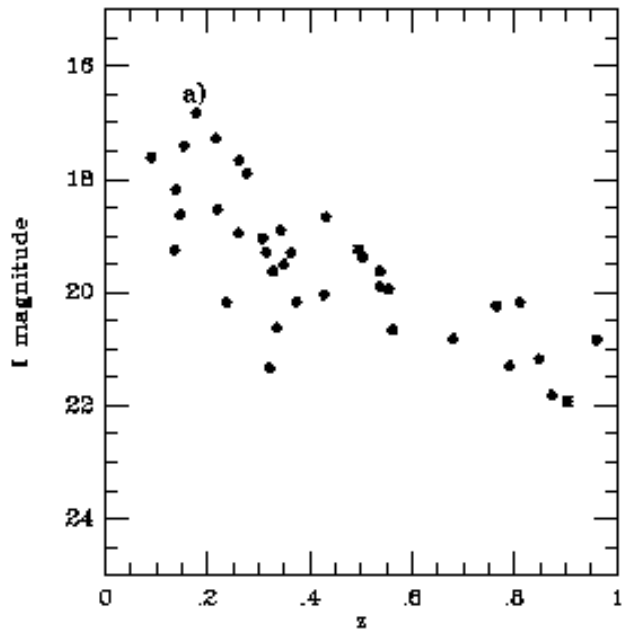
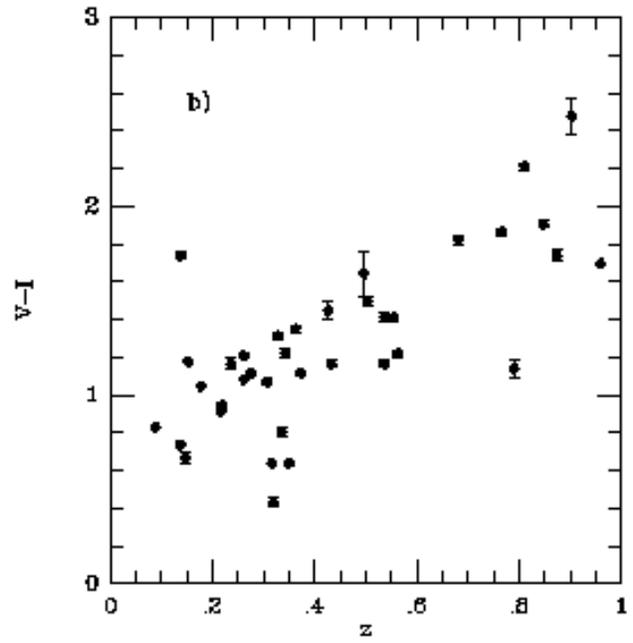
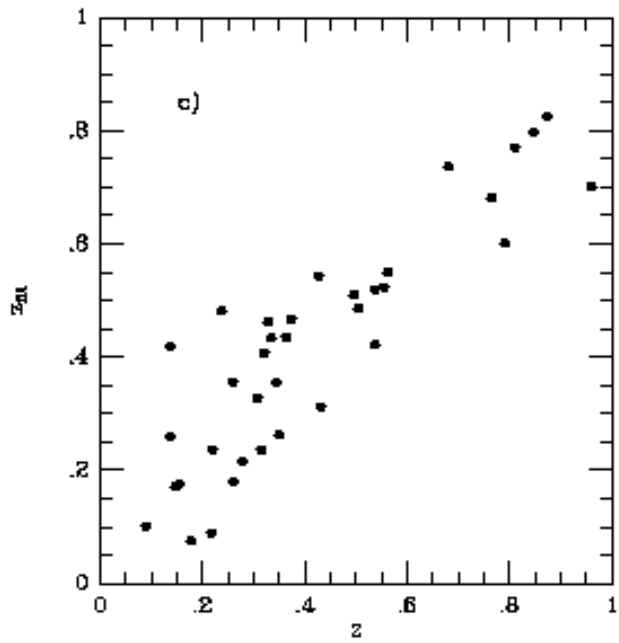
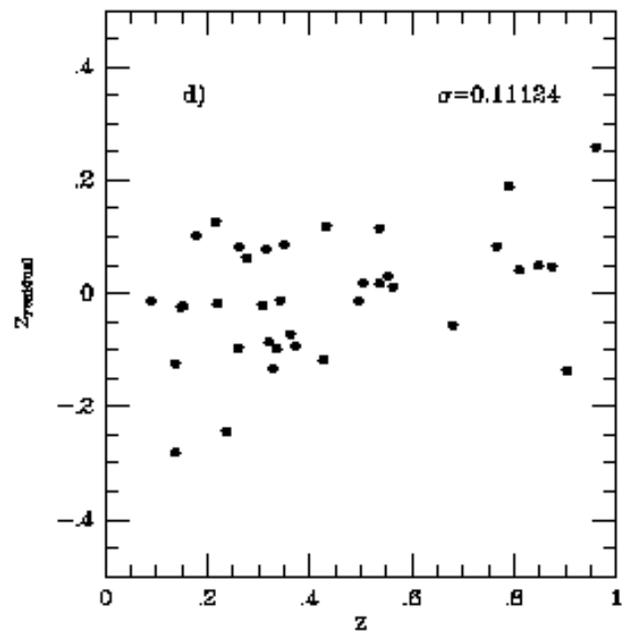
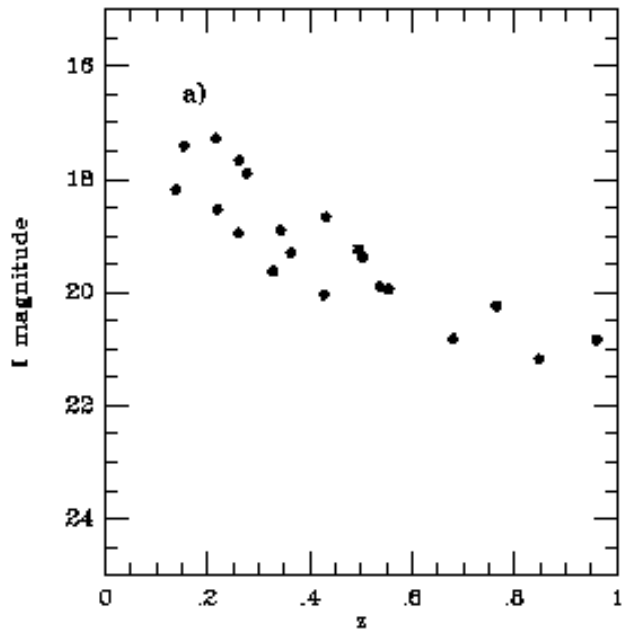
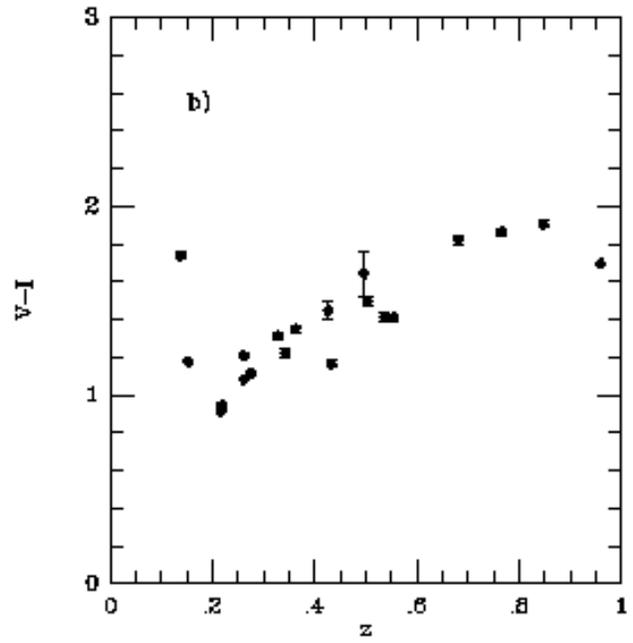
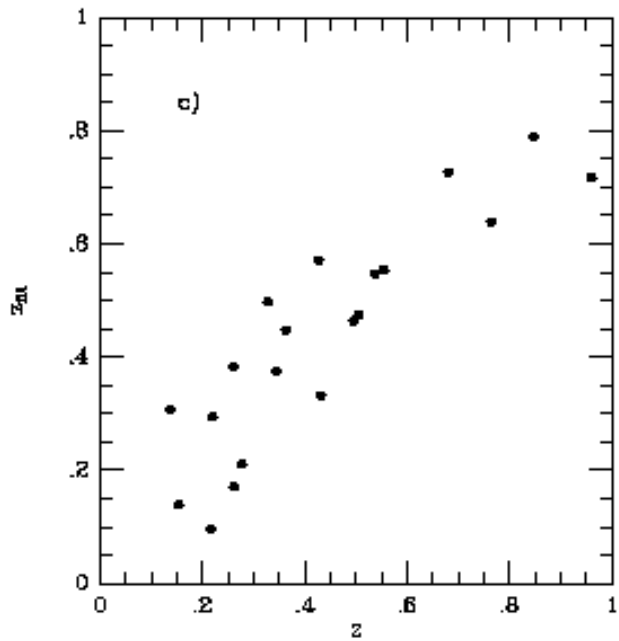
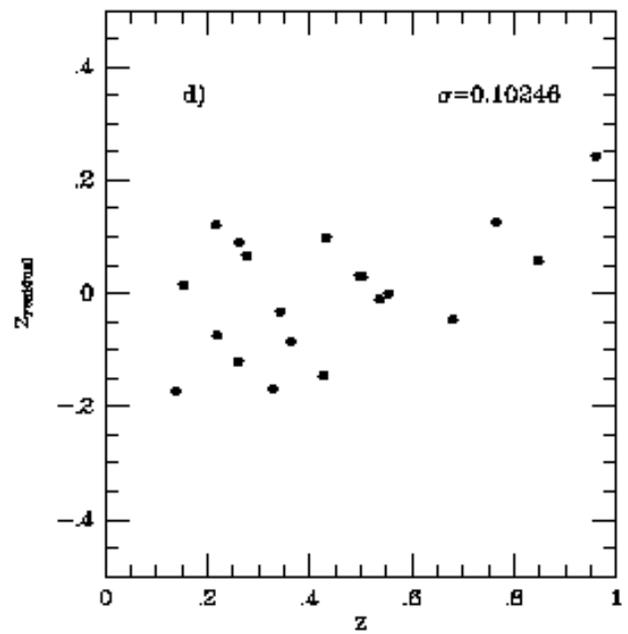
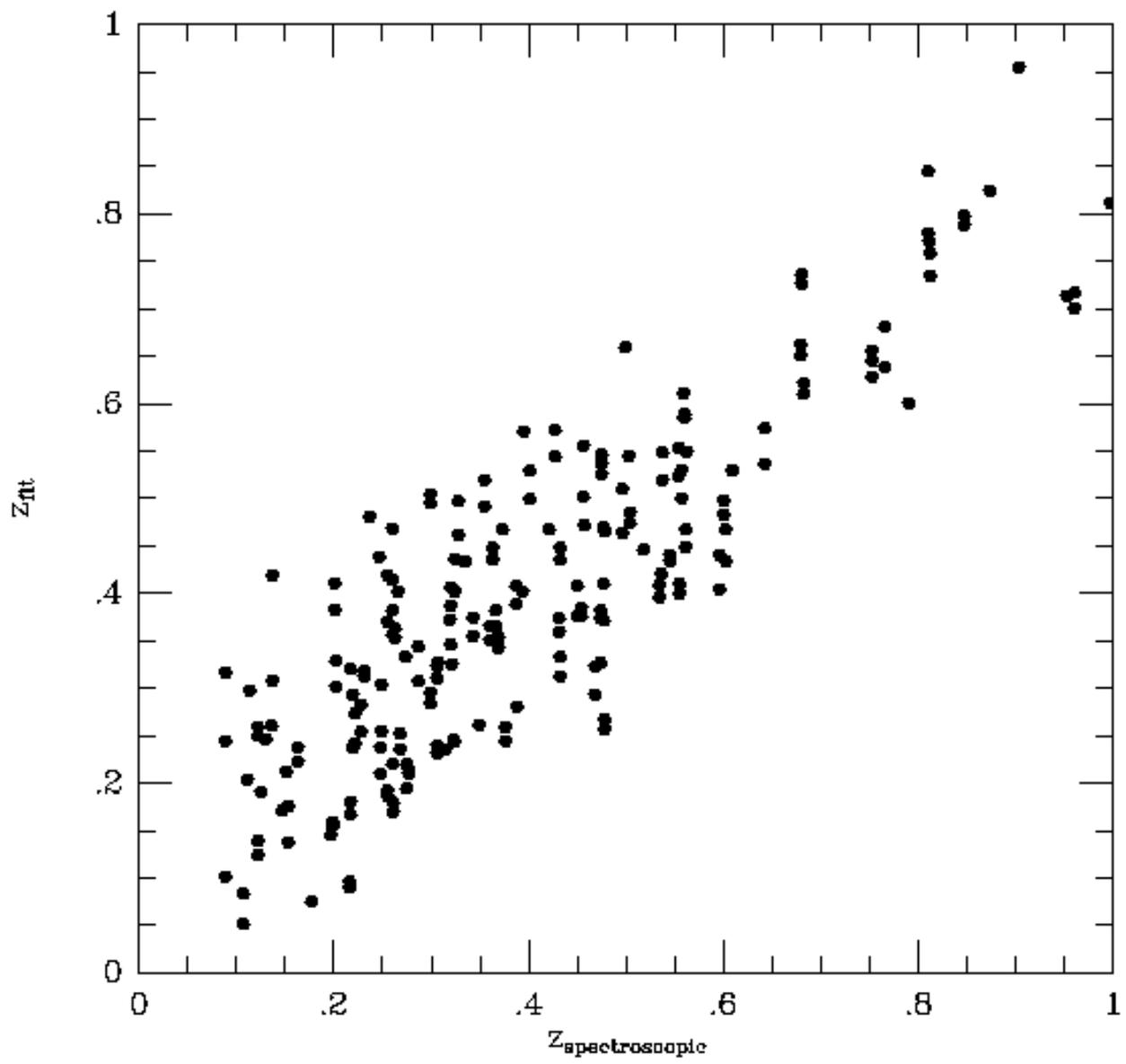
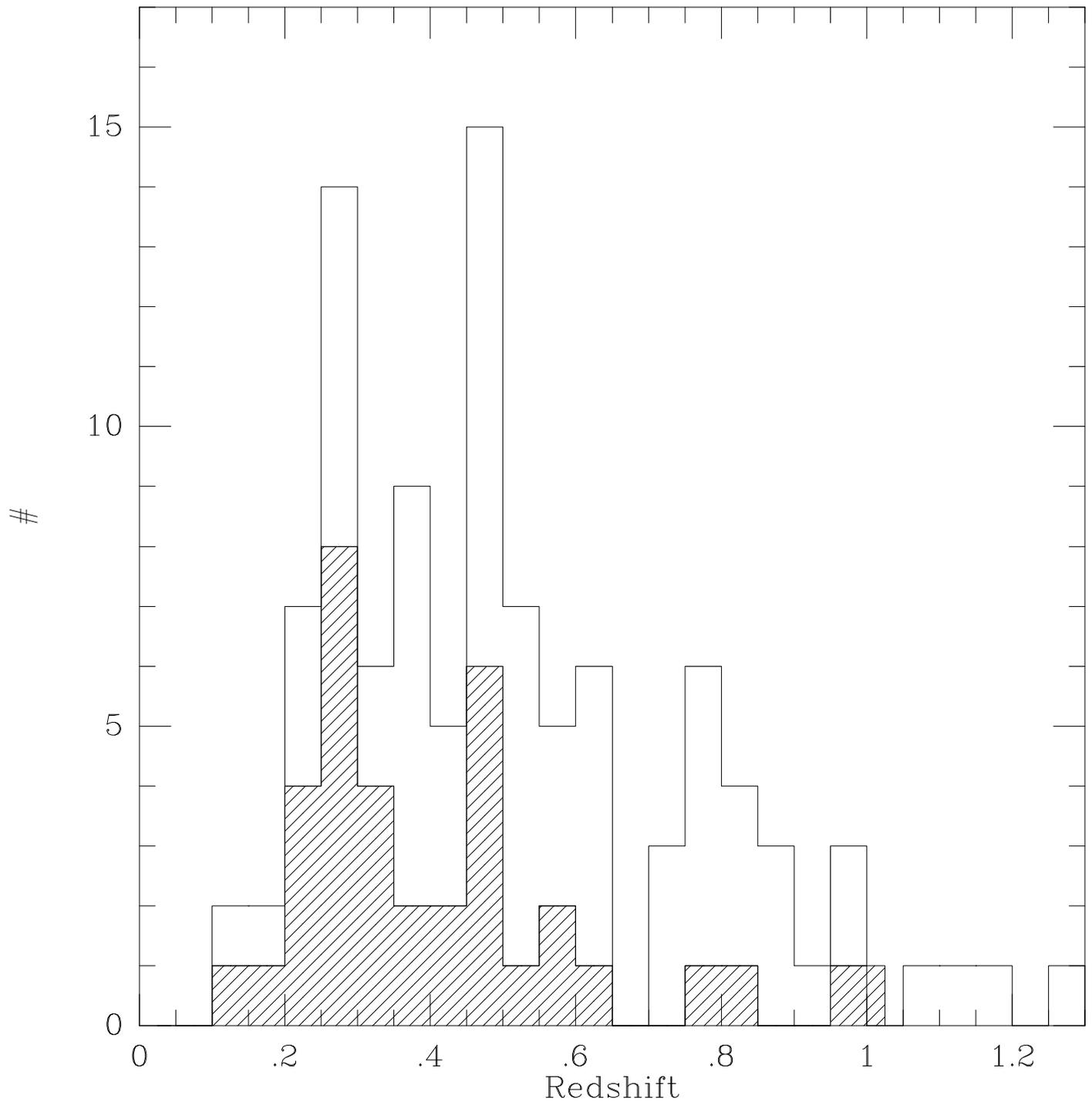
\end{document}